\documentclass[prd,
a4paper,
nofootinbib,
onecolumn,
reprint,
eqsecnum,
]{revtex4}
\usepackage{multirow}
\usepackage{graphicx}
\usepackage{amsmath}
\usepackage{amsfonts}
\usepackage{amssymb}
\usepackage{slashed} 
\usepackage{color}
\usepackage{hyperref}
\usepackage{units}
\def\eq#1{{Eq.~(\ref{#1})}}
\def\eqs#1#2{{Eqs.~(\ref{#1})--(\ref{#2})}}

\def\eg{{\it e.g.}}

\definecolor{oucrimsonred}{rgb}{0.6, 0.0, 0.0}
\definecolor{persianblue}{rgb}{0.11, 0.22, 0.73}
\definecolor{forestgreen}{rgb}{0.13,0.35,0.13}
 \hypersetup{colorlinks, citecolor=oucrimsonred, linkcolor=persianblue, urlcolor=oucrimsonred}
\newcommand{\be}{\begin{equation}}
\newcommand{\ee}{\end{equation}}
\newcommand{\bea}{\begin{eqnarray}}
\newcommand{\eea}{\end{eqnarray}}
\newcommand{\nn}{\nonumber}

\begin{document}
%%%%%%%%%%%%%%%%%%%%%%%%%%%%%%%%%%%%%%%%%%%%%%%%%%%%%%%%%%%  FRONT PAGE
\title[]{Phenomenological consequences \\ 
of an interacting multicomponent dark sector
}
\date{\today}
\author{Jan Tristram Acu\~{n}a$^{\ast\dag\ddag}$}
\author{Marco Fabbrichesi$^{\dag}$}
\author{Piero Ullio$^{\ast\dag\ddag}$}
\affiliation{$^{\ast}$Scuola Internazionale Superiore di Studi Avanzati (SISSA), via Bonomea 265, 34136 Trieste, Italy}
\affiliation{$^{\dag}$INFN, Sezione di Trieste, via  Valerio 2, 34127 Trieste, Italy }
\affiliation{$^{\ddag}$Institute for Fundamental Physics of the Universe (IFPU), via Beirut 2,  34151 Trieste, Italy}
%%%%%%%%%%%%%%%%%%%%%%%%%%%%%%%%%%%%%%%%%%%%%%%%%%%
\begin{abstract}
\noindent  
We consider a dark sector model containing stable fermions charged under an unbroken $U(1)$ gauge interaction, with a massless dark photon as force carrier, and interacting with ordinary matter via scalar messengers. We study its early Universe evolution by solving a set of coupled Boltzmann equations that track the number density of the different species, as well as entropy and energy exchanges between the dark and visible sectors. Phenomenologically viable realizations include: i) a heavy (order 1 TeV or more) lepton-like dark fermion playing the role of the dark matter candidate, with various production mechanisms active depending on the strength of the dark-visible sector portal; ii) light (few GeV to few tens of GeV) quark-like dark fermions, stable but with suppressed relic densities; iii) an extra radiation component in Universe due to dark photons, with temperature constrained by cosmic microwave background data, and in turn preventing dark fermions to be lighter than about 1~GeV. Extra constraints on our scenario stem from dark matter direct detection searches: the elastic scattering on nuclei is driven by dipole or charge radius interactions mediated by either Standard Model or dark photons, providing long-range effects which, however, are not always dominant, as usually assumed in this context. Projected sensitivities for next-generation detectors cover a significant portion of the viable parameter space and are competitive with respect to the model-dependent constraints derived from the magnetic dipole moments of leptons and cooling of stellar systems.
\end{abstract}
%%%%%%%%%%%%%%%%%%%%%%%%%%%%%%%%%%%%%%%%%%%%%%%%%%%%%%%%%%%%%%%%%%%
\maketitle
%%%%%%%%%%%%%%%%%%%%%%%%%%%%%%%%%%%%%%%%%%%%%%%%%%%%%%%%%%%%%%%%%%%

%%%%%%%%%%%%%%%%%%%%%%%%%%%%%%%%%%%%%%%%%%%%%%%%%%%
\section{Motivations and synopsis}
\label{sec:intro}
%%%%%%%%%%%%%%%%%%%%%%%%%%%%%%%%%%%%%%%%%%%%%%%%%%%

The existence of a multicomponent \textit{dark sector} has been extensively discussed in the literature (see~\cite{Deliyergiyev:2015oxa,Alexander:2016aln} for two recent reviews). Such framework generally includes many new states with no direct interactions with the Standard Model (SM) particles, but possibly interacting among themselves by means of new forces. Motivations for this construction have been put forward in a variety of different contexts, ranging, e.g., from beyond SM physics in connection to collider data and flavor anomalies, to explaining the nature of the dark matter component of the Universe, and to addressing possible shortcomings in the SM of cosmology.

In particular, regarding the dark matter problem, any non-relativistic stable dark state can potentially contribute to the Universe's matter budget. Because of the secluded nature of the dark sector which prevents large couplings to ordinary matter, these states automatically satisfy observational properties for dark matter, mostly derived under the assumption that the only relevant interaction between dark and ordinary matter is gravity. On the other hand, given the complexity of the dark sector, the phenomenology of dark matter candidates in this context could be richer than simply looking at gravitational effects. For example, dark matter itself could be multicomponent or in composite forms; dark sector interactions may lead to macroscopic effects and, for instance, impact on the paradigm in the SM of cosmology that dark matter should be described as a collisionless fluid. 

In this paper we illustrate the interplay among different effects occurring when the dark sector contains several species. More explicitly, we will discuss the early Universe's thermal history  in such a scenario and the generation of dark matter and other stable relics. One peculiarity is the fact that there are two reservoirs of states, ordinary and dark, and their temperatures are not necessarily the same. Therefore, a set of coupled Boltzmann equations, tracking at the same time the number density of the different species and the energy exchanges between the two sectors, needs to be considered.

To investigate explicitly this issue, we must first commit ourselves to a specific model of the dark sector (which we do in  section \ref{sec:model} by considering a rather minimal setup). The choice of  model provides an explicit  spectrum of states within the dark sector, the interaction strengths among the dark states, and the strength of the portal interaction between the dark and SM states. These must be supplied in order to extract definite predictions.  In particular, we shall assume that the dark force is long range, that is mediated by an unbroken $U(1)$ gauge interaction. Regarding the particle content, besides the force carrier, a massless dark photon, we introduce a set of stable dark fermions charged under the $U(1)$. One of these may account for most of dark matter in the Universe since it is rather heavy, at the TeV scale or above, and passes upper limits from self-interaction effects~\cite{Ackerman:mha,CyrRacine:2012fz,Agrawal:2016quu}. The others are much lighter, have suppressed relic abundances, but concur in determining the ratio between dark and visible photon temperatures at late times; such ratio is constrained by cosmic microwave background (CMB) data, given that dark photons contribute as an extra radiation component to the Universe's dynamics. In this respect, the role of portal interactions between dark and visible sectors is also important: we consider scalar messengers mediating Yukawa-like interactions. The latter are also crucial for selecting the mechanism for dark matter generation and final relic densities. Such interplay is discussed in detail in section \ref{sec:relic}.

Direct detection, namely the attempt to measure nuclear recoils induces by dark matter scatterings, is one of the main tools to test a given dark matter scenario. In our framework, the direct-detection cross section is mostly driven, via loop induced magnetic dipole and charge radius interactions, by the massless mediators, SM and dark photons. While long-range interactions are present and boost the recoil spectrum at low recoil energies, the correlated contact terms are also contributing to the cross section and may be dominant (contrary to standard lore that contact interactions can be neglected in the presence of long-range effects). These aspects are illustrated in section \ref{sec:direct}, bridging also between astrophysical, cosmological, and high-energy observables and relative constraints, demonstrating once more the diversity of the phenomenological implications of introducing such a multicomponent dark sector.

%%%%%%%%%%%%%%%%%%%%%%%%%%%%%%%%%%%%%
\section{A  model of the dark sector}
\label{sec:model}
%%%%%%%%%%%%%%%%%%%%%%%%%%%%%%%%%%%%%%%%%%%%%%%%%%%

Several dark sector models have been studied in the literature and they are usually classified~\cite{Alexander:2016aln}  according to the  portal through which they  interact with ordinary matter. We  consider a model  consisting of dark  fermions that are, by definition, singlets under the SM gauge interactions. These dark fermions interact with the visible sector through a portal provided by scalar messengers which carry both SM and dark-sector charges. These scalars  are phenomenologically akin to the sfermions  of supersymmetric models.

In general, we can have  as many dark fermions as there are in the SM;  they can  be classified conveniently  according  to whether they couple (via the corresponding messengers) to quarks ($q_L$, $u_R$, $d_R$)  or leptons ($l_L$, $e_R$): we denote  the former (hadron-like) $Q$ and the latter (lepton-like) $\chi$.  The Yukawa-like interaction Lagrangian can  be written as~\cite{Gabrielli:2013jka,Gabrielli:2016vbb}:
\be
 \mathcal{L} \supset -g_L \left(\phi^\dag_L \bar{\chi}_R l_L + S_L^{U\dag} \bar{Q}^U_R q_L + S_L^{D\dag} \bar{Q}^D_R q_L\right)
- g_R \left(\phi^\dag_R \bar{\chi}_L e_R + S_R^{U\dag} \bar{Q}^U_L u_R + S_R^{D\dag} \bar{Q}^D_L d_R\right) + \text{h.c.} \label{LLRR} \, .
\ee
The $L$-type scalars are doublets under SU(2)$_L$, while the $R$-type scalars are singlets under SU(2)$_L$. The $S_{L,R}$ messengers carry color indices (unmarked in (\ref{LLRR})), while the messengers $\phi_{L,R}$ are color singlets.  The Yukawa coupling strengths  are parameterized by $\alpha_{L,R}\equiv g_{L,R}^2/(4 \pi)$; they can be different for different fermions and as many as the  SM fermions. 

In order to generate chirality-changing processes, we must  have the mixing terms
 \be
\mathcal{L} \supset -\lambda_s S_0\left(H^\dag \phi_R^\dag \phi_L + \tilde{H}^\dag S_R^{U\dag} S_L^U + H^\dag S_R^{D\dag} S^D_L\right) + \text{h.c.} 
 \, , \label{mix} 
\ee
where $H$ is the SM Higgs boson, $\tilde{H}=i\sigma_2 H^\star$, and $S_0$ a scalar singlet of the dark sector. 
After both   $S_0$ and $H$   take a vacuum expectation value (VEV)  ($\mu_S$  and $v$---the electroweak VEV---respectively), the Lagrangian in \eq{mix} gives rise to  the mixing between right- and left-handed states.

Dark sector states interact by means of  an unbroken $U(1)_D$ gauge symmetry;  the corresponding massless gauge boson is the dark photon $\gamma_D$ whose coupling strength we denote by $\alpha_D\equiv  g_D^2/(4 \pi)$. We  assign different dark $U(1)_D$ charges to the various dark sector fermions to ensure,  by charge conservation, their stability. There is no kinetic mixing between the ordinary  and the dark photon~\cite{Holdom:1985ag,delAguila:1995rb}. The latter is a distinctive feature of models in which the dark photon is, and remains, massless as opposed to those in which the gauge symmetry is broken and the dark photon is massive.  While there is no tree-level coupling between dark fermions and SM photons, and between ordinary matter and dark photons, the  mixing in \eq{mix}  leads, through one-loop diagrams and therefore operators of dimension larger than four, to an effective coupling of ordinary matter to the dark photon as well as of the dark fermions to the ordinary photon.

When the dark sector scalar $S_0$ and the Higgs boson acquire VEVs,  the scalar messengers must be rotated to identify the physical states.  Considering first the lepton sector, while before the rotation all $\phi$ states have the same mass $m_\phi$, after the rotation we find the mass eigenstates (labeled by $\pm$)
\be
\phi_{\pm} \equiv \frac{1}{\sqrt{2}}\left(\phi_{Le} \pm \phi_R\right) \,,
\ee
with masses $m_{\phi_{\pm}} = m_\phi  \sqrt{1 \pm \eta_s}$, where we defined the mixing parameter:
\be
\eta_s \equiv \frac{\lambda_s \mu_S v}{m_\phi^2}\,.
\ee
We must have $\eta_s < 1$ in order for the $\phi_-$ state to be physical. In the new basis, the  interaction terms in \eq{LLRR} in the lepton sector  is given  by
\be
\label{intdiag}\mathcal{L}^{(lep)} \supset -g_L \phi_{L\nu}^\dag \left(\bar{\chi}_R \nu_L\right) - \frac{g_L}{\sqrt{2}}\left(\phi_+^\dag + \phi_-^\dag\right)\left(\bar{\chi}_R e_L\right)-\frac{g_R}{\sqrt{2}}\left(\phi_+^\dag - \phi_-^\dag\right)\left(\bar{\chi}_L e_R\right) + \text{h.c.}  \, .
\ee
The picture in the hadronic sector is perfectly specular; in the following we will indicate generically with $m_S$ the mass for the eigenstates $S^U_{Ld}$ and $S^D_{Lu}$ before the rotation, and keep $\eta_s$ as mixing parameter for the physical eigenstates:
\be
S^U_{\pm} \equiv \frac{1}{\sqrt{2}}\left(S^U_{Lu} \pm S^U_R\right) \quad \mbox{and} \quad
S^D_{\pm} \equiv \frac{1}{\sqrt{2}}\left(S^D_{Ld} \pm S^D_R\right) \, .
\ee

Looking at (\ref{intdiag}), we can see that for $\chi$ to be a stable dark-sector species, its mass must be at most $m_{\phi_-} + m_e$. Similarly, for a dark-sector species $Q$, the mass must be no heavier than $m_{S_-} + m_q$, where $m_q$ is the mass of the SM species corresponding to $Q$. This sets an upper bound for the mixing $\eta_s$:
\be
\label{mixinglimit}\eta_s < 1 - \left(\frac{M_{\chi,Q}}{m_{\phi,S}}\right)^2  \,,  
\ee
where $M_{\chi,Q}$ stands for the mass of the heaviest stable dark-sector species and $m_{\phi,S}$ for the mass parameter of the corresponding messenger. We assume that $M_{\chi,Q}$ is much heavier than any SM species. The upper bound in \eq{mixinglimit} also guarantees that the scalar messengers are heavier than the dark fermion into which can thus decay.

This model can be considered as a template for many models of the dark sector with the scalar messenger as stand-in for more complicated portals.  It is a simplified version of the model in \cite{Gabrielli:2013jka}, which might  provide a natural solution to the SM flavor-hierarchy problem. It has been  used to predict new decays for the Higgs boson~\cite{Gabrielli:2014oya,Biswas:2015sha,Biswas:2017lyg}, neutral Kaons~\cite{Fabbrichesi:2017vma} and the  $Z$-boson~\cite{Fabbrichesi:2017zsc} as well as  invisible decays for the neutral $K$- and $B$-mesons~\cite{Barducci:2018rlx}. 

Models of self-interacting dark matter charged under Abelian or non-Abelian gauge groups and interacting through the exchange of massless as well as massive  particles have a long history~\cite{Goldberg:1986nk,Holdom:1986eq,Gradwohl:1992ue,Carlson:1992fn,Foot:2004pa,Feng:2008mu,Ackerman:mha,Feng:2009mn,ArkaniHamed:2008qn,Kaplan:2009de,Buckley:2009in,Hooper:2012cw,Aarssen:2012fx,Cline:2012is,Tulin:2013teo,Gabrielli:2013jka,Baldi:2012ua,CyrRacine:2012fz,Cline:2013zca,Chu:2014lja,Boddy:2014yra,Buen-Abad:2015ova,Agrawal:2016quu}.  We have relied in particular on   \cite{Ackerman:mha,Feng:2008mu,Feng:2009mn,Agrawal:2016quu}---the constraints of which we recover in our framework   where dark matter is only a component among the many of the dark sector within the specific underlining model defined by \eqs{LLRR}{mix}. Interacting dark matter can form bound states. The phenomenology of such  atomic dark matter \cite{Kaplan:2009de} has been discussed in the literature (see \cite{CyrRacine:2012fz}  and references therein). In this paper, we shall only consider the case in which  these bound states, if they exist, are  mostly ionized. 

%%%%%%%%%%%%%%%%%%%%%%%%%%%%%%%%%%%%%%%%%%%%%%%%%%%
\subsection{Constraining  the model}
%%%%%%%%%%%%%%%%%%%%%%%%%%%%%%%%%%%%%%%%%%%%%%%%%%%
\label{model-constraints}

Several limits on the parameter space of the model are known from high-energy physics and tests in astrophysical and cosmological environments. We list below the most severe constraints and the relative implications for mass parameters and coupling constants, as a preliminary outline of the regions in parameter space which will be relevant in the analysis of dark matter candidates within this framework. These constraints will be discussed further in Section~\ref{sec:direct}, when examining  current limits and projected sensitivities from dark matter direct detection experiments. 

Contrary to the case of a massive dark photon, constraints from flavor and precision physics, as well as radiative emission in astrophysical bodies, come from one-loop order corrections providing the coupling to SM fermions. Under the assumption of CP conservation in the dark sector, the limits quoted below are mostly derived from the effective magnetic moment of SM fermions with respect to the dark photon or the ordinary photon, induced by dark fermion - scalar messenger loops. Since a change in the chirality of the fermions is required, the limits are strongly dependent to the mixing $\eta_s$. Depending on the process under consideration, the experimental limits only constrain particular combinations of couplings and masses in the dark sector. At this level, it is then more useful to quote results for Yukawa couplings and dark-sector masses for specific flavors, rather than taking them to be universal as in \eq{LLRR}.

\begin{itemize}
\item \underline{Precision physics}: Magnetic dipole moments of leptons provide a deeper insight on the parameter space. From the experimental measurement of the electron magnetic dipole moment~\cite{Hanneke:2008tm}, we find:
\be
\frac{(m_{\phi^-}^e)^2}{m_{\chi}^e} \frac{0.01}{\eta_s \sqrt{\alpha_L^e \alpha_R^e}}   \gtrsim \unit[2 \times 10^3]{TeV}, \label{el}
\ee
where $m_\chi^e$ stands for the mass of corresponding dark fermion. A comparable limit can be found from the experimental measurement of the muon magnetic dipole moment~\cite{Bennett:2006fi}: 
\be
\frac{(m_{\phi^-}^\mu)^2}{m_{\chi}^\mu} \frac{0.01}{\eta_s \sqrt{\alpha_L^\mu \alpha_R^\mu}}   \gtrsim \unit[4 \times 10^2]{TeV}. \label{mu}
\ee
Since the measurement of the tau magnetic dipole moment is experimentally challenging, the corresponding limit is much less relevant, at about the GeV level. 

Except for tau-like dark sector species, these limits point to lepton-like scalar messengers at a heavy scale, say 10 TeV or above, and lepton-like dark fermions significantly lighter, say at 1~TeV or below - unless the couplings $\alpha_L$ or $\alpha_R$ gets suppressed, or the mixing parameter $\eta_s$ is small.

\item \underline{Collider physics}: Direct searches for charged scalar particles at the LHC~\cite{Aaboud:2017vwy} set a limit \cite{Barducci:2018rlx}
\be
m_S^i \gtrsim 940 \; \mbox{GeV}\, ,
\ee
for the messenger mass related to the dark fermions $Q^U$ and $Q^D$, while \cite{Sirunyan:2018nwe} have set constraints on the mass of sleptons, which give the following lower bound on the mass of lepton-like scalar messengers:
\begin{eqnarray}
m_\phi^e \gtrsim \unit[290]{GeV}.
\end{eqnarray}
The limit increases to 1.5 TeV if more families are included. No limits exist for the masses of the dark fermions from events in which they are produced because they are SM singlets and do not interact directly with the detector.

\item  \underline{Astrophysics probes}: Dark sector species can change the energy transport in astrophysical environments. Constraints for models with a massless dark photon from astrophysics have been discussed in \cite{Hoffmann:1987et,Dobrescu:2004wz,Giannotti:2015kwo}. The most stringent limit comes from stellar cooling in globular clusters by dark-photon Bremsstrahlung emission of electrons scattering on $^4$He nuclei; for a standard choice of environmental parameters, and an upper value of 10~erg~g$^{-1}$~s$^{-1}$ on the extra cooling rate by exotic processes~\cite{Raffelt:1990yz}, we find:
\be
\frac{(m_{\phi^-}^e)^2}{m_{\chi}^e} \frac{1}{\eta_s} \frac{0.1}{\sqrt{\alpha_D}} \frac{0.01}{\sqrt{\alpha_L^e \alpha_R^e}}
\gtrsim \unit[3 \times 10^3]{TeV}.
\ee
This limit applies specifically to the Yukawa coupling to electrons and the corresponding messenger state, and affects regions in parameter space analogous to the limit in \eq{el}. When considering, instead, extra cooling effects in supernovae, the most relevant process is the dark photon emission in nucleon-nucleon Bremsstrahlung. From the neutrino signal of supernova 1987A one can deduce:
\be
\frac{(m_S^i)^2}{m_{Q^i}} \frac{0.001}{\eta_s \sqrt{\alpha_D \alpha_L^i \alpha_R^i}} \gtrsim \unit[2.4 \times 10^2]{TeV}\, . \label{sn1987a}
\ee
The above limit applies to the Yukawa couplings of $u$ and $d$ quarks and the corresponding messenger states. This sets an impact on the parameter space analogous to the leptonic sector, except that, for quark-like dark fermions, we will also explore the possibility of larger mass splittings with respect to the messenger states, with $m_{Q}$ even at the GeV scale.

\item  \underline{Self-interactions for dark matter particles}:
As already anticipated, our scenario gets severely constrained for light dark matter candidates because of the long-range self-interactions induced by the $U(1)_D$ gauge symmetry. The most severe observational limits come from the impact on the dark matter density distribution in collapsed dark matter structures, rather than effects in the early Universe or the early stages of structure formation~\cite{Ackerman:mha,Feng:2009mn,CyrRacine:2012fz}. Bounds have been derived from the dynamics in merging clusters, such as the Bullet Cluster~\cite{Clowe:2006eq}, the tidal disruption of dwarf satellites along their orbits in the host halo, and kinetic energy exchanges among dark matter particles in virialized halos. Among these limits, the latter turns out to be the most constraining: energy exchanges through dark matter self-interactions tend to isotropize dark matter velocity distributions, while there are galaxies whose gravitational potentials show a triaxial structure with significant velocity anisotropy. A limit has been derived by estimating an isotropization timescale (via hard scattering and cumulative effects of many interactions, with Debye screening taken into account) and comparing that timescale to the estimated age of the object~\cite{Feng:2009mn}: a refinement of this limit involves tracking the evolution of the velocity anisotropy due to the energy transfer~\cite{Agrawal:2016quu}. The ellipticity profile inferred for the galaxy NGC720, according to Ref~~\cite{Agrawal:2016quu} (see Fig. 4) sets a limit of about:
\be
m_\chi \left(\frac{0.01}{\alpha_D}\right)^{2/3}  \gtrsim \unit[300]{GeV}
\label{SIlimit}
\ee
where $m_\chi$ here stands for the dark matter mass --- anticipating that we will focus on a lepton-like dark fermion as dark matter candidate --- and the $\alpha_D$ scaling quoted this equation is approximate and comes from the leading $m_\chi$ over $\alpha_D$ scaling in the expression for the isotropization timescale. Note that the limit quoted here is subject to a number of uncertainties and assumptions; it is less stringent than earlier results, such as the original bound quoted from~\cite{Ackerman:mha}, as well about a factor of 3.5 weaker than~\cite{Feng:2009mn} (see also, e.g.,~\cite{Feng:2009hw,Lin:2011gj}). On the other hand, results on galaxies from N-body simulations in self-interacting dark matter cosmologies~\cite{Peter:2012jh}, taking into account predicted ellipticities and dark matter densities in the central regions, seem to go in the direction of milder constraints, at about the same level or slightly weaker than the value quoted in \eq{SIlimit}. This result is also subject to uncertainties, such as the role played by the central baryonic component of NGC720.

As benchmark avoiding self-interaction constraints we will consider cases with dark matter mass about 1 TeV and $\alpha_D \simeq 10^{-2}$.

\end{itemize}

%%%%%%%%%%%%%%%%%%%%%%%%%%%%%%%%%%%%%%%%%%%%%%%%%%%
\subsection{Reference framework and parameter space}
%%%%%%%%%%%%%%%%%%%%%%%%%%%%%%%%%%%%%%%%%%%%%%%%%%%

Taking into account the emerging picture, we will consider a scenario with: {\sl i)} scalar messengers as the heaviest states in the dark-sector, {\sl ii)} a lepton-like dark fermion $\chi$ playing the role of dark matter, lighter than scalar messengers but at a comparable mass scale, and {\sl iii)} two dark fermions $Q^U$ and $Q^D$ coupled to the quarks, which are much lighter than $\chi$ and representative of the light dark sector (we shall see that the masses of the light dark species turn out to be indirectly constrained by CMB limits on exotic radiation components). Unless comparing to specific observables, to keep the model numerically tractable --- but also without losing any of the main trends --- we will adopt a set of simplifying assumptions. We restrict ourselves to the case in which all messenger states have a degenerate mass spectrum defined by a single mass parameter $m_\phi = m_S$ and a single mixing parameter $\eta_s$. For simplicity, the Yukawa couplings of all the dark fermions are also taken to be equal, and with $\alpha_L=\alpha_R$. The extra parameters we need to deal with are the mass of the dark matter candidate $m_\chi$, the common mass $m_Q$ for the two light quark-like dark fermions and the dark photon coupling $\alpha_D$.

The remainder of the paper is devoted to additional constraints coming from the thermal history of the Universe and dark matter searches.

%%%%%%%%%%%%%%%%%%%%%%%%%%%%%%%%%%%%%%%%%%%%%%%%%%%
\section{Thermal history and relic density}
\label{sec:relic}
%%%%%%%%%%%%%%%%%%%%%%%%%%%%%%%%%%%%%%%%%%%%%%%%%%%
\subsection{General picture}
\label{subsec:generalpic}
The aim is to compute the cosmological relic density for the stable species in the dark sector. The technical calculation, via a set of coupled Boltzmann equations, is discussed in the next section. However, it is useful to illustrate first a few features characterizing our setup. 

\begin{figure}[t]
\centering
\includegraphics[scale=0.3]{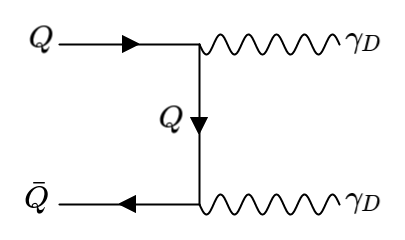}
\includegraphics[scale=0.3]{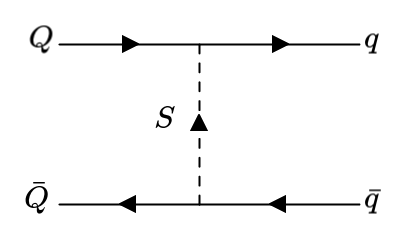}
\caption{\small \label{fig:dfdiagrams} The Feynman diagrams giving the dominant contributions to the total pair annihilation rate of hadron-like dark sector fermions; diagrams contributing to the process for lepton-like dark fermions are analogous.}
\end{figure}

The lightest fermions of given dark charge, lepton-like or hadron-like, are stable, and their number density in the early Universe heat bath changes through processes involving pair productions and pair annihilations; initially in equilibrium (\textit{chemical} equilibrium; see the discussion below for a clarification on this point), they decouple in the non-relativistic regime. Thus, they have a relic density which can be approximated by the celebrated ``WIMP miracle" formula: 
\begin{eqnarray}
  \label{eq:rot}
  \Omega_{\chi,Q} h^2 \sim 0.1 \left(\frac{\unit[2.5 \times 10^{-9}]{GeV^{-2}}}{\langle v \sigma_{\chi\bar{\chi},Q\bar{Q}} \rangle}\right) \,,
\end{eqnarray}
where $\langle v \sigma_{\chi\bar{\chi},Q\bar{Q}} \rangle$ is the thermal average of the pair annihilation cross section for either $\chi$ or $Q$, including all kinematically allowed final states. However, there are two elements which make the computation in the case at hand more involved than in other WIMP setups. First, while one usually deals only with SM final states, the pair annihilation may involve both particles belonging to the dark sector and to the SM sector; the leading processes are into two dark photons and a pair of SM fermion-antifermion of the corresponding type, as shown in Fig.~\ref{fig:dfdiagrams} for the $Q\bar{Q}$ initial state.
Assuming that $s$-wave processes dominate, the thermal average of the pair annihilation cross sections, in the limit of small temperature corrections and massless final states, are approximately given by:
\be
\langle v \sigma_{\chi\bar{\chi},Q\bar{Q}\rightarrow \gamma_D\gamma_D} \rangle \sim \frac{\alpha_D^2}{m_{\chi,Q}^2} \quad \mbox{and} \quad
\langle v \sigma_{\chi\bar{\chi},Q\bar{Q}\rightarrow f\bar{f}} \rangle \sim \frac{\alpha_L^2}{m_{\phi,S}^2}\left(\frac{m_{\chi,Q}}{m_{\phi,S}}\right)^2 \label{eq:cs} .
\ee
Substituting these approximate expressions into \eq{eq:rot}, one can find the preferred mass ranges for which $\Omega_{\chi}$ is at the level of the cosmological dark matter abundance, while $\Omega_{Q}$ is instead negligible (fulfilling the scheme emerging from the set of constraints discussed in the previous section). Taking $\alpha_D$ and $ \alpha_L$ to be $O(10^{-2})$, and messenger scalars lying around $\unit[10]{TeV}$, we find that $\Omega_\chi h^2 \sim 0.1$ if $m_{\chi}$ is in the 1-10~TeV range; $\chi$s predominantly annihilate into dark photons (SM fermions) if $\left(\alpha_L/\alpha_D\right)^2\left(m_\chi/m_\phi\right)^4$ is much less than (greater than) unity. Requiring that $\Omega_{Q}$ is at most 1\% of the Universe's matter density, we find as a conservative upper bound on the masses of the hadron-like species $m_{Q} \lesssim \unit[100]{GeV}$;  $Q$s predominantly annihilate into dark photons. 

\begin{figure}[ht!]
\centering
\includegraphics[scale=0.4]{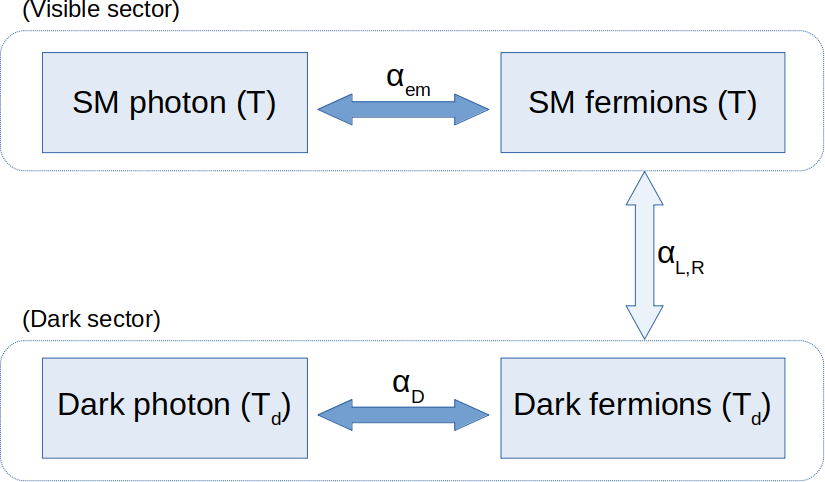}
\caption{\small \label{fig:schematic} Schematic diagram of the interactions between the different reservoirs of states. $\alpha_{em}\equiv e^2/(4 \pi)$ and $\alpha_D \equiv g_D^2/(4 \pi)$ are, respectively, the electromagnetic and dark photon interaction strengths. The coupling $\alpha_{L,R}$ are defined by the Lagrangian in \eq{LLRR}. Ordinary and dark photons do not talk directly to each other.}
\end{figure}

The second point we need to pay attention to is the fact that ``thermal bath" effects, neglected so far, can actually have a significant impact on the overall picture. Analogously to the photon in the SM sector, the dark photon is crucial in keeping dark sector particles at a common temperature
via, \eg, the large energy exchanges in Compton-like dark fermion - dark photon elastic scatterings. These elastic scattering processes maintain \textit{kinetic} equilibrium within the dark sector. Moreover, being a stable massless particle, the dark photon can potentially give a sizable contribution to the budget for the energy density in radiation in the Universe, even at epochs, such as recombination, at which extra radiation components are tightly constrained. The general picture is given schematically in Fig.~\ref{fig:schematic}.  Assuming that the $U(1)_D$ coupling $\alpha_D$ is perturbative but still sufficiently large, dark photon interactions (or, eventually, a chain of processes involving additional interactions with other mediators/forces in the dark sector) enforce that all dark sector particles in the thermal bath have a common temperature $T_d$. Analogously, Compton scattering between SM photons and SM particles maintains kinetic equilibrium within the visible sector. However, the temperature $T$ of the visible sector may be different from $T_d$.

%%%%%%%%%%%%%%%%%%%%%%%%%%%%%%%%%%%%%%%%%%%%%%%%%%%
%%%%%%%%%%%%%%%%%%%%%%%%%%%%%%%%%%%%%%%%%%%%%%%%%%%

In the regime at which messenger scalars are non-relativistic, and with their number densities suppressed, the communication between visible and dark sectors (both at the level of particle number-changing processes and elastic scatterings) is mostly regulated by the Yukawa-like interactions in \eq{LLRR}. Let us first turn off the portal interactions, \textit{i.e.} $\alpha_L=\alpha_R=0$. In this case the thermal bath in the visible and dark sectors evolve independently, and one can track $T$ and $T_d$ by imposing entropy conservation separately in each of the two sectors, see, \eg\,~\cite{Feng:2008mu}. The cooling process goes as the inverse of the scale factor plus a correction due to the change in effective number of relativistic degrees of freedom, when particles becoming non-relativistic transfer their entropy to lighter, relativistic states of the corresponding sector. We define the temperature ratio between dark and visible sectors at a given time $t$ to be
\be
\xi(t) \equiv \frac{T_d(t)}{T(t)}\,
\ee
and consider some initial time $t_0$, with the temperature in the visible sector denoted by $T_0$, at which two sectors are already decoupled. Assuming that entropy densities in the dark and visible sectors, which are respectively given by:
\begin{eqnarray}
\label{sdsv}s_d = \frac{2\pi^2}{45}g_{*S_d}(T_d) T_d^3\quad\quad\text{and}\quad\quad s_v = \frac{2\pi^2}{45}g_{*S_v}(T) T^3,
\end{eqnarray}
are separately conserved in a comoving volume, one finds that the temperature ratio at the CMB epoch is given by:
\be
\label{xicmbent0}\xi_{{\rm CMB}} =
 \left[\frac{g_{*S_d}(\xi_0 T_0)}{g_{*S_d}(\xi_{{\rm CMB}} T_{{\rm CMB}})}\frac{g_{*S_v}(T_{\rm CMB})}{g_{*S_v}(T_0)}\right]^{1/3}\xi_0 \,,
\ee
where $g_{*S_v}(T)$ counts the number of internal degrees of freedom (fermionic species are weighted by 7/8) for all SM particles that are relativistic at temperature $T$, and $g_{*S_d}(T_d)$  is the analogous quantity in the dark sector. Evaluating this ratio is relevant since this is the epoch at which extra radiation components are most severely constrained by cosmological observables. The limit is usually given in terms of $N_{eff}$, the effective number of neutrino-like species, \textit{i.e.} fully relativistic fermions with two internal degrees of freedom, and with a temperature which is a factor of $(4/11)^{1/3}$ cooler than photons. $N_{eff}$ is related to the radiation energy density by: 
\be
\rho_r(t)
\equiv \rho_\gamma(T(t))\left[1+\frac{7}{8}\left(\frac{4}{11}\right)^{4/3} N_{eff}(t)\right] \, .
\ee
The Planck satellite has measured $N_{eff}$ at the CMB epoch to be \cite{Aghanim:2018eyx}: $N_{eff} = 3.27\pm 0.15$, 68\% CL.  Subtracting out the contribution from the three standard model neutrinos~\cite{deSalas:2016ztq} $N_{eff}^{\rm SM} = 3.046$, and assuming that the dark photon is the only dark sector relativistic state at the CMB epoch, giving rise to the extra radiation component $\rho_{r,d}(T_d(t)) = \rho_\gamma(T(t))\,  \xi^4(t)$, we can translate the upper limit on $N_{eff}$ from Planck into a limit on the temperature ratio at the CMB epoch; one finds:
\be
   \xi_{\rm CMB} < 0.54,  \quad 68\% \;{\rm CL}
   \label{eq:xilimitcmb}
\ee   
The $2\sigma$ and $3\sigma$ upper limits are, respectively, about 0.59 and 0.63. Our reference dark sector framework consists of: the dark photon, one lepton-like Dirac fermion $\chi$, and $N_Q$ light hadron-like Dirac fermions $Q$ being relativistic at the initial time $t_0$. From \eq{xicmbent0} we obtain $g_{*S_d}(\xi_0 T_0)/g_{*S_d}(\xi_{{\rm CMB}} T_{{\rm CMB}}) = (7N_Q+11)/4$. Even for a single family of dark hadrons ($N_Q = 2$) we find $\xi_{{\rm CMB}} \approx 0.61\, \xi_0$, in tension with the limit quoted in (\ref{eq:xilimitcmb})
if $\xi_0=1$ (namely $T = T_d$ at $t=t_0$).  As we increase the number of light species in the dark-sector, this problem gets more severe. A possible way out is to relax the initial condition. In principle the picture with decoupled sectors can be extrapolated to $T_0$ as high as, say, the reheating temperature. One can then assume an initial temperature mismatch between the two sectors, with a cooler dark sector (\textit{i.e.} $\xi_0<1$), and thus the dark photon contribution to the radiation component of the Universe can be made small relative to the visible sector contribution. Similar conclusions (for various implementations of the dark-sector portal) were reached in, e.g.,  \cite{Hodges:1993yb,Berezhiani:1995am,Dobrescu:2004wz,Berezhiani:2000gw,Feng:2008mu,Ackerman:mha,Vogel:2013raa}.

On the other hand, when the messenger portal is turned back on, allowing for non-vanishing Yukawa couplings $\alpha_L$ and $\alpha_R$, energy (and entropy) can be exchanged between visible and dark sectors. Regardless of what is assumed for $\xi_0$,  even if the system is not initially in kinetic equilibrium, for couplings sufficiently large, we expect it to relax to a maximum entropy configuration with the two temperature in the two sectors that will tend to become equal. This brings back the problem of satisfying the bound on extra radiation component associated to the dark photon at the CMB epoch, and will effectively translate on an upper bound on the Yukawa couplings. Since $\alpha_L$ and $\alpha_R$ both enter in the discussion for kinetic and chemical equilibrium, these two aspects have to be considered at the same time, as we will do with the set of coupled Boltzmann equations that we introduce in the next subsection and solve numerically.

%%%%%%%%%%%%%%%%%%%%%%%%%%%%%%%%%%%%%%%%%%%%%%%%%%%
\subsection{Boltzmann equations}
%%%%%%%%%%%%%%%%%%%%%%%%%%%%%%%%%%%%%%%%%%%%%%%%%%%
Having highlighted above that SM and dark sector states may have, in general, different temperatures, $T$ and $T_d$ respectively, it is useful to keep track of them separately. Hence, in what follows, we adopt the following notation: $i_d$ will generically indicate a species in the dark sector, while species in the visible sector will be denoted by $i_v$; $i$ will, in general, stand for any species in either sector. To track the distribution function of a state $i_d$, we follow \cite{Gondolo:1990dk,bernstein1988kinetic} and consider the generic Boltzmann equation
 \begin{eqnarray}
L[f_{i_d}] = C[f_{i_d}] \, ,
\end{eqnarray}
where $f_{i_d}$ is the occupation number for the particle $i_d$, $L$ is the Liouville operator tracking the evolution in the Friedmann-Robertson-Walker (FRW)  background, and $C$ is the collision operator. 
The Liouville operator takes the form 
\begin{eqnarray}
L[f_{i_d}] = E_{i_d}\left(\frac{\partial f_{i_d}}{\partial t} - H\vec{p}\cdot \frac{\partial f_{i_d}}{\partial \vec{p}}\right) \, ,
\end{eqnarray}
where $\vec{p}$ is the physical momentum of $i_d$ and $H$ is the Hubble rate. In the early Universe, the Hubble rate is dominated by radiation components coming from the visible and dark sectors. The first Friedmann equation tells us that
\begin{eqnarray}
\label{hubb}H^2(t) \approx \frac{4\pi^3}{45M_{Pl}^2}\left[g_{*v}(T) T^4+g_{*d}(T_d) T_d^4\right] \, .
\end{eqnarray}
In the dilute limit, the collision operator acting on $f_{i_d}$ is driven by $2 \rightarrow 2$ processes, such as $i_d + j \leftrightarrow k + l$. It is then obtained by summing terms of the form: 
\bea
 C_{i_d+j \leftrightarrow k + l}[f_{i_d}(p_{i_d})] &=& \frac{1}{2}\int d\Pi_j(p_j) d\Pi_{k}(p_k) d\Pi_{l}(p_l) (2\pi)^4 \delta^{(4)}(p_{i_d}+p_j-p_k-p_l) \nn \\
\nonumber & & \times \Big\{-\Big\vert\mathcal{M}(i_d + j \rightarrow k + l)\Big\vert^2 f_{i_d}(p_{i_d}) f_j(p_j)[1 \pm f_{k}(p_k)][1 \pm f_{l}(p_l)]\\
& &+ \Big\vert\mathcal{M}(k + l \rightarrow i_d + j)\Big\vert^2 f_{k}(p_k) f_{l}(p_l)[1 \pm f_{i_d}(p_{i_d})][1 \pm f_{j}(p_j)]\Big\}  \, ,\label{integrandgeneric}
\eea
where $d\Pi_j(p_j) \equiv d^3 \vec{p}_j /[(2\pi)^3 \,2\,E_j(p_j)]$ are the usual phase-space integration factors.

When tracking chemical equilibrium, \textit{i.e.} the evolution of the number density of $i_d$, only inelastic processes are relevant. Given the structure of our model, the relevant number changing processes for $\chi$ and $Q$ states (for $T_d$ not too large) are all in the form of particle-antiparticle pair annihilation or creation (see Fig.~\ref{fig:dfdiagrams}), namely
\begin{eqnarray}
C^{(in)}[f_{id}] = \sum_{j_v} C_{i_d + \bar{i}_d \leftrightarrow j_v + \bar{j}_v}[f_{id}] + \sum_{j_d \neq i_d}C_{i_d + \bar{i}_d \leftrightarrow j_d + \bar{j}_d}[f_{id}].
\end{eqnarray} 
The expression for $C^{(in)}[f_{i_d}]$ can be simplified under the standard set of assumptions: (i) {\bf CP invariance} in the process $i_d + \bar{i}_d \rightarrow j_d + \bar{j}_d$, so that $\vert\mathcal{M}_{\rightarrow}\vert^2 = \vert\mathcal{M}_{\leftarrow}\vert^2$ (strictly true in our model); (ii) {\bf dilute limit}, with $f_i \ll 1$, $1 \pm f_i \approx 1$, and equilibrium distributions with occupation numbers approximated as 
\begin{eqnarray}
\label{mb_occupation}f_i^{(eq)} = f_i^{(eq)}(E_i,T) \approx \exp\left(-\frac{E_i - \mu_i}{T}\right)\,; 
\end{eqnarray}
and (iii) {\bf kinetic equilibrium} among dark sector states as enforced by elastic scatterings on the dark photon. Following from (ii), one can safely assume that standard model states follow equilibrium distributions and, using conservation of energy, formally rewrite their occupation numbers in terms of thermal distributions for the dark sector states in the form
\be
f_{i_v} f_{\bar{i}_v} = f_{i_v}^{(eq)}(E_{i_v},T)f_{\bar{i}_v}^{(eq)}(E_{\bar{i}_v},T) = \exp\left(-\frac{E_{i_v}+E_{\bar{i}_v}}{T}\right)
= \exp\left(-\frac{E_{i_d}+E_{\bar{i}_d}}{T}\right) = f_{i_d}^{(eq)}(E_{i_d},T)f_{\bar{i}_d}^{(eq)}(E_{\bar{i}_d},T)\,.
\ee
Note that we have $T$ rather than $T_d$ in the last expression. As for (iii), this implies that, for any dark sector state, one may assume that there is an overall scaling -- only dependent on time -- of the  occupation numbers of dark sector species with respect to equilibrium distributions: 
\begin{eqnarray}
f_{i_d}(E_{i_d},t) \simeq \frac{n_{i_d}(t)}{n_{i_d}^{(eq)}(t)}f_{i_d}^{(eq)}(E_{i_d},T_d(t)) \equiv A_{i_d}(t) \,f_{i_d}^{(eq)}(E_{i_d},T_d(t))\,,
\label{shape}
\end{eqnarray}
with $n_{i_d}$ and $n_{i_d}^{(eq)}$ being the number densities of $i_d$ obtained by integrating $f_{i_d}$ and $f_{i_d}^{(eq)}$, respectively.

To find the evolution equations for the number densities $n_{i_d}$ of the relevant dark-sector fermions, we take the zeroth-order moment of the Boltzmann equation to obtain
\begin{eqnarray}
\nonumber\dot{n}_{i_d} + 3Hn_{i_d} &=& \sum_{i_v}\left[-\langle\sigma v\rangle_{i_d\bar{i}_d \rightarrow i_v\bar{i}_v}(T_d) n_{i_d}^2 + \langle\sigma v\rangle_{i_d\bar{i}_d \rightarrow i_v\bar{i}_v}(T)n_{i_d,eq}^2 (T)\right]\\
 &+& \sum_{j_d \neq i_d}\left[-\langle\sigma v\rangle_{i_d\bar{i}_d \rightarrow j_d\bar{j}_d}(T_d) n_{i_d}^2 + \langle\sigma v\rangle_{j_d\bar{j}_d \rightarrow i_d\bar{i}_d}(T_d) n_{j_d}^2\right] \, . \label{number_density}
\end{eqnarray}
It is understood that the sum over $j_d$ includes the dark photon. The thermally averaged cross section $\langle\sigma v\rangle$ in (\ref{number_density}) is defined, in terms of the corresponding M\o ller cross section, as
\begin{eqnarray}
\label{sigmavdef}\langle\sigma v\rangle_{i\bar{i} \rightarrow j\bar{j}}(\tilde{T}) \equiv \frac{\int \frac{d^3\vec{p}_1}{(2\pi)^3}\frac{d^3\vec{p}_2}{(2\pi)^3}~\left(\sigma v\right)_{i\bar{i} \rightarrow j\bar{j}}f_{i}^{(eq)}(p_1; \tilde{T})f_{\bar{i}}^{(eq)}(p_2; \tilde{T})}{\int \frac{d^3\vec{p}_1}{(2\pi)^3}\frac{d^3\vec{p}_2}{(2\pi)^3}~f_{i}^{(eq)}(p_1; \tilde{T})f_{\bar{i}}^{(eq)}(p_2; \tilde{T})}.
\end{eqnarray}
Looking at (\ref{number_density}), there are three independent variables: $t$, $T$, and $T_d$. In the standard approach, one closes the system by assuming entropy conservation; this leads to a time-temperature relation. In our current set-up, however, the two sectors are allowed to exchange energy and entropy, and thus the entropy of either sector is neither conserved. Nevertheless, the time evolution of the entropies of both sectors will allow us to obtain a well-posed ODE system.

In tracking the entropy of both sectors, we first need to introduce the definition of entropy of species $i$ in terms of the occupation number $f_i$. This is given by
\begin{eqnarray}
s_i = -\int \frac{d^3 \vec{p}}{(2\pi)^3}\left(f_i \ln f_i - f_i\right).
\end{eqnarray}
Its evolution can be obtained by differentiating $s_i$ with respect to time, and then using Boltzmann equation. We have
\begin{eqnarray}
\label{sievol}\dot{s}_i + 3Hs_i = -\int\frac{d^3 \vec{p}}{(2\pi)^3}~C[f_i]\ln f_i.
\end{eqnarray}
We then take the sum of (\ref{sievol}) over dark-sector species. Using kinetic equilibrium and dilute limit assumptions, one obtains:
\be
\dot{s}_d + 3Hs_d = \frac{1}{T_d}\sum_{i_d}\int\frac{d^3 \vec{p}}{(2\pi)^3}~E~C^{(in)}[f_{i_d}] + \frac{1}{T_d}\sum_{i_d}\int\frac{d^3 \vec{p}}{(2\pi)^3}~E~C^{(el)}[f_{i_d}]\\
- \sum_{i_d}\ln A_{i_d}(t) \int\frac{d^3 \vec{p}}{(2\pi)^3}~C^{(in)}[f_{i_d}],
\ee
where $A_{i_d}$ has been defined in \eq{shape} above, and $C^{(el)}[f_{i_d}]$ is the elastic part of the collision operator. Similarly, for $s_v$, we have:
\begin{eqnarray}
\dot{s}_v + 3Hs_v &=& \frac{1}{T}\sum_{i_v}\int\frac{d^3 \vec{p}}{(2\pi)^3}~E~C^{(in)}[f_{i_v}] + \frac{1}{T}\sum_{i_v}\int\frac{d^3 \vec{p}}{(2\pi)^3}~E~C^{(el)}[f_{i_v}]
\end{eqnarray}
In the sum over dark-sector/visible sector species, we only take those processes that involve the transfer of entropy from one sector to the other. To proceed further, it is appropriate to digress into the discussion of the elastic part of the collision operator. It encodes the processes of type $i + B \leftrightarrow i + B$, where $i$ is some species scattering from bath particles $B$, which also contribute to the entropy transfer between the two sectors. As demonstrated in \cite{Binder_2016}, it can be written as
\begin{eqnarray}
C^{(el)}[f_i] = \sum_B C_{i+B \leftrightarrow i + B}[f_i]
\end{eqnarray}
where $C_{i+B \leftrightarrow i + B}[f_i]$ is a Fokker-Planck type operator, given by 
\be
C_{i+B \leftrightarrow i + B}[f_i] \equiv \frac{\partial}{\partial \vec{p}_i}\cdot\left[\gamma_{iB}(E_i,T_B)\left(E_i T_B \frac{\partial f_i}{\partial \vec{p}_i} + \vec{p}_i f_i\right) \right]
= \frac{\partial}{\partial \vec{p}_i}\cdot\left[\gamma_{iB}(E_i,T_B) E_i \frac{\partial f_i}{\partial \vec{p}} \right]\left(T_B - T_i\right). 
\label{cibib}
\ee
In obtaining this expression, it is assumed that the momentum transfer between $i$ and $B$ is much smaller than the typical momentum of either species. The momentum transfer rate can be shown to be given by:
\be
\gamma_{iB}(E_i,T_B) = \frac{1}{48\pi^3 E_i^2 T_B \left(1 - |\vec{v}_i|^2/3\right)}\int_{m_B}^\infty dE_B~f_B(E_B,T_B)~\frac{p_B}{\sqrt{E_i^2 E_B^2 - m_i^2 m_B^2}}\left[\frac{1}{16}\int_{-4p_{CM}^2}^0 dt~\vert\mathcal{M}\vert^2 (-t)\right],
\label{gammaeitb}
\ee
where $4p_{CM}^2 \equiv [s_0 - (m_i+m_B)^2][s_0-(m_i-m_B)^2]/s_0$ and $s_0 \equiv m_i^2 + m_B^2 + 2E_i E_B$. Using (\ref{cibib}), we have
\be
\int \frac{d^3\vec{p}_i}{(2\pi)^3}~E_i~C^{(el)}[f_i] = -\left(T_B - T_i\right)\int \frac{d^3\vec{p}_i}{(2\pi)^3}~\gamma_{iB}\left(E_i,T_B\right)~\vec{p}_i\cdot\frac{\partial f_i}{\partial \vec{p}_i}
= 3n_i\left(T_B - T_i\right) \langle\gamma_{iB}\rangle(T_i,T_B),
\ee
where we identify the thermal average of the momentum transfer rate
\be
\langle\gamma_{iB}\rangle(T_i,T_B) \equiv \frac{\int \frac{d^3\vec{p}_i}{(2\pi)^3}~\gamma_{iB}\left(E_i,T_B\right)~\vec{p}_i\cdot\frac{\partial f_i}{\partial \vec{p}_i}}{\int \frac{d^3\vec{p}_i}{(2\pi)^3}~\vec{p}_i\cdot\frac{\partial f_i}{\partial \vec{p}_i}}
= \frac{\int_{m_i}^\infty dE_i~\left(E_i^2 - m_i^2\right)^{3/2}\gamma_{iB}\left(E_i,T_B\right)~f_i^{(eq)}(E_i,T_i)}{\int_{m_i}^\infty dE_i~\left(E_i^2 - m_i^2\right)^{3/2}~f_i^{(eq)}(E_i,T_i)}\,.
\label{gammarateave}
\ee
\begin{figure}[t]
\centering
\includegraphics[scale=0.3]{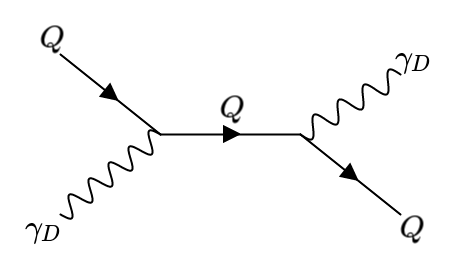}
\includegraphics[scale=0.3]{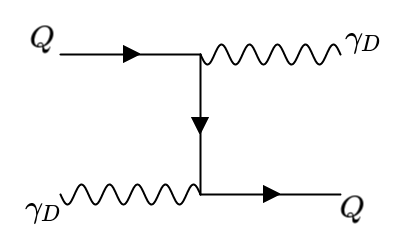}
\includegraphics[scale=0.3]{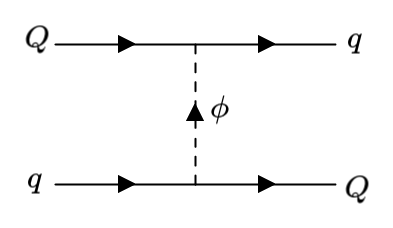}
\caption{\small \label{fig:elastic} The Feynman diagrams for the elastic amplitudes of the dark sector fermions with dark photons (left and center) and with SM fermions  (right). Only the contribution of the diagram on the right is included in the numerical solutions since the  two Compton-like diagrams  are by assumption in equilibrium.}
\end{figure}
At this point we would like to emphasize the following: if the species $i$ were non-relativistic, $T_i \ll  m_i$, one could ignore the dependence of $\gamma_{iB}$ on energy, and the thermal average may be safely replaced as $\langle\gamma_{iB}\rangle(T_i,T_B) \approx \gamma_{iB}(E_i = m_i,T_B)$. For instance, this applies for the case of scatterings of non-relativistic DM particles from a bath of relativistic SM species (this is the limit applied, \textit{e.g.}, in \cite{Bringmann:2006mu}). In our scenario, however, we would also like to account for entropy transfers from the dark-sector to the visible sector; this situation corresponds to the case where the dark-sector species act as bath particles for scatterings of visible sector species. When the scattering species are relativistic, one needs to take into account the energy dependence of $\gamma_{iB}$ and perform the thermal average at each step in the numerical solution of the system of coupled differential equations; further details about the technical implementation of this term are given  Appendix \ref{app:elastic}.\\
\newline
\noindent 
We are now in the position to write down the evolution equations for the entropies in the visible and dark sectors; 
we have (see also the analogous set of equations in~\cite{Foot:2014uba})
\begin{eqnarray}
\nonumber\dot{s}_v + 3Hs_v &\approx& \frac{1}{T}\sum_{i_v} \sum_{i_d}\left[-\langle\sigma v E\rangle_{i_v\bar{i}_v \rightarrow i_d\bar{i}_d}(T) n_{i_v,eq}^2(T) + \langle\sigma v E\rangle_{i_d\bar{i}_d \rightarrow i_v\bar{i}_v}(T_d)n_{i_d}^2\right]\\
&&-3\sum_{i_v}\sum_{i_d}\langle\gamma_{i_vi_d}\rangle(T,T_d)\left(\frac{T-T_d}{T}\right)n_{i_v,eq}(T) \, ,\nn \\
\nonumber\dot{s}_d + 3Hs_d &\approx& \frac{1}{T_d}\sum_{i_d}\sum_{i_v}\Big[-\left(\langle\sigma v E\rangle_{i_d\bar{i}_d \rightarrow i_v\bar{i}_v}(T_d) - \langle\sigma v\rangle_{i_d\bar{i}_d \rightarrow i_v\bar{i}_v}(T_d)~T_d \ln A_{i_d}\right)n_{i_d}^2\\
\nonumber && \qquad \qquad \quad +\left(\langle\sigma v E\rangle_{i_v\bar{i}_v \rightarrow i_d\bar{i}_d}(T) - \langle\sigma v\rangle_{i_v\bar{i}_v \rightarrow i_d\bar{i}_d}(T)~T_d \ln A_{i_d}\right) n_{i_v,eq}^2(T)\Big]\\
&&+3\sum_{i_d}\sum_{i_v}\langle\gamma_{i_di_v}\rangle(T_d,T)\left(\frac{T-T_d}{T_d}\right)n_{i_d} \, , \label{entropy_density}
\end{eqnarray}
where we have introduced yet another thermal average
\begin{eqnarray}
\label{sigmavEdef}
\langle\sigma vE\rangle_{i\bar{i} \rightarrow j\bar{j}}(\tilde{T}) \equiv \frac{\int \frac{d^3\vec{p}_1}{(2\pi)^3}\frac{d^3\vec{p}_2}{(2\pi)^3}~\left(\sigma v\right)_{i\bar{i} \rightarrow j\bar{j}}\left[E_{i}(p_1)+E_{\bar{i}}(p_2)\right]f_{i}^{(eq)}(p_1; \tilde{T})f_{\bar{i}}^{(eq)}(p_2; \tilde{T})}{\int \frac{d^3\vec{p}_1}{(2\pi)^3}\frac{d^3\vec{p}_2}{(2\pi)^3}~f_{i}^{(eq)}(p_1; \tilde{T})f_{\bar{i}}^{(eq)}(p_2; \tilde{T})}.
\end{eqnarray}
From~\eq{entropy_density} it is transparent that if $T=T_d$ at early times the entropy exchange processes balance out, as expected from the condition of thermal equilibrium. Also, once the dark sector particles decouple, the entropies of the two sectors are separately conserved. The approach to kinetic equilibrium between the two sectors will then be relevant if we start with an initial temperature asymmetry and when the heavy dark sector species are still relativistic. 

We choose to solve the system of coupled differential equations using the scale factor $a$ as the independent variable. Using \eq{sdsv}, we rewrite the evolution equations for the entropies as evolution equations for the temperatures:
\begin{eqnarray}
\nonumber\frac{d(\ln T)}{d(\ln a)}&=& -\frac{1}{h_{*S_v}(T)}+\frac{s_v(T)}{3T\,H(T,T_d)\,h_{*S_v}(T)}\sum_{i_v} \sum_{i_d}\left[-\langle\sigma v E\rangle_{i_v\bar{i}_v \rightarrow i_d\bar{i}_d}(T) Y_{i_v,eq}^2(T) + \langle\sigma v E\rangle_{i_d\bar{i}_d \rightarrow i_v\bar{i}_v}(T_d)~Y_{i_d}^2\right]\\
&&-\frac{1}{H(T,T_d)\, h_{*S_v}(T)}\sum_{i_v}\sum_{i_d}\langle\gamma_{i_vi_d}\rangle(T,T_d)\left(\frac{T-T_d}{T}\right)Y_{i_v,eq}(T) \, , \nn \\
\nonumber\frac{d(\ln T_d)}{d(\ln a)} &=& -\frac{1}{h_{*S_d}(T_d)}+\frac{s_v^2(T)}{3T_d\,H(T,T_d)\,s_d(T_d)\,h_{*S_d}(T_d)}\sum_{i_d}\sum_{i_v}\Big\{-\big[\langle\sigma v E\rangle_{i_d\bar{i}_d \rightarrow i_v\bar{i}_v}(T_d)  \\
&& \nonumber 
 - \langle\sigma v\rangle_{i_d\bar{i}_d \rightarrow i_v\bar{i}_v}(T_d)~T_d \ln A_{i_d}\big]Y_{i_d}^2
 + \big[\langle\sigma v E\rangle_{i_v\bar{i}_v \rightarrow i_d\bar{i}_d}(T) - \langle\sigma v\rangle_{i_v\bar{i}_v \rightarrow i_d\bar{i}_d}(T)~T_d \ln A_{i_d}\big] Y_{i_v,eq}^2(T)\Big\}\\
&& + \frac{s_v(T)}{H(T,T_d)\,s_d(T_d)\,h_{*S_d}(T_d)}\sum_{i_d}\sum_{i_v}\langle\gamma_{i_di_v}\rangle(T_d,T)\left(\frac{T-T_d}{T_d}\right)Y_{i_d} \, ,
\label{dlntdlna}
\end{eqnarray}
where we have written explicitly that the Hubble rate $H$ depends both on $T$ and $T_d$, see \eq{hubb}, we have defined
\begin{eqnarray}
h_{*S_v}(T) \equiv 1 + \frac{1}{3}\frac{d(\ln g_{*S_v})}{d(\ln T)} \quad \text{and} \quad h_{*S_d}(T_d) \equiv 1 + \frac{1}{3}\frac{d(\ln g_{*S_d})}{d(\ln T_d)}\,,
\end{eqnarray}
and have normalized all number densities to the entropy density in the visible sector, defining $Y_i \equiv n_i/s_v$, with $i$ being any species -- in the visible sector or in the dark sector. For such variables and again using the scale factor $a$ as independent variable, the Boltzmann equation (\ref{number_density}) takes the form:  
\begin{eqnarray}
\nonumber
\frac{d Y_{i_d}}{d(\ln a)} &=& - 3 Y_{i_d} \left[1 + h_{*S_v}(T) \frac{d(\ln T)}{d(\ln a)}\right]
+\frac{s_v(T)}{H(T,T_d)}\Big\{\sum_{i_v}\big[-\langle\sigma v\rangle_{i_d\bar{i}_d \rightarrow i_v\bar{i}_v}(T_d)~Y_{i_d}^2 \\
 && + \langle\sigma v\rangle_{i_d\bar{i}_d \rightarrow i_v\bar{i}_v}(T)~Y_{i_d,eq}^2 (T)\big]
+ \sum_{j_d \neq i_d}\big[-\langle\sigma v\rangle_{i_d\bar{i}_d \rightarrow j_d\bar{j}_d}(T_d)~Y_{i_d}^2 + \langle\sigma v\rangle_{j_d\bar{j}_d \rightarrow i_d\bar{i}_d}(T_d)~Y_{j_d}^2\big]\Big\} \, .
\label{dlnydlna}
\end{eqnarray}
Equations (\ref{dlntdlna}) and (\ref{dlnydlna}) constitute the closed system of differential equations to be solved. 

%%%%%%%%%%%%%%%%%%%%%%%%%%%%%%%%%%%%%%%%%%%%%%%%%%
\subsection{Numerical results}
\label{sec:relic_numeric}
%%%%%%%%%%%%%%%%%%%%%%%%%%%%%%%%%%%%%%%%%%%%%%%%%%%
As mentioned at the end of Section \ref{sec:model}, we will consider a dark sector framework with the following fermionic content: (i) one lepton-like dark fermion $\chi$, with mass $m_\chi$, which acts as our dark matter candidate, and (ii) two hadron-like states, with masses $m_{Q^U}$ and $m_{Q^D}$, that are lighter than $\chi$. The evolution of the number density of each dark sector fermionic species is governed by Eq. (\ref{dlnydlna}). Regarding scalar messengers, we assume them to be degenerate in mass such that they are specified by a single mass parameter $m_S$, and a universal mixing $\eta_S$. Meanwhile, the other parameters relevant for the discussion are: (i) the $U(1)_{\rm D}$ gauge coupling $\alpha_D$, and (ii)  the Yukawa-like couplings $\alpha_L$ and $\alpha_R$, which are taken to be equal for simplicity. Despite the model residing in a seven-dimensional parameter space, main trends can be illustrated on benchmark cases. In particular, unless explicitly stated, we will start illustrating the framework by focusing on the following choice of parameters:
\begin{eqnarray}
m_{Q^U} = \unit[10]{GeV},\quad m_{Q^D} = \unit[20]{GeV},\quad m_S = \unit[10]{TeV},\quad \alpha_D = 10^{-2},\quad \text{and}\quad \eta_S = 0.
\end{eqnarray}  
We will then vary the Yukawa-like coupling $\alpha_L$ and properly adjust $m_\chi$, so that the relic density of $\chi$ approximately matches the dark matter density in the Universe as measured from cosmological observations.
\begin{figure}[t]
\centering
\includegraphics[scale=0.45]{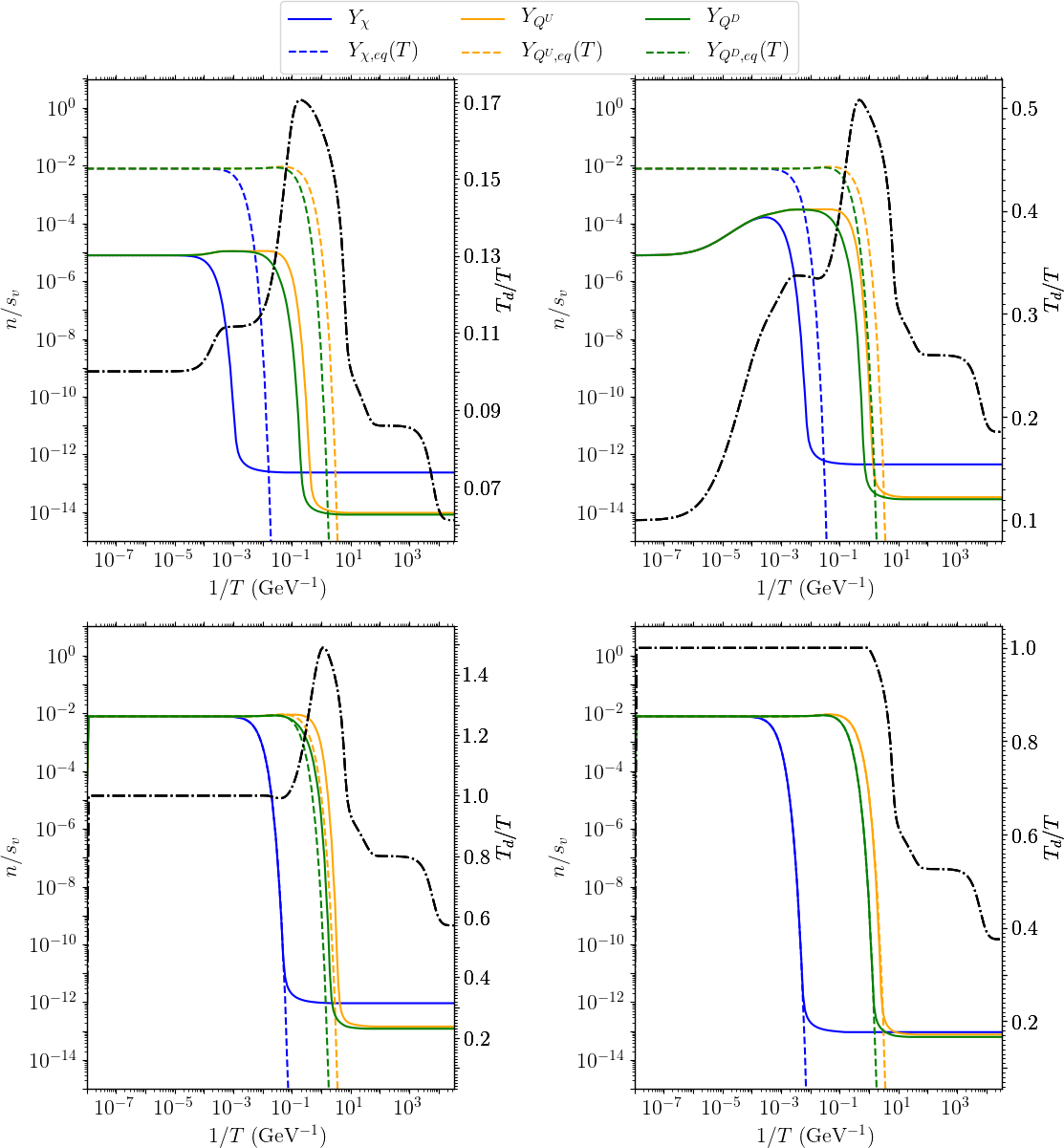}
\caption{\small \label{fig:relicsall} Solutions of the Boltzmann equations for four different benchmark point in parameter space, as specified in Table \ref{table:boltzvals}. These are representative of the four regimes labelled region I, II, III, and IV (from left to right and top to bottom) and described in the text. The solid lines track $Y_{i_d}$, the comoving number density normalized over the visible sector entropy, for each fermionic dark species $i_d$. The dashed lines indicate the value of $Y_{i_d}$ if $i_d$ were in chemical equilibrium with the visible sector heat bath at temperature $T$. The dash-dotted line shows the evolution of $\xi \equiv T_d/T$, the ratio of the dark-to-visible sector temperatures.} 
\end{figure}
\begin{table}[h!]
\begin{tabular}{|c|c|l|c|c|c|c|}
\hline
Region & Coupling & \multicolumn{1}{c|}{} & Species & Mass (GeV) & Relic density ($\Omega h^2$) & Temp. ratio at CMB \\ \hline
\multirow{3}{*}{I} & \multirow{3}{*}{$\alpha_L = 10^{-11}$} & \multirow{3}{*}{} & $\chi$ & $1850 $ & 0.1183 & \multirow{3}{*}{$0.0613$} \\ \cline{4-6}
 &  &  & $Q^U$ & 10 & $2.573 \times 10^{-5}$ &  \\ \cline{4-6}
 &  &  & $Q^D$ & 20 & $4.457 \times 10^{-5}$ &  \\ \hline
\multirow{3}{*}{II} & \multirow{3}{*}{$\alpha_L = 1.75 \times 10^{-8}$} & \multirow{3}{*}{} & $\chi$ & $1000$ & 0.1221 & \multirow{3}{*}{0.1856} \\ \cline{4-6}
 &  &  & $Q^U$ & 10 & $8.948 \times 10^{-5}$ &  \\ \cline{4-6}
 &  &  & $Q^D$ & 20 & $1.520 \times 10^{-4}$ &  \\ \hline
\multirow{3}{*}{III} & \multirow{3}{*}{$\alpha_L = 10^{-4}$} & \multirow{3}{*}{} & $\chi$ & $480 $ & 0.1192 & \multirow{3}{*}{0.5712} \\ \cline{4-6}
 &  &  & $Q^U$ & 10 & $3.866 \times 10^{-4}$ &  \\ \cline{4-6}
 &  &  & $Q^D$ & 20 & $6.499 \times 10^{-4}$ &  \\ \hline
\multirow{3}{*}{IV} & \multirow{3}{*}{$\alpha_L = 0.35$} & \multirow{3}{*}{} & $\chi$ & $5000 $ & 0.1239 & \multirow{3}{*}{0.3757} \\ \cline{4-6}
 &  &  & $Q^U$ & 10 & $2.039 \times 10^{-4}$ &  \\ \cline{4-6}
 &  &  & $Q^D$ & 20 & $3.372 \times 10^{-4}$ &  \\ \hline
\end{tabular}
\caption{\label{table:boltzvals}\small Numerical values of the couplings and masses used to generate the plots in Fig. (\ref{fig:relicsall}), as well as their corresponding results for the relic densities and temperature ratio at CMB. We have chosen the couplings and masses such that $\chi$ would give a relic density that is close to the measured value of the matter density: $\Omega h^2 = 0.1186$. In all cases, we have taken $\alpha_D = 10^{-2}$ and $m_s = \unit[10]{TeV}$.}
\end{table}
In Fig. \ref{fig:relicsall} we present results for the numerical solution of the Boltzmann code for four different sets of pairs $(\alpha_L, m_\chi)$. In each panel a solid line shows, as a function of the inverse of the temperature in the visible sector $T$, the evolution of the number density for $\chi$,  $Q^U$ and $Q^D$, normalized to the entropy density in the visible sector; such evolution is followed from an initial time $t_0$, with initial temperature $T_0 = \unit[10^8]{GeV}$, to some low temperature at which all comoving number densities are frozen to their relic values. $Y_{i_d}$ for each dark fermion species $i_d$ is compared to the corresponding $Y_{i_d,eq}(T)$, namely the ratio between the equilibrium number density $n_{i_d,eq}(T_d)$ -- assuming $T_d=T$ -- and again $s_v(T)$, which is shown as a dashed line. This comparison is relevant since the case of $Y_{i_d}$ tracking  $Y_{i_d,eq}$ corresponds to the species $i_d$ being in chemical equilibrium as well as kinetic equilibrium between visible and dark sectors. In each panel we also show, with a dash-dotted line, the temperature ratio between dark and visible sectors; values of $\xi(t)=T_d/T$ can be read on the vertical scale on the right-hand side -- notice that, to show more clearly its variation over time, the range of values displayed is adjusted in each panel (while the displayed range for $Y_{i_d}$, on the left-hand side of each panel, is kept fixed). Following the general discussion in Section~\ref{subsec:generalpic}, for all benchmark models considered in the plot, it is assumed that at $t_0$ the dark sector is significantly colder than the visible sector, starting the numerical solution with $\xi(t_0) = 0.1$. 

\begin{figure}[ht!]
\centering
\includegraphics[scale=0.45]{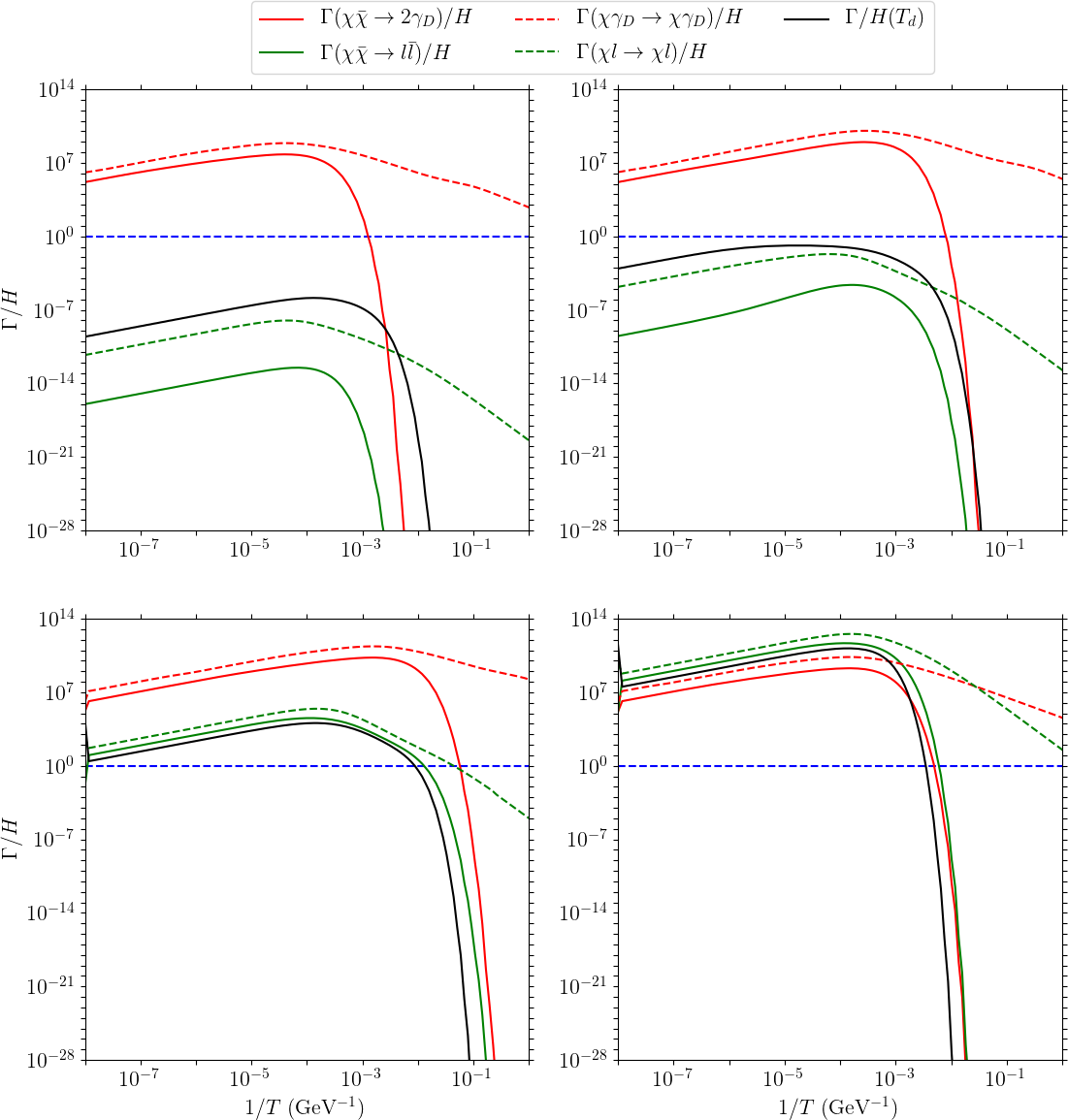}
\caption{\small \label{fig:rates} Rates (normalized over the Hubble constant) of annihilation (solid lines) and scattering processes (dashed lines) involving the lepton-like heavy dark fermion (see \eq{number_density}) into dark photons ($\gamma_D$) and SM states. The rate of entropy exchange is also plotted (see \eq{entropy_density}), and is labelled by $\Gamma/H (T_d)$. The plot refers to the same four benchmark models displayed in Fig. \ref{fig:oh_and_ratio} and specified in Table \ref{table:boltzvals}, as representative of regions I, II, III, and IV in the parameter space (from left to right and top to bottom).} 
\end{figure}

In the four panels of Fig. \ref{fig:relicsall}, going from left to right and top to bottom, $\alpha_L$ is progressively increased from a relatively small value for which the entropy exchanges between dark and visible sector are inefficient at any time, up to a regime at which kinetic equilibrium between the two sectors is reached at the very beginning of the numerical solution and maintained at temperatures lower than the chemical decoupling temperature of the lightest dark fermion. The values of the couplings, the dark fermion mass spectrum, as well as the results of the relic densities of the three dark fermions, and the value $\xi_{\rm CMB}$ of the temperature ratio at the CMB epoch, are given in Table~\ref{table:boltzvals}. To explain trends in Fig. \ref{fig:relicsall}, considering the same benchmark cases and focusing on $\chi$, in Fig~\ref{fig:rates} the effective interaction rates for relevant processes in Eqs.~(\ref{dlntdlna}) and (\ref{dlnydlna}) are compared to the Universe's expansion rate $H$ (as usual, as a rule of thumb, a given process is efficient only when the ratio is larger than one). The pair annihilation rates into dark photons and/or SM leptons, which are shown separately, drive chemical decoupling; the role of $\chi$ in restoring and maintaining kinetic equilibrium can be sketched from the effective energy transfer rate from dark fermion annihilations and $\chi$ elastic scattering on SM leptons, i.e. the combinations one obtains when factorizing out $Y_{i_d}^2/H$ and $Y_{i_d}/H$ in, respectively, the second and third term on the r.h.s. of Eq.~(\ref{dlntdlna}). In the same plot we also show that, for all benchmark models, the scattering rate of $\chi$ on dark photons is much larger than $H$ at any temperature, justifying the assumption of kinetic equilibrium among dark sector states.

The four panels in Figs. \ref{fig:relicsall} and \ref{fig:rates} correspond to four different regimes in the parameter space. These are:
\begin{itemize}
\item \underline{Region I} (top-left plots) This is the regime in which the portal between dark and visible sectors is virtually absent, and the pair annihilation into dark photons enforces chemical equilibrium of fermions in the dark sector at large temperatures. In this case the relic density of $\chi$ can be estimated as the thermal freeze-out of a non-relativistic species from the dark sector, which is analogous to the freeze-out of a standard WIMP from the visible sector: $Y_\chi$ at freeze-out can be shown to be
\begin{eqnarray}
Y_{\chi,\rm f.o.} \simeq \frac{\xi_{\rm f.o.} (m_\chi/T_d)_{\rm f.o.}}{\langle \sigma v\rangle_{\gamma_D} m_\chi M_{Pl}},
\end{eqnarray}
with the dark-sector freeze-out temperature being about
\begin{eqnarray}
\left(\frac{m_\chi}{T_d}\right)_{\rm f.o.} \simeq \ln\left(\xi_{\rm f.o.}^2 \langle\sigma v\rangle_{\gamma_D} m_\chi M_{Pl}\right) + \frac{1}{2}\ln \ln\left(\xi_{\rm f.o.}^2 \langle\sigma v\rangle_{\gamma_D} m_\chi M_{Pl}\right)\,.
\end{eqnarray}
The relic density of $\chi$ is then
\begin{eqnarray}
\label{darkrelic}\Omega_\chi h^2 = \left. \Omega_\chi h^2\right|_{\xi_{\rm f.o.} = 1} \frac{\xi_{\rm f.o.} (m_\chi/T_d)_{\rm f.o.}}{\left.(m_\chi/T_d)_{\rm f.o.}\right|_{\xi_{\rm f.o.} = 1}}\,.
\end{eqnarray}
In this regime, the evolution of $\xi$ is obtained by assuming that the entropies of the dark and visible sectors are separately conserved. Note that due to the Universe's expansion, both $T_d$ and $T$ decrease; the temperature ratio $\xi = T_d/T$ increases whenever $T_d$ decreases slower than $T$. This occurs when a dark species becomes non-relativistic and heats up the dark photon plasma. The ratio reaches a peak at around $T = m_{Q^U}$, and then decreases since SM photons are heated up by SM degrees of freedom becoming non-relativistic, and, especially, at the QCD phase transition when quarks and gluons are transformed into bound-state hadrons.

\item \underline{Region II} (top-right plots): The moderate increase in $\alpha_L$ is still insufficient to reach kinetic equilibrium between the two sectors. The effective energy transfer rate and the elastic scattering rates are still smaller than $H$ at all temperatures, see Fig.~\ref{fig:rates}. Nevertheless, the entropy leakage between the two sectors cannot be ignored, as one can see in the partial readjustment of $\xi$ in the top-right panel of Fig. \ref{fig:relicsall}. Meanwhile in this regime, the relic density for $\chi$, while still mostly determined by $\chi$ pair annihilations into dark photons, is also dictated by pair annihilations of visible sector particles populating the dark-sector with more dark fermions. This scheme is reminiscent of the freeze-in production mechanism for feebly interacting massive particles (FIMPs) \cite{Hall:2009bx}. As an approximate expression, 
Eq.~(\ref{darkrelic}) still applies, with however a slight increase in the thermal bath reservoir within which the freeze out of the thermal component is taking place and a shift in $\xi_{\rm f.o.}$.  

\item \underline{Region III} (bottom-left plots): This is the regime in which $\alpha_L$ is large enough to enforce kinetic equilibrium between the two sectors from the very first steps of the numerical solution, up to the freeze out temperature of the dark matter component (but -- for the specific parameter choice displayed -- not up to the temperature at which the light fermions become non-relativistic). It is however still too small for the $\chi$ pair annihilation into SM leptons to play a role in setting the dark matter relic density; the annihilation into dark photons is still the dominant channel and the standard WIMP formula, \eq{eq:rot} applies. Notice that the peak in the temperature ratio exceeds  unity, since the light fermions become non-relativistic after kinetic decoupling. It follows that this is the benchmark case with largest $\xi_{\rm CMB}$, slightly above the 1~$\sigma$ bound from Planck.  

\item \underline{Region IV} (bottom-right plots): This scenario is similar to region III, except that, concerning the relic density of $\chi$, $\alpha_L$ is sufficiently large for SM lepton-anti-lepton pairs to be the dominant final state in the annihilation rate driving the WIMP rule-of-thumb formula  \eq{eq:rot}. In the case at hand, $\alpha_L$ is also large enough to ensure kinetic equilibrium between dark and visible sectors at all temperatures at which dark fermions are relativistic, hence $\xi$ becomes 1 immediately after $t_0$ and is not increasing further. 
\end{itemize}

\begin{figure}[t]
\centering
\includegraphics[scale=0.5]{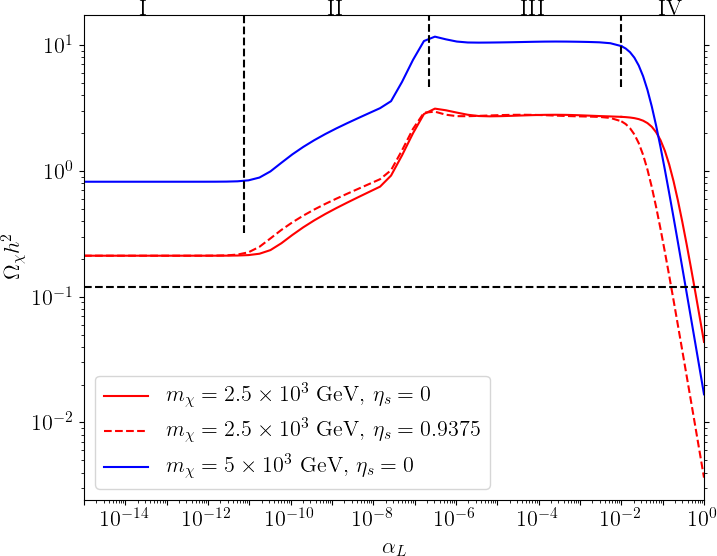}\quad\includegraphics[scale=0.5]{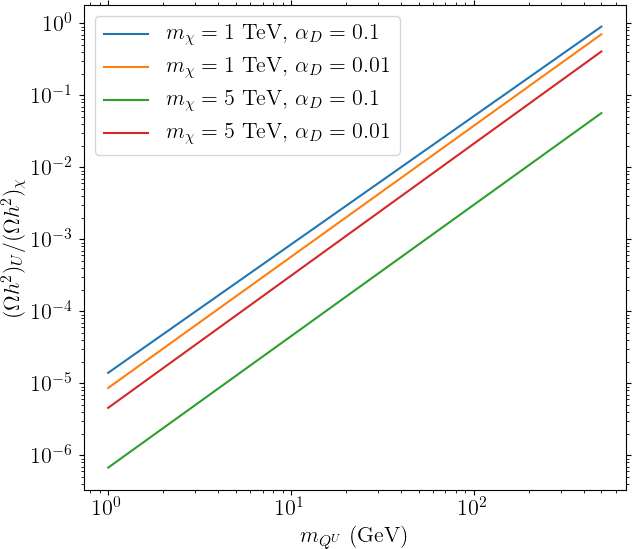}
\caption{\small \label{fig:oh_and_ratio} Left panel: Relic density $\Omega_\chi h^2$ of the DM candidate $\chi$ vs $\alpha_L$, for fixed $\alpha_D = 10^{-2}$ and initial temperature ratio $\xi_0 = 0.1$. One can see the effect of changing $m_\chi$ and the effect of changing the left-right mixing for messenger scalars. The four marked regions are discussed in the text. Right panel: Relic density of $Q^U$ relative to the relic density of $\chi$, as a function of $m_{Q^U}$. As expected, $\Omega_U h^2$ increases with $m_{Q^U}$ since the annihilation cross section to dark photons goes as $m_{Q^U}^{-2}$. The Yukawa coupling is taken everywhere to be $\alpha_L = 0.1$.}
\end{figure}

The four regions are also shown in the left panel of Fig. \ref{fig:oh_and_ratio}, where the relic density of $\chi$ is plotted as a function of $\alpha_L$. We have kept $m_\chi$, $\alpha_D$, and $\eta_S$ fixed for each curve. As expected from the previous discussion, $\Omega_\chi h^2$ is not necessarily a monotonic function of $\alpha_L$ and so there are multiple values of $\alpha_L$ giving the same relic density. In regions I and III, $\Omega_\chi h^2$ is independent of $\alpha_L$, since in both regimes it is the annihilation to dark photons that determines the relic density of $\chi$. Note that region I is the regime where the dark sector out of kinetic equilibrium with respect to the visible sector at all times, while region III is the regime where kinetic equilibrium holds until, at least, the chemical freeze-out of $\chi$. Region II is the transition region between I and III: since the energy/entropy transfers and freeze-in effects become more efficient as $\alpha_L$ increases, $\Omega_\chi h^2$ increases as well. Region IV is the regime in which annihilations into SM leptons become dominant: following from Eqs. (\ref{eq:rot}) and (\ref{eq:cs}), we have $\Omega_\chi h^2 \propto \alpha_L^{-2}$. The change in $m_\chi$ produces a vertical shift of regions I, II, and III. This follows from the fact that, for these regimes, the relic density of $\chi$ is determined by the annihilation to dark photons, and thus $\Omega_\chi h^2 \propto m_\chi^2$. The trend changes for region IV; we have $\Omega_\chi h^2 \propto m_\chi^{-2}$. We also include the case where the left-right mixing between scalar messengers is maximal, \textit{i.e.} $\eta_S = 1 - (m_\chi/m_\phi)^2$; this makes one of the scalar messengers lighter. A lighter scalar messenger increases the rate of processes enforcing kinetic equilibrium, which slightly changes the transition in region II; it also increases the annihilation rate to SM fermions, leading to the transition from region III to IV at a smaller $\alpha_L$, as well as it leads to a decrease in the relic density in region IV. The value $\alpha_{L*}$ of the transition between regions III and IV can be roughly estimated by imposing that the annihilation cross section to SM species is about the same as the annihilation to dark photons; this leads to
\begin{eqnarray}
\alpha_{L*} \simeq \alpha_D \left(\frac{m_S}{m_\chi}\right)^2\left(1-\eta_s\right) \, .
\end{eqnarray}
For instance, if $\eta_s = 0$, $\alpha_D = 10^{-2}$, and $m_S/m_\chi = 2$ (as in the blue curve in the left panel of Fig.\ \ref{fig:oh_and_ratio}), we have $\alpha_{L*} \simeq 4 \times 10^{-2}$.

In the right panel of Fig. \ref{fig:oh_and_ratio} we explore how the relic density of the lighter dark fermions change with their masses. Given the constraints on light particles with long-range interactions in DM halos, such relic densities must be much suppressed compared to $\Omega_\chi h^2$, at the level of about 1\% or lower. In general, the lighter the dark fermion, the more efficient the pair production/annihilation is into dark photons; since chemical decoupling is regulated by this final state, the relic density decreases accordingly. The right panel of Fig. \ref{fig:oh_and_ratio} indeed shows the expected scaling $\Omega_U h^2 \propto m_{Q^U}^2$, for each choice of the parameters $\alpha_D$ and $m_\chi$. In general a contribution to the matter density below one percent can be obtained for $m_{Q} \lesssim \unit[500]{GeV}$. For example,  for $\alpha_D =$   0.1 and $m_\chi = \unit[1]{TeV}$, this upper value is \unit[40]{GeV}, while for $m_\chi = \unit[5]{TeV}$  the upper value shifts up to \unit[195]{GeV}. For $\alpha_D =$   0.01, the upper values are \unit[50]{GeV} and \unit[70]{GeV}, for $m_\chi = \unit[1]{TeV}$ and  $m_\chi = \unit[5]{TeV}$, respectively. The ratio of the relic densities of $Q^U$ over $\chi$ is weakly dependent on $\alpha_L$. 

\begin{figure}[t]
\centering
\includegraphics[scale=0.5]{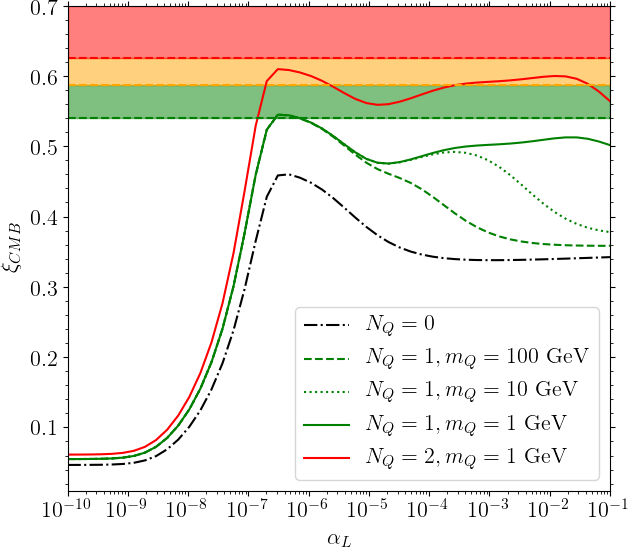}\includegraphics[scale=0.5]{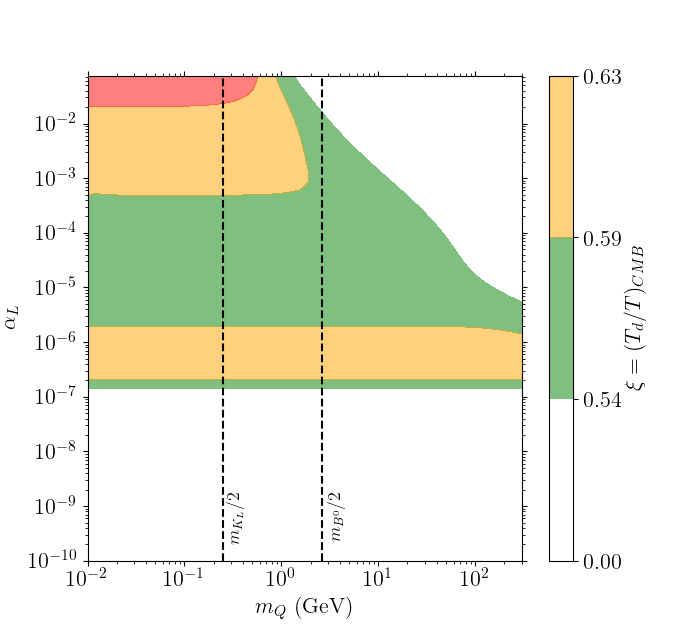}
\caption{\small \label{fig:aLmQscan} (Left) Plot of $\xi_{\rm CMB}$ versus $\alpha_L$, for varying $m_Q$ at fixed $N_Q = 1$, and for varying $N_Q$ at fixed $m_Q = \unit[1]{GeV}$. The colored regions correspond to 2$-\sigma$ (green), 3$-\sigma$ (orange), and $>3$-$\sigma$ (red) bands. Here we have taken $m_s = \unit[10]{TeV}$, $m_\chi = \unit[1]{TeV}$, and $\alpha_D = 10^{-2}$. (Right) Contour plot of $\xi_{\rm CMB}$ on the $\alpha_L-m_Q$ plane, taking the same values of $m_s, m_\chi$ and $\alpha_D$ as in the left panel. Each colored regions correspond to 2$-\sigma$ (green), 3$-\sigma$ (orange), and $>3$-$\sigma$ (red) bands. The remaining regions correspond to $\xi_{CMB}$ that are not excluded at $1\sigma$ by the current CMB limit on $N_{eff}$. The vertical lines correspond to half the masses of the neutral mesons $K_L$ and $B^0$, which could decay into a particle-anti-particle pair of dark quarks. (e.g. see \cite{Barducci:2018rlx}).}
\end{figure}

In the left panel of Fig. \ref{fig:aLmQscan}, we show the temperature ratio at the CMB as a function of $\alpha_L$, for fixed $\alpha_D$, $m_\chi$, and $m_\phi$, while $N_Q$, the number of light dark quarks, and $m_Q$, the common mass of the dark quarks, are allowed to change individually. On the right panel of Fig. \ref{fig:aLmQscan}, we present a contour plot of $\xi_{\rm CMB}$ in the $m_Q-\alpha_L$ plane, for fixed $\alpha_D = 10^{-2}$, $m_\chi = \unit[1]{TeV}$, and $N_Q = 2$. The contour plot has been generated by performing a scan of $m_Q$ from $\unit[10]{MeV}$ to $\unit[300]{GeV}$, and $\alpha_L$ values from $10^{-10}$ to $10^{-1}$. All results in Fig. \ref{fig:aLmQscan} are obtained in numerical solutions of the Boltzmann code assuming as initial temperature ratio $\xi_0=0.1$. As previously mentioned in Sec. \ref{subsec:generalpic}, bounds on $N_{eff}$ constrain extra contributions to the amount of radiation energy density. This constraint translates to an upper bound on the temperature ratio at CMB, given by Eq. (\ref{eq:xilimitcmb}). 

There are a few features emerging from Fig. \ref{fig:aLmQscan}. As expected, at any given $\alpha_L$, the ratio $\xi_{\rm CMB}$ increases as the number of light species $N_Q$ increases. In particular, in the limit of vanishing Yukawa coupling $\alpha_L$, \textit{i.e.} when the two sectors do not communicate with each other, $\xi_{\rm CMB}$ depends on $N_Q$ only. For our reference model, the scaling is $\xi_{\rm CMB} \propto (7N_Q + 11)^{1/3}$. This follows from the fact that entropy is injected into the dark sector bath when dark species become non-relativistic. Since the CMB epoch occurs at relatively late times, $\xi_{\rm CMB}$ does not depend on $m_Q$. Recall also that in this limit, $\xi_{\rm CMB} \propto \xi_0$ and we are assuming a rather small $\xi_0$. 

Starting from a vanishingly small $\alpha_L$, entropy exchanges between visible and dark sectors, that tend to equilibrate the mismatch $\xi_0$ in the initial temperatures, become more efficient as we increase $\alpha_L$. This leads to increasing $\xi_{\rm CMB}$. In the left panel of Fig. \ref{fig:aLmQscan} this is the rising branch at $\alpha_L \lesssim 10^{-6}$. The largest increase is obtained at some intermediate $\alpha_L$ for which kinetic equilibrium is reached at early times, but is not maintained at the epoch at which $\chi$ or the light dark fermions become non-relativistic. When these particles become non-relativistic, they transfer their entropies mainly to dark photons, which makes $\xi(t)$ become larger than 1 at some intermediate temperatures. 

If instead $\alpha_L$  is large enough to maintain kinetic equilibrium when dark fermions become non-relativistic, entropy injections are shared by the SM degrees of freedom and the result is a decrease in $\xi_{\rm CMB}$.  At the same time the reverse effect occurs: SM states becoming non-relativistic and injecting entropy into the dark sector, rather than just heating SM photons, with then an increase in $\xi_{\rm CMB}$. The efficiency in these two-direction exchanges clearly depends on all parameters regulating kinetic equilibrium between the two sectors, including $m_\chi$, $m_Q$, and the messenger masses $m_\phi$ and $m_S$, as well as on the parameters setting the temperatures at which the dark fermions become non-relativistic (regulated also by $m_\chi$ and $m_Q$). In the left panel of the figure, we show in particular the $\alpha_L$ dependence of $\xi_{\rm CMB}$ for different values $m_Q$ and $N_Q=1$, while the case $N_Q=2$ is illustrated for a sample value in the left panel and in the full range $m_Q \in (\unit[10]{MeV},\unit[300]{GeV})$ in the right panel. As the entropy transfer is particularly large at the QCD phase transition, at a temperature of about $\unit[150]{MeV}$~\cite{Drees:2015exa}, it is crucial whether, at this epoch, $Q$ are relativistic and/or visible and dark sectors are in kinetic equilibrium.  

As seen from Fig.~\ref{fig:aLmQscan}, the CMB limits on $N_{eff}$ turn out to be a very severe constraint on the content of light fermions in the dark sector. Assuming an $\alpha_L$ of at least $10^{-2}$, a favorable situation in order to satisfy the CMB limits at 1-$\sigma$ level would be to keep $N_Q \leq 2$ and take $m_Q$ to be at least \unit[5]{GeV}. Future tighter constraints on $N_{eff}$ will impact on the parameter space even more severely.

%%%%%%%%%%%%%%%%%%%%%%%%%%%%%%%%%%%%%%
\section{Direct detection searches}
\label{sec:direct}
\begin{figure}[t]
\centering
\includegraphics[scale=0.2]{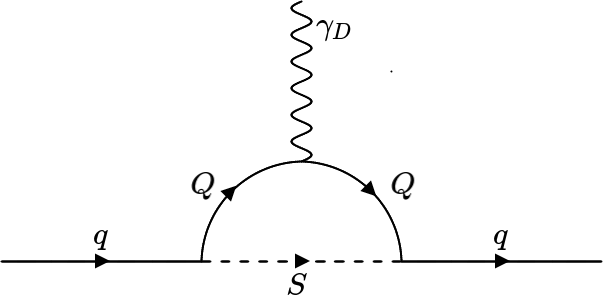}\quad\quad\quad\quad\quad\includegraphics[scale=0.2]{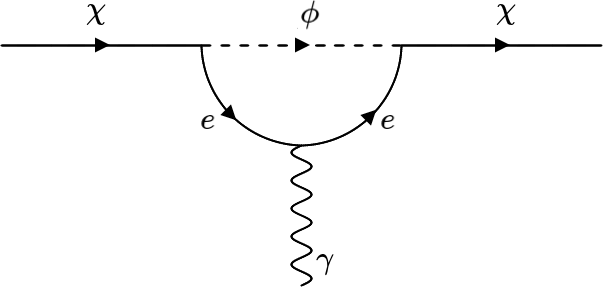}\\
~\\
\includegraphics[scale=0.2]{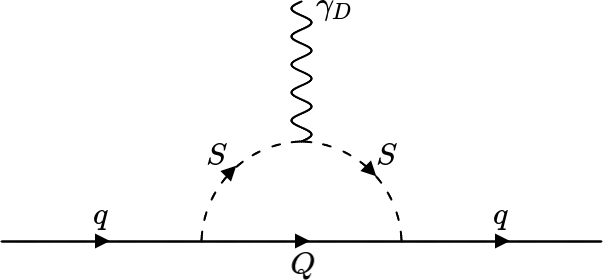}\quad\quad\quad\quad\quad\includegraphics[scale=0.2]{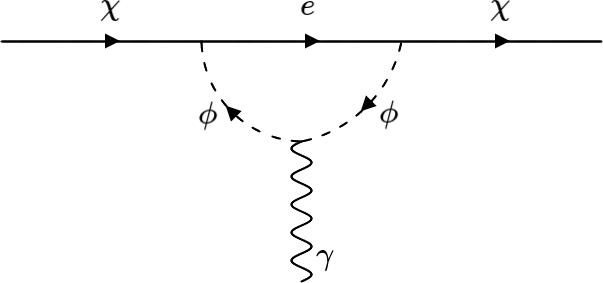} 
\caption{\small \label{fig:loopvertex} Feynman diagrams for the leading (one-loop) contributions to the the coupling between quarks and the dark photon  (left) and  (lepton-like) dark fermions and the ordinary photon (right).}
\end{figure}

%%%%%%%%%%%%%%%%%%%%%%%%%%%%%%%%%%%%%%
Direct searches test the interactions of dark matter particles with ordinary matter. As a preliminary step to project direct detection limits into our framework, we need to write down the effective coupling between dark leptons and quarks. Scattering processes are mostly driven by massless mediators: SM and dark photons. Since  there is no kinetic mixing between the SM and the dark photon, the leading contributions appear at one-loop order, as shown in Fig.\ \ref{fig:loopvertex}. 

Computing the diagrams in Fig. (\ref{fig:loopvertex}) yields the following dimension 5 (magnetic dipole) and dimension 6 (charge-radius) effective operators\footnote{We assume for simplicity that  $CP$ invariance  is respected in the dark-sector and there are no electric dipole moments.}:
\begin{eqnarray}
\label{l5dip}\mathcal{L}_5 &\supset& g_D\frac{d_{M,\gamma_D}^{(q)}}{2\Lambda_{D,\gamma_D}^{(q)}}\left(\bar{q}\sigma^{\mu\nu}q\right)X_{\mu\nu} + e\frac{d_{M,\gamma}^{(\chi)}}{2\Lambda_{D,\gamma}^{(\chi)}}\left(\bar{\chi}\sigma^{\mu\nu}\chi\right)F_{\mu\nu}\\
\label{l6dip}\mathcal{L}_6 &\supset& -g_D\frac{c_{CR,\gamma_D}^{(q)}}{[\Lambda_{CR,\gamma_D}^{(q)}]^2}\left(\bar{q}\gamma^\nu q\right)\partial^\mu X_{\mu\nu} - e\frac{c_{CR,\gamma}^{(\chi)}}{[\Lambda_{CR,\gamma}^{(\chi)}]^2}\left(\bar{\chi}\gamma^\nu \chi\right)\partial^\mu F_{\mu\nu}
\end{eqnarray}
where $F_{\mu\nu}$ and $X_{\mu\nu}$ are, respectively, the field strength associated with the SM photon and the dark photon. The dipole and charge-radius couplings, denoted by $d_M/\Lambda_D$ and $c_{CR}/[\Lambda_{CR}]^2$, respectively, carry additional labels. These additional labels specify: the fermion they are associated with, and the massless gauge boson such fermion is coupled to. In the discussion below we will both show results referring to a generic framework in which dipole and charge-radius couplings are treated independently of each other, as well as focus on our specific framework; in the latter case, they are given in terms of our model parameters and strong correlations appear. In particular, assuming universal couplings and $g_L = g_R$, we have
\begin{eqnarray}
\frac{d_{M,\gamma}^{(\chi)}}{\Lambda_{D,\gamma}^{(\chi)}}=\frac{\alpha_L}{4\pi}\frac{m_\chi}{m_{\phi_-}^2}F_{D,\gamma}^{(\chi)}\left(m_l, m_{\phi_-}, m_{\phi_+}\right)&,& \quad\quad \frac{c_{CR,\gamma}^{(\chi)}}{[\Lambda_{CR,\gamma}^{(\chi)}]^2}=\frac{\alpha_L}{4\pi}\frac{1}{m_{\phi_-}^2}F_{CR,\gamma}^{(\chi)}\left(m_l, m_{\phi_-}, m_{\phi_+}\right) \nonumber \\
\frac{d_{M,\gamma_D}^{(q)}}{\Lambda_{D,\gamma_D}^{(q)}}=\frac{\alpha_L}{4\pi}\frac{m_Q}{m_{S_-}^2}F_{D,\gamma_D}^{(q)}\left(m_Q, m_{S_-}, m_{S_+}\right)&,& \quad\quad \frac{c_{CR,\gamma_D}^{(q)}}{[\Lambda_{CR,\gamma_D}^{(q)}]^2}=\frac{\alpha_L}{4\pi}\frac{1}{m_{S_-}^2}F_{CR,\gamma_D}^{(q)}\left(m_Q, m_{S_-}, m_{S_+}\right).
\label{eq:M&CR}
\end{eqnarray}
The exact expressions for the functions $F_D$ and $F_{CR}$, which are either of order 1 or logarithmically-enhanced, are given in Appendix \ref{appendix:loop}. Here we just quote useful approximate expressions assuming  that $m_l \ll m_{\phi_-}$ and $m_Q \ll m_{S_-}$:
\begin{eqnarray}
F_{D,\gamma}^{(\chi)}\left(m_l, m_{\phi_-}, m_{\phi_+}\right) &\simeq& -\left(1 + \frac{m_{\phi_-}^2}{m_{\phi_+}^2}\right) \nonumber \\
F_{CR,\gamma}^{(\chi)}\left(m_l, m_{\phi_-}, m_{\phi_+}\right) &\simeq& -\left[\frac{1}{3}\ln\left(\frac{m_{\phi_-}}{m_l}\right) - \frac{1}{4}\right]-\left[\frac{1}{3}\ln\left(\frac{m_{\phi_+}}{m_l}\right) - \frac{1}{4}\right]\left(\frac{m_{\phi_-}}{m_{\phi_+}}\right)^2 \nonumber \\
F_{D,\gamma_D}^{(q)}\left(m_Q, m_{S_-}, m_{S_+}\right) &\simeq& \left[4\ln\left(\frac{m_{S_-}}{m_Q}\right) - 2\right] - \left[4\ln\left(\frac{m_{S_+}}{m_Q}\right) - 2\right]\left(\frac{m_{S_-}}{m_{S_+}}\right)^2
\nonumber \\
F_{CR,\gamma_D}^{(q)}\left(m_Q, m_{S_-}, m_{S_+}\right) &\simeq& \frac{1}{36}\left[13 - 12 \ln\left(\frac{m_{S_-}}{m_Q}\right)\right] + \frac{1}{36}\left[13 - 12 \ln\left(\frac{m_{S_+}}{m_Q}\right)\right]\left(\frac{m_{S_-}}{m_{S_+}}\right)^2.
\end{eqnarray}
If we plug-in typical values of dark-visible couplings and particle masses in the dark sector (where we take $m_{\phi_+} = m_{S_+} = \unit[10^3]{TeV}$), we have
\begin{eqnarray}
\frac{d_{M,\gamma}^{(\chi)}}{\Lambda_{D,\gamma}^{(\chi)}} &\simeq& \left(\unit[-1.59 \times 10^{-8}]{GeV^{-1}}\right)\left(\frac{\alpha_L}{10^{-1}}\right)\left(\frac{m_\chi}{\unit[200]{GeV}}\right)\left(\frac{\unit[10]{TeV}}{m_{\phi_-}}\right)^2 \nonumber \\
\frac{c_{CR,\gamma}^{(\chi)}}{[\Lambda_{CR,\gamma}^{(\chi)}]^2} &\simeq& \left(\unit[-4.26 \times 10^{-10}]{GeV^{-2}}\right)\left(\frac{\alpha_L}{10^{-1}}\right)\left(\frac{\unit[10]{TeV}}{m_{\phi_-}}\right)^2
\nonumber \\
g_D^2\frac{d_{M,\gamma_D}^{(q)}}{\Lambda_{D,\gamma_D}^{(q)}} &\simeq& \left(\unit[2.56 \times 10^{-9}]{GeV^{-1}}\right)\left(\frac{\alpha_D}{10^{-2}}\right)\left(\frac{\alpha_L}{10^{-1}}\right)\left(\frac{m_Q}{\unit[10]{GeV}}\right)\left(\frac{\unit[10]{TeV}}{m_{S_-}}\right)^2 \nonumber \\
g_D^2 \frac{c_{CR,\gamma_D}^{(q)}}{[\Lambda_{CR,\gamma_D}^{(q)}]^2} &\simeq& \left(\unit[-1.94 \times 10^{-11}]{GeV^{-2}}\right)\left(\frac{\alpha_D}{10^{-2}}\right)\left(\frac{\alpha_L}{10^{-1}}\right)\left(\frac{\unit[10]{TeV}}{m_{S_-}}\right)^2.
\end{eqnarray}
A further correlation is with the additional contribution to the magnetic dipole moment of SM leptons predicted in our framework; this involves a single class of loop diagrams in which the virtual messenger scalars couple with the SM photon, which is analogous with the bottom-right diagram in Fig. (\ref{fig:loopvertex}) with the particles $\chi$ and $e$ exchanged. Such contribution is given by
\begin{eqnarray}
\frac{d_{M,\gamma}^{(l)}}{\Lambda_{D,\gamma}^{(l)}} = \frac{\alpha_L}{4\pi}\frac{m_\chi}{m_{\phi_-}^2}F_{D,\gamma}^{(l)}\left(m_\chi, m_{\phi_-}, m_{\phi_+}\right)
\label{eq:M&CR2}
\end{eqnarray}
where again $F_D$ is order one and can be approximated as
\begin{eqnarray}
F_{D,\gamma}^{(l)}\left(m_\chi, m_{\phi^-}, m_{\phi^+}\right) \approx \begin{cases}-2\left\{1 + 8\left(\frac{m_\chi}{m_{\phi^-}}\right)^2\left[1-\ln\left(\frac{m_{\phi^-}}{m_\chi}\right)\right]\right\}\left(\frac{m_{\phi^+}}{m_{\phi^-}} - 1\right),~m_\chi \ll m_{\phi^-} \lesssim m_{\phi^+}\\
~\\
-\frac{1}{3} + \frac{m_{\phi^-}^2}{m_{\phi^+}^2} - \frac{1}{3}\left(1-\frac{m_\chi}{m_{\phi^-}}\right),~m_\chi \lesssim m_{\phi^-} \ll m_{\phi^+}.
\end{cases}
\end{eqnarray}
In the following we will first analyze separately the cases in which DM-nucleus scattering is mediated by: {\sl (a)} the SM photon, and {\sl (b)} the dark photon. Case (a) has already been discussed in the literature, and some results are reproduced here (see, e.g., \cite{Sigurdson:2004zp,Banks:2010eh,Barger:2010gv,Heo:2009vt,DelNobile:2017fzy}; see  \cite{Fornengo:2011sz} for another possible long-range interaction for dark matter). Case (b), in which nuclei carry a (dark) magnetic moment, is explored here for the first time. We discuss the differential recoil rates, exclusion curves and projected sensitivities that one obtains  considering each of the two massless mediators. Since both cases (a) and (b) involve dipole-vector interactions between DM and nucleons, one expects a term in the scattering amplitude which scales as the inverse of the momentum transfer, giving it an enhancement in the recoil rate at small recoil energies. On the other hand, we demonstrate here that one cannot naively conclude that the latter is the dominant effect and neglect other terms. While the enhancement is indeed present, it may get dominant over other terms only at extremely small recoil energies. It follows that, for what regards the phenomenology of the model, dimension 5 operators are not always playing the main role.  

\subsection{Direct detection analysis: an overview}
The direct detection differential recoil rate, namely the number of scattering events per unit time, detector mass and recoil energy, can be generally written as
\begin{eqnarray}
\label{drder}\frac{dR}{dE_R} = \sum_T c_T \frac{\rho_0}{m_\chi m_T}\int_{|\vec{v}|\geq v_{min}}d^3\vec{v}~|\vec{v}\,|\,f(\vec{v})\frac{d\sigma_T}{dE_R}.
\end{eqnarray}
In this equation the product of $|\vec{v}\,|$, the modulus of the velocity of the DM particle in the detector frame, times the local DM particle number density, expressed in terms of the ratio between the local DM density $\rho_0$ and the DM mass $m_\chi$, gives the flux of DM particles in the detector at given $|\vec{v}\,|$. Such flux is weighted over the velocity distribution for DM particles in the detector frame $f(\vec{v})$ and convolved with the DM-nucleus differential cross section $d\sigma_T/dE_R$. The sum in the equation is over target nuclear isotopes $T$, with mass $m_T$ and relative abundance $c_T$. The integral includes any $|\vec{v}\,|$ large enough to give a recoil energy $E_R$, i.e., larger than $v_{min} = |\vec{q}\,|/(2\mu_{\chi T})$, where $\mu_{\chi T}$ is the target nucleus-DM reduced mass, $\mu_{\chi T} = m_\chi m_T /(m_\chi + m_T)$, and the momentum transfer $|\vec{q}\,|$ is related to the value of the recoil energy via $E_R = |\vec{q}\,|^2/(2m_T)$. In what follows, for the astrophysical dependent quantities $\rho_0$ and $f(\vec{v})$, we just refer to the standard assumptions in the direct detection community: a local DM halo density of $\unit[0.3]{GeV/cm^3}$ and a Maxwellian velocity distribution in the Galactic frame, with standard values of the velocity dispersion, and of the circular and escape velocities at the position of the Sun. While results are mildly dependent on these assumptions, they do not affect in any way the general discussion.   

The DM-nucleus differential cross section $d\sigma_T/dE_R$ is derived in steps. Given the coupling of DM with quarks, one retrieves the effective coupling of DM with nucleons. The general formalism developed to describe non-relativistic EFT interactions goes as follows: the non-relativistic reduction of the Lagrangian density for the elastic scattering of a heavy DM particle on a proton or a neutron at rest can be written in terms of a set of 15 hermitian, leading-order operators (see. e.g.,~\cite{Fitzpatrick:2012ix,Anand:2014kea}), \textit{i.e.}:
\begin{eqnarray}
\label{lnerftgeneric}\mathcal{L}_{\rm NREFT} = \sum_{i=1}^{15}  \sum_{N=p,n}  c_i^{(N)}{\mathcal O}^{(N)}_i \left(\vec{q}\,,\vec{v}^\perp,\vec{S}_\chi,\vec{S}_N \right).
\end{eqnarray} 
Each ${\mathcal O}^{(N)}_i$ is built out of a different contraction of four three-vectors: the momentum transfer $\vec{q}$; the transverse component of the DM particle velocity $\vec{v}^\perp$ ($\vec{v}^\perp \cdot  \vec{q}=0$); the spin of the DM particle and of the nucleon, respectively $\vec{S}_\chi$ and $\vec{S}_N$. The second step is mapping the single-nucleon interactions into nuclear interactions; the general structure for the differential cross section takes the form:
\begin{eqnarray}
\frac{d\sigma_T}{dE_R} = \frac{m_T}{2\pi |\vec{v}\,|^2}\sum_{\alpha=1}^8 \sum_{\tau,\tau^\prime=0,1} S_\alpha^{(\tau\tau^\prime)}\left(\left|\vec{v}_\perp\right|^2,\left|\vec{q}\,\right|^2\right)\widetilde{W}_\alpha^{(\tau\tau^\prime)}(\left|\vec{q}\,\right|^2) \, ,
\end{eqnarray}
where the proton-neutron basis has been replaced by the isospin basis, $\tau$ and $\tau^\prime$ are isospin indices, $S_\alpha$ are the dark-matter response functions containing contractions of ${\mathcal O}^{(N)}_i$ terms and depend on the coefficients appearing in (\ref{lnerftgeneric}), $\vec{v}_\perp \equiv \vec{v} + \vec{q}/(2\mu)$, and $\widetilde{W}_\alpha$ are the nuclear response functions which are essentially form factors accounting for the composite structure of the nucleus. 

Once we have the differential recoil rate, the expected number of direct detection events can be computed using \citep{Workgroup:2017lvb}: 
 \be
 N_p = MT_E \int_0^\infty \phi(E_R) \frac{dR}{dE_R}dE_R \, ,
 \ee
where $M$ is the mass of the detector, $T_E$ is the exposure time, and $\phi(E_R)$ is the efficiency curve specific to a particular experiment. We can then use the data on the observed number of scattering events in a direct detection experiment, to constrain DM-nucleon interactions. To obtain the usual exclusion curves with some specified confidence level $1 - \alpha$, one must, in principle, obtain the confidence interval $[0,N_{p*}]$ from the posterior probability distribution of $N_p$, given the observed number of events $N_o$. A fixed value of $N_{p*}$ corresponds to a contour in the space of parameters that we are trying to constrain. Alternatively, to obtain exclusion plots, we use here the likelihood ratio test. First compute the Poisson likelihood functions:
\begin{eqnarray}
\mathcal{L} \left(N_o, b\vert N_p\right)=\frac{\left(b+N_p\right)^{N_o}}{N_o!}e^{-(b+N_p)} \, ,
\end{eqnarray}
where $b$ is the number of background events, and then obtain the test statistic
\begin{eqnarray}
\lambda \equiv -2 \ln\frac{\mathcal{L}\left(N_o = 0, b\vert N_p\right)}{\mathcal{L}\left(N_o, b\vert N_p\right)}.
\end{eqnarray}
The test statistic $\lambda$ follows a half-chi-squared distribution with one degree of freedom. The exclusion region will then correspond to those values of $N_p$, which give probabilities above the confidence level: for 90\% CL, we reject those values of $N_p$ which give $\lambda \lesssim -1.64$. In what follows we shall use {\sc DDCalc} \cite{Workgroup:2017lvb,Athron:2018hpc}, a package written specifically for dark-matter direct detection calculations, including the calculation of differential recoil rates and likelihoods needed for obtaining parameter constraints at some specified confidence level. We will apply the procedure above to compare against the latest results from the XENON collaboration, which has produced the strongest upper limits in the DM particle mass range of interest for our framework~\cite{Aprile:2018dbl}, and to infer projected sensitivities of one of the proposed next-generation direct detection experiments, the DARWIN experiment~\cite{Schumann_2015}, as representative of nearly final target for the direct detection field. In both cases we have checked that our results match closely published results when the DM nucleus interaction is assumed to be mediated by the standard spin independent operator.

\subsection{SM photon-mediated processes}

We consider first interactions mediated by SM photons (abbreviated as $\gamma m$ in the following). The dipole and charge radius effective coupling between dark leptons and SM quarks can be readily extracted from the effective operators in Eqs.~(\ref{l5dip}) and (\ref{l6dip}): 
\begin{eqnarray}
\label{smeft}\mathcal{L}_{\gamma} = e^2\left\{\frac{d_{M,\gamma}^{(\chi)}}{\Lambda_{D,\gamma}^{(\chi)}}~\frac{1}{q^2}\left(\bar{\chi}i\sigma^{\mu\nu}q_\nu \chi\right) + \frac{c_{CR,\gamma}^{(\chi)}}{[\Lambda_{CR,\gamma}^{(\chi)}]^2}\left(\bar{\chi}\gamma^\mu \chi\right)\right\}~\left(\frac{2}{3}\bar{u}\gamma_\mu u - \frac{1}{3}\bar{d}\gamma_\mu d\right),
\end{eqnarray}
where $q^\mu$ is the transfer four-momentum. We map the quark operators to the nucleon operators by using the form factors in \citep{Bishara:2017pfq}. We have
\begin{eqnarray}
\label{qvectonuc}\bar{q}(k_2) \gamma^\mu q(k_1) &\rightarrow& \bar{N}(k_2)\left[F_1^{(q/N)}(q^2) \gamma^\mu + \frac{i}{2m_N}F_2^{(q/N)}(q^2)\sigma^{\mu\nu}q_\nu\right]N(k_1),
\end{eqnarray} 
where $N=n,p$, and the $F^{(q/N)}_i$ coefficients are QCD matrix elements. Applying (\ref{qvectonuc}) to the quark vector current in (\ref{smeft}) we get
\begin{eqnarray}
\frac{2}{3}\bar{u}\gamma_\mu u - \frac{1}{3}\bar{d}\gamma_\mu d \; \rightarrow \; \bar{p}\gamma_\mu p + \frac{1}{2m_p}\left[\frac{2}{3}F_2^{(u/p)} - \frac{1}{3}F_2^{(d/p)}\right]\left(\bar{p}i\sigma_{\mu\alpha}q^\alpha p\right) + \frac{1}{2m_n}\left[\frac{2}{3}F_2^{(u/n)} - \frac{1}{3}F_2^{(d/n)}\right](\bar{n}i\sigma_{\mu\alpha}q^\alpha n).
\end{eqnarray}
Following the prescription for mapping dark-matter-nucleon operators to their non-relativistic counterparts  \cite{Fitzpatrick:2012ix,DelNobile:2018dfg},  the effective, non-relativistic DM-nucleon interaction is
\begin{eqnarray}
\nonumber\mathcal{L}_\gamma &=& e^2\frac{d_{M,\gamma}^{(\chi)}}{\Lambda_{D,\gamma}^{(\chi)}}\,\left[\frac{2m_p}{|\vec{q}\,|^2}{\mathcal O}_5^{(p)}\right]\\
\label{smneft}~~~~~& &+ e^2\frac{d_{M,\gamma}^{(\chi)}}{\Lambda_{D,\gamma}^{(\chi)}}\left[-\frac{1}{2m_\chi}{\mathcal O}_1^{(p)}\right] + e^2\frac{c_{CR,\gamma}^{(\chi)}}{[\Lambda_{CR,\gamma}^{(\chi)}]^2}{\mathcal O}_1^{(p)} + e^2\frac{d_{M,\gamma}^{(\chi)}}{\Lambda_{D,\gamma}^{(\chi)}}F^{(N)}\left[\frac{2}{m_N}{\mathcal O}_4^{(N)} - \frac{2m_N}{|\vec{q}\,|^2}{\mathcal O}_6^{(N)}\right].
\end{eqnarray}
Here we have adopted the standard operator numbering
\begin{eqnarray}
\label{nrops} {\mathcal O}_1^{(N)} \equiv 1_\chi 1_N,\quad {\mathcal O}_4^{(N)} \equiv \vec{S}_\chi\cdot\vec{S}_N,\quad {\mathcal O}_5^{(N)} \equiv i\vec{S}_\chi\cdot\left(\frac{\vec{q}}{m_N}\times \vec{v}^\perp\right),\quad {\mathcal O}_6^{(N)} \equiv \left(\vec{S}_\chi\cdot\frac{\vec{q}}{m_N} \right)\left(\vec{S}_N\cdot\frac{\vec{q}}{m_N} \right),
\end{eqnarray}
where ${\mathcal O}_1$ and ${\mathcal O}_4$ are the operators commonly labelled as, respectively, \textit{spin-independent} and \textit{spin-dependent} couplings, and
\begin{eqnarray}
F^{(p)} \equiv 1 - \frac{2}{3}F_2^{(u/p)} + \frac{1}{3}F_2^{(d/p)} \approx -0.772,\quad\quad F^{(n)} \equiv -\frac{2}{3}F_2^{(u/n)} + \frac{1}{3}F_2^{(d/n)} \approx 1.934.
\end{eqnarray}
Notice that we have organized the terms in (\ref{smneft}) in powers of $|\vec{q}\,|$; the first line is of order $1/|\vec{q}\,|$, while the second line is of order $|\vec{q}\,|^0$. Looking at (\ref{smneft}), we see that the $\gamma m$ dipole interaction gives an ${\mathcal O}_5$ contribution, which is long-range and coherent, a ${\mathcal O}_1$ contribution, which is a contact term and coherent, and other short-range, incoherent contributions. On the other hand, the $\gamma m$ CR interaction gives only a contact, coherent ${\mathcal O}_1$ contribution. We summarize these information in Table \ref{table:summint}. 

\begin{table}[t]
\begin{tabular}{|c|c|c|c|c|c|}
  \hline Mediator&DM-nucleon operator& & &coherent&incoherent\\
  \hline \multirow{4}{*}{SM photon} &
  \multirow{2}{*}{$(\bar{\chi}i\sigma^{\mu\nu}q_\nu \chi)(\bar{N}\gamma_\mu N)$} & & long-range & ${\mathcal O}_5$ & none\\ \cline{4-6}
  ~&~&~& contact & ${\mathcal O}_1$ & ${\mathcal O}_4, {\mathcal O}_6$\\ \cline{2-6}
  ~&\multirow{2}{*}{$(\bar{\chi}\gamma^\mu \chi)(\bar{N}\gamma_\mu N)$} & & long-range & none & none\\ \cline{4-6}
  ~&~&~& contact & ${\mathcal O}_1$ & none\\ \cline{1-6}
  \hline \multirow{4}{*}{Dark photon} &
  \multirow{2}{*}{$(\bar{N}i\sigma^{\mu\nu}q_\nu N)(\bar{\chi}\gamma_\mu \chi)$} & & long-range & none & ${\mathcal O}_3$\\ \cline{4-6}
  ~&~&~& contact & ${\mathcal O}_1$ & ${\mathcal O}_4, {\mathcal O}_6$\\ \cline{2-6}
  ~&\multirow{2}{*}{$(\bar{N}\gamma^\mu N)(\bar{\chi}\gamma_\mu \chi)$} & & long-range & none & none\\ \cline{4-6}
  ~&~&~& contact & ${\mathcal O}_1$ & none\\\hline
\end{tabular}
\caption{\small \label{table:summint} Types of DM-nucleon interactions mediated by Standard Model and dark photons, and classifications of the non-relativistic operators generated from such interactions. For a given relativistic operator in the second column, the corresponding non-relativistic interactions are listed as effectively long-range/contact (\textit{i.e.} of order $1/|\vec{q}\,|$ or $|\vec{q}\,|^0$) and coherent/incoherent.}
\end{table}

\begin{figure}[t!]
\centering
\includegraphics[scale=0.45]{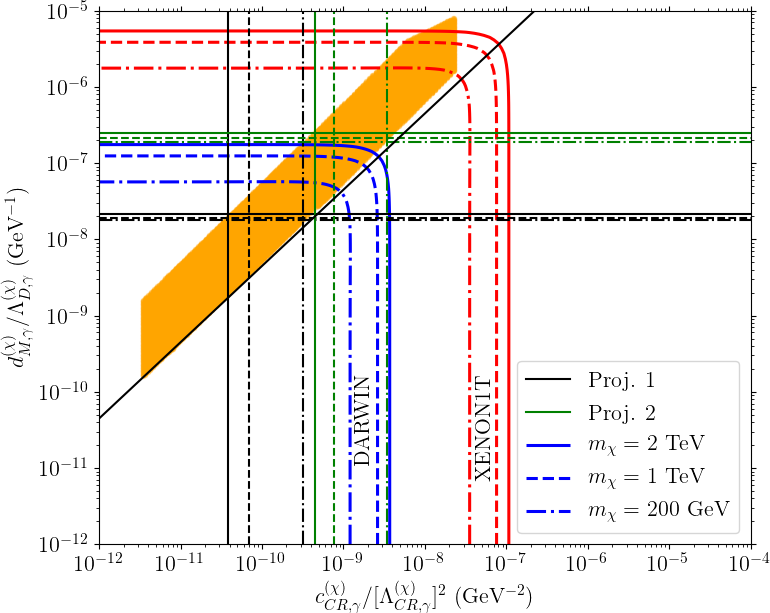}\quad\includegraphics[scale=0.45]{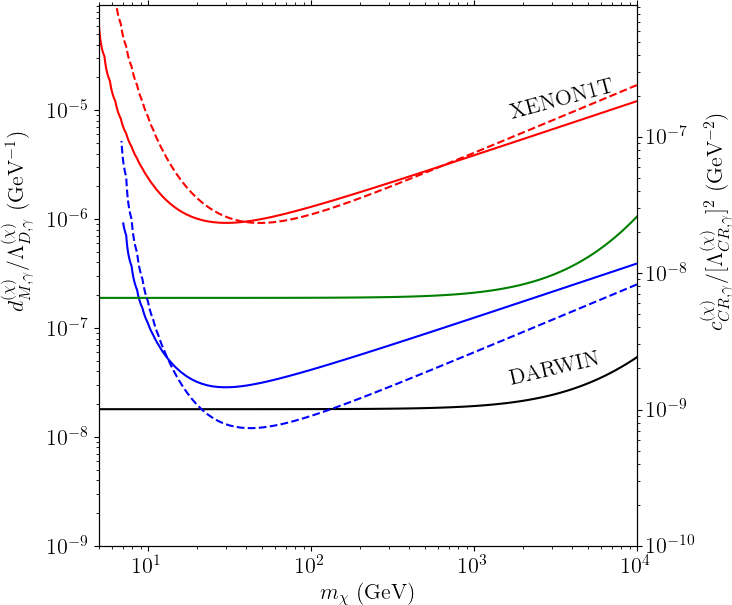}
\caption{\small \label{fig:smm_dmcr} (Left) 90\% confidence level exclusion curves from XENON1T data and the projected sensitivity curves for DARWIN for a few values of the dark matter mass $m_\chi$ in the plane dipole coupling versus CR coupling. The diagonal line gives a visual guidance to separate the regime in which, for a Xenon target and typical detector setups, direct detection rates are dominated by $\gamma m$ dipole interactions or $\gamma m$ CR interactions. The orange region is the area spanned by a large sample of models within our dark sector setup. The horizontal and vertical lines represent a projection of the muon magnetic dipole moment limit into a limit on, respectively, the dipole and CR coefficients, within our framework and for two representative cases: a model with large mixing for scalar messenger (Proj.~1) and one with small mixing (Proj.~2), see the text for details; the intersection points only should be compared with the result for XENON1T and DARWIN. (Right) Exclusion and projected sensitivity curves (90\% CL) versus dark matter mass in case of either $\gamma m$ dipole interactions only (solid lines, referring to the vertical scale displayed on the left-hand side) or $\gamma m$ CR interactions only (dashed lines, referring to the vertical scale displayed on the right-hand side); also shown are the limits on the $\gamma m$ dipole coefficient derived within our framework and the same parameter choices as in the left panel.}
\end{figure}

Coherent terms are likely to provide the largest contributions to the recoil spectrum. Depending on the relative size of the corresponding couplings, the recoil spectrum can either be dominated by dipole or charge-radius interactions. We address this issue by treating first the two couplings, namely $d_{M,\gamma}^{(\chi)}/\Lambda_{D,\gamma}^{(\chi)}$ and $c_{CR,\gamma}^{(\chi)}/[\Lambda_{CR,\gamma}^{(\chi)}]^2$, as independent free parameters. In the right panel of Fig.~\ref{fig:smm_dmcr}, assuming that only one of them is non-zero, we show the 90\% confidence level exclusion curve from XENON1T data and the projected sensitivity curve for DARWIN as a function of the dark matter mass $m_\chi$. Solid lines refer to the case when the $\gamma m$ CR interaction is switched off, with values of $\gamma m$ dipole coupling shown on vertical axis on the left-hand side; on the other hand, dashed lines assume that $\gamma m$ dipole interactions are negligible, with values of the $\gamma m$ CR coupling displayed on the scale on the right-hand side. In the left panel of Fig.~\ref{fig:smm_dmcr}, we show instead exclusion and sensitivity curves in the plane $d_{M,\gamma}^{(\chi)}/\Lambda_{D,\gamma}^{(\chi)}$ versus $c_{CR,\gamma}^{(\chi)}/[\Lambda_{CR,\gamma}^{(\chi)}]^2$ for a few representative values of the DM mass $m_\chi$: \unit[200]{GeV} (dot-dashed lines), \unit[1]{TeV} (dashed lines), and \unit[2]{TeV} (solid lines). In this plot the solid diagonal line, which runs through the area where exclusion and sensitivity curves bend, approximately marks the separation between the \textit{dipole-dominated} (region above the line) and the \textit{CR-dominated} regimes (region below the line). In fact, looking at the expression for the recoil rate contribution from $\gamma m$ dipole interactions, this is mostly driven by the long-range and coherent ${\mathcal O}_5$ operator and can be approximated as:
\begin{eqnarray}
\label{smm_dipole}\left(\frac{dR}{dE_R}\right)_{dip, \gamma} \simeq {\mathcal C}\,
\alpha_{em}^2\left(\frac{d_{M,\gamma}^{(\chi)}}{\Lambda_{D,\gamma}^{(\chi)}}\right)^2 \frac{4s_\chi\left(s_\chi+1\right)}{3}\frac{1}{4\pi E_R}Z^2\,;
\end{eqnarray} 
the $\gamma m$ CR contribution is instead of the form:
\begin{eqnarray}
\label{smm_CR}\left(\frac{dR}{dE_R}\right)_{CR, \gamma} \simeq {\mathcal C}\,
\alpha_{em}^2\left(\frac{c_{CR,\gamma}^{(\chi)}}{[\Lambda_{CR,\gamma}^{(\chi)}]^2}\right)^2 \frac{m_T}{2\pi v^2}Z^2\,.
\end{eqnarray} 
Using Xe as nuclear target, and considering experiments which lose sensitivity below a recoil energy of few keV, we find:
\be
\label{smm_dipole_dominated}
\left. \left(\frac{dR}{dE_R}\right)_{dip, \gamma} \Biggm/ \left(\frac{dR}{dE_R}\right)_{CR, \gamma} \right|_{E_R\simeq 5\,{\rm keV}} \gtrsim 1
\quad \Longrightarrow \quad  \frac{d_{M,\gamma}^{(\chi)}}{\Lambda_{D,\gamma}^{(\chi)}}  \Biggm/
\frac{c_{CR,\gamma}^{(\chi)}}{[\Lambda_{CR,\gamma}^{(\chi)}]^2} \gtrsim 50\,{\rm GeV}
\ee
which is about the delimiter shown in the plot. 

There are additional information displayed in Fig.~\ref{fig:smm_dmcr}. The orange polygonal region in the left panel denotes the pairs of dipole-CR coefficients obtainable in our model assuming $\alpha_L = 10^{-1}$, $m_\chi \in \left[\unit[200]{GeV}, \unit[2]{TeV}\right]$, $m_{\phi_-} \in \left[\unit[1]{TeV}, \unit[100]{TeV}\right]$, 
$m_{\phi_+} \in \left\{\unit[11]{TeV}, \unit[10^3]{TeV}\right\}$,  and $m_\chi \leq m_{\phi_-} \leq m_{\phi_+}$. As it can be seen, there are models in our framework that are excluded by XENON1T data, while DARWIN will cut deeper into the parameter space. The full region is within the area delimited by the condition in~(\ref{smm_dipole_dominated}). Hence we can infer that within our framework, for what concerns $\gamma m$ interactions, the dipole term contributes more to the direct detection rate than the CR term, although the latter can be relevant as well. Note that this statement depends on the type of the nuclear target and on the range of recoil energies at which the experiment is sensitive.

Finally, in Fig.~\ref{fig:smm_dmcr} we try to compare the direct detection limits and projected sensitivities with other constraints. There is no other process in which the operators introduced in \eq{smeft} are tested at a significant level, and hence a model independent comparison is not possible. On the other hand, as described above, within our framework the loop diagrams giving rise to these interactions are closely related to the loop diagrams contributing to the magnetic dipole moments of leptons, which in turn are providing among the tightest constraints on our model, recall the discussion in Section~\ref{model-constraints}. For reference, we consider the case in which the dark matter particle $\chi$ is coupled to muons (stronger constraints would follow in case of coupling to electrons; the limits get essentially irrelevant in case of coupling to tau leptons). The relation between coefficients of the different operators is simply:           
\begin{eqnarray}
\frac{d_{M,\gamma}^{(\chi)}}{\Lambda_{D,\gamma}^{(\chi)}} = \frac{F_{D,\gamma}^{(\chi)}\left(m_\mu, m_{\phi_-}, m_{\phi_+}\right)}{F_{D,\gamma}^{(\mu)}\left(m_\chi, m_{\phi_-}, m_{\phi_+}\right)}\frac{d_{M,\gamma}^{(\mu)}}{\Lambda_{D,\gamma}^{(\mu)}},\quad \frac{c_{CR,\gamma}^{(\chi)}}{[\Lambda_{CR,\gamma}^{(\chi)}]^2} = \frac{1}{m_\chi}\frac{F_{CR,\gamma}^{(\chi)}\left(m_\mu, m_{\phi_-}, m_{\phi_+}\right)}{F_{D,\gamma}^{(\mu)}\left(m_\chi, m_{\phi_-}, m_{\phi_+}\right)}\frac{d_{M,\gamma}^{(\mu)}}{\Lambda_{D,\gamma}^{(\mu)}}\,.
\end{eqnarray}
Comparing against the experimental measurement of the muon magnetic dipole moment~\cite{Bennett:2006fi}, we find:
\begin{eqnarray}
\frac{d_{M,\gamma}^{(\mu)}}{\Lambda_{D,\gamma}^{(\mu)}} \leq \unit[1.80 \times 10^{-8}]{GeV^{-1}}\,.
\end{eqnarray}
We project this limit into a limit on the $\gamma m$ dipole and $\gamma m$ CR coefficients (hence comparing at this level against direct detection) choosing two representative set of values for the masses of the corresponding scalar messenger: in the first --- to which we refer as projection 1 --- we choose a large mixing configuration $\left(m_{\phi_-}, m_{\phi_+}\right) = \left(\unit[10]{TeV}, \unit[10^3]{TeV}\right)$, while in the other --- to which we refer as projection 2 --- we consider a small mixing case $\left(m_{\phi_-}, m_{\phi_+}\right) = \left(\unit[10]{TeV}, \unit[11]{TeV}\right)$. In the left panel of Fig.~\ref{fig:smm_dmcr} derived limits on $\gamma m$ dipole and $\gamma m$ CR coefficients are shown, respectively, as horizontal and vertical lines; the line-style reflects again the three sample choices for $m_\chi$ and the position of the crossing point of horizontal and vertical lines for the same model configuration should be compared to the corresponding direct detection curves (crossing points correspond to physical models in our framework, and, as expected, they all lie in the dipole dominated region). We see that in general, within our framework, the muon magnetic dipole moment limit is more constraining than the current direct detection limit. On the other hand, future detectors will be more sensitive to smaller dark matter dipole moments. Note that the effective dipole operator requires a change in the chirality of the external fermion, either through a sizable $\eta_s$ or a mass insertion on the external leg. When $\eta_s$ is sufficiently small, \textit{i.e.} $\eta_s \ll m_\mu/m_\chi$, the muon dipole is proportional to $m_\mu$ while the dark matter dipole is proportional to $m_\chi$: in this case the muon dipole tends to be much smaller than the dark matter dipole. The projected limits on the $\gamma m$ dipole are also shown in the right panel of Fig.~\ref{fig:smm_dmcr}; given that physical models in our framework have a direct detection rate mostly driven by $\gamma m$ dipole interactions to a first approximation the displayed limits can be compared to the direct detection curves shown in this plot is the case $c_{CR,\gamma}^{(\chi)}/[\Lambda_{CR,\gamma}^{(\chi)}]^2=0$, reinforcing the picture just described.

\subsection{Dark photon-mediated processes}
The same procedure outline above can be applied to compute the recoil rate in case of processes that are dark photon-mediated (in the following: $\gamma_D m$); we start with the effective SM quark-dark lepton interaction:
\begin{eqnarray}
\label{quarkdmeff}\mathcal{L}_{\gamma_D} = g_D^2 \frac{d_{M,\gamma_D}^{(q)}}{\Lambda_{D,\gamma_D}^{(q)}}~\frac{1}{q^2}\left(\bar{q}i\sigma^{\mu\nu}q_\nu q\right)\left(\bar{\chi}\gamma_\mu \chi\right) + g_D^2 \frac{c_{CR,\gamma_D}^{(q)}}{[\Lambda_{CR,\gamma_D}^{(q)}]^2}\left(\bar{q}\gamma^\mu q\right)\left(\bar{\chi}\gamma_\mu \chi\right).
\end{eqnarray}
Borrowing the terminology from the previous section, we identify the first and second terms in (\ref{quarkdmeff}) as $\gamma_D m$ \textit{dipole} and \textit{charge-radius} (CR) interactions, respectively. We then map the quark vector and tensor currents to nucleonic operators. The non-relativistic reduction of the effective DM-nucleon interaction yields
\begin{eqnarray}
\nonumber\mathcal{L}_{\gamma_D} &=& g_D^2 \frac{d_{M,\gamma_D}^{(N)}}{\Lambda_{D,\gamma_D}^{(N)}}\left[-\frac{2m_N}{|\vec{q}\,|^2}{\mathcal O}_3\right]\\
\label{darkmeft}&&+g_D^2\left[\frac{d_{M,\gamma_D}^{(N)}}{\Lambda_{D,\gamma_D}^{(N)}} \frac{1}{2m_N} + \frac{d_{M,\gamma_D,1}^{(N)}}{\Lambda_{D,\gamma_D}^{(N)}}~\frac{1}{m_N} \right] {\mathcal O}_1 + g_D^2 \frac{c_{CR,\gamma_D}^{(N)}}{[\Lambda_{CR,\gamma_D}^{(N)}]^2}{\mathcal O}_1+g_D^2\frac{d_{M,\gamma_D}^{(N)}}{\Lambda_{D,\gamma_D}^{(N)}}\left[\frac{2}{m_\chi}{\mathcal O}_4 - \frac{2m_N^2}{m_\chi\,|\vec{q}\,|^2}{\mathcal O}_6\right],
\end{eqnarray}
where ${\mathcal O}_1, {\mathcal O}_4, $ and ${\mathcal O}_6$ are defined in \eq{nrops}, and
\begin{eqnarray}
\label{o3defn}{\mathcal O}_3 &\equiv& i\vec{S}_N\cdot\left(\frac{\vec{q}}{m_N}\times \vec{v}^\perp\right).
\end{eqnarray}
Using the numerical values of QCD matrix elements obtained from lattice calculations \citep{Bishara:2017pfq}, the coefficients in \eq{darkmeft} are in the form
\begin{eqnarray}
\label{coeffsN}\frac{d_{M,\gamma_D}^{(N)}}{\Lambda_{D,\gamma_D}^{(N)}} \equiv f_T\left\langle\frac{d_{M,\gamma_D}^{(q)}}{\Lambda_{D,\gamma_D}^{(q)}}\right\rangle,\quad \frac{d_{M,\gamma_D,1}^{(N)}}{\Lambda_{D,\gamma_D}^{(N)}} \equiv f_{T1}\left\langle\frac{d_{M,\gamma_D}^{(q)}}{\Lambda_{D,\gamma_D}^{(q)}}\right\rangle,\quad \frac{c_{CR,\gamma_D}^{(N)}}{[\Lambda_{CR,\gamma_D}^{(N)}]^2} \equiv f_1\left\langle\frac{c_{CR,\gamma_D}^{(q)}}{[\Lambda_{CR,\gamma_D}^{(q)}]^2}\right\rangle,
\end{eqnarray}
with
\begin{eqnarray}
f_T = 0.59\pm 0.023,\quad\quad f_{T1} = 0.79,\quad\quad f_1 = 3,
\end{eqnarray}
and angle brackets denoting weighted averages that can be safely removed if $\gamma_D m$ dipole and CR coefficients are about the same for all light quarks.
Analogously to the previous case, we organized the terms in \eq{darkmeft} in powers of $|\vec{q}\,|$, with the first line of order $1/|\vec{q}\,|$ and the second of order $|\vec{q}\,|^0$. In the non-relativistic reduction, the $\gamma_D m$ dipole interaction has led to: (i) a long-range, incoherent ${\mathcal O}_3$ term, (ii) a contact, coherent ${\mathcal O}_1$ term, and (iii) other short-range, incoherent terms; the $\gamma_D m$ CR interaction has generated only one leading operator corresponding to a contact, coherent ${\mathcal O}_1$ term. A summary with relativistic operators and the corresponding non-relativistic reductions is given in Table \ref{table:summint}.

\begin{figure}[t]
\centering
\includegraphics[scale=0.46]{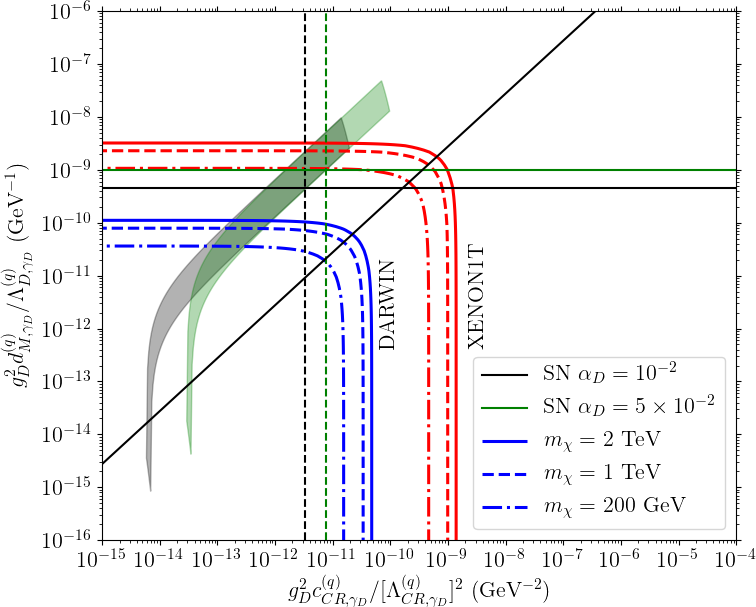}\quad\includegraphics[scale=0.46]{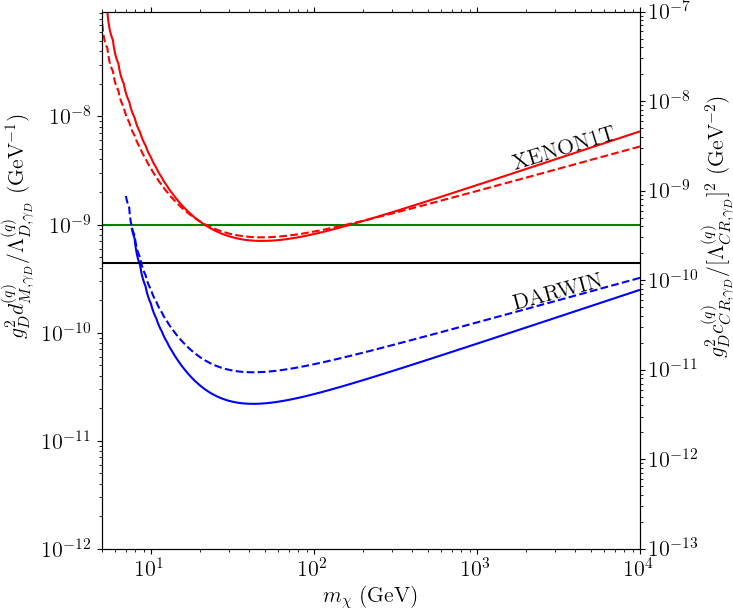}
\caption{\small \label{fig:dmm_dmcr} 90\% CL exclusion curves from XENON1T data and projected sensitivity curves for DARWIN in case of $\gamma_D$-mediated DM-nucleus scatterings, in the plane dipole-CR coefficients (left panel) or when assuming that only one of the two coefficients is non-zero (right panel, solid lines and the vertical scale on the left side refer to the $\gamma_D m$ dipole operator, while dashed lines and the vertical scale on the right side refer to the $\gamma_D m$ CR operator). Model-independent supernova cooling limit on the $\gamma_D m$ dipole for quarks are displayed as horizontal solid lines for two representative values of $\alpha_D$: $10^{-2}$ (black) and $5 \times 10^{-2}$ (green). Also shown in the left panel are two results specific for our dark sector framework: vertical dashed lines represent the projection of the supernova limit on the $\gamma_D m$ CR coefficient - the intersection points with horizontal lines should be compared against direct detection results; the coloured regions correspond to two representative scans in the model parameter space, see the text for details.}
\end{figure}

Similarly to what has been done above for the $\gamma m$ case, we consider first the $\gamma_D m$ dipole and CR couplings to a quark as two independent coefficients, without any reference to our scheme. In the left panel of Fig.~\ref{fig:dmm_dmcr}, 90\% confidence level exclusion curve from XENON1T data and projected sensitivity curves for DARWIN are shown in the plane $g_D^2 \,d_{M,\gamma_D}^{(q)}/\Lambda_{D,\gamma_D}^{(q)}$ versus $g_D^2 \,c_{CR,\gamma_D}^{(q)}/[\Lambda_{CR,\gamma_D}^{(q)}]^2$ for a few sample values of the dark matter mass: $m_\chi = \unit[200]{GeV}, \unit[1]{TeV}, \unit[2]{TeV}$. In the right panel they are shown instead versus mass, assuming that only one among the two coefficients are different from zero. The solid diagonal line in the left panel marks again the separation between the dipole-dominated region and CR-dominated region, as we can check looking at the expressions for the differential recoil rate. As in the previous case, CR interactions contributes with the coherent term in the form
\begin{eqnarray}
\label{dmm_CR}\left(\frac{dR}{dE_R}\right)_{CR, \gamma_D} \simeq  {\mathcal C}^\prime\,
\left(g_D^2 \frac{c_{CR,\gamma_D}^{(q)}}{[\Lambda_{CR,\gamma_D}^{(q)}]^2}\right)^2 \frac{m_T}{2\pi v^2}~f_1^2 A^2\,,
\end{eqnarray} 
where $A$ is the atomic number of the target nucleus. On the other hand and contrary to the previous case, for $\gamma_D m$ dipole interaction we cannot a priori assume that the long-range effects dominate: given that the long-range ${\mathcal O}_3$ term is incoherent, we need to keep also the short-range coherent ${\mathcal O}_1$ term, getting
\be
\label{dmm_dipole}
\left(\frac{dR}{dE_R}\right)_{dip, \gamma_D} \simeq  {\mathcal C}^\prime\,
\left(g_D^2 \frac{d_{M,\gamma_D}^{(q)}}{\Lambda_{D,\gamma_D}^{(q)}}\right)^2\frac{f_T^2}{2\pi}\left[\frac{2}{3\,E_R} \left\langle\vec{S}_N^2\right\rangle+
\frac{4\,m_T}{v^2\,m_N^2}\left\langle\left(\vec{L}\cdot\vec{S}_N\right)^2\right\rangle + 
\frac{A^2\,m_T}{v^2\,m_N^2}  \left(\frac{1}{2}+\frac{f_{T1}}{f_T}\right)^2\right],
\ee
where $S_N$ is the spin operator for the valence nucleon (which is usually relevant for odd-even nuclei) and $L$ is the angular momentum associated with the internal motion of the valence nucleon. Among the three contributions on the right-hand-side, although the first has a $m_N/E_R$ enhancement, this has to compete with the large $A^3$ and $1/v^2$ enhancements in the third term; moreover, the second term is most often sub-leading compared to the third given that $\langle(\vec{L}\cdot\vec{S}_N)^2\rangle \approx l^2_{max}$, where $l_{max}$ is the maximum angular quantum number attained by the valence nucleon, typically much less than $A$. Comparing first and third contributions, one finds that the long-range $1/E_R$ enhancement takes over only at recoil energies  
\begin{eqnarray}
E_R \lesssim \left(\unit[1.5 \times 10^{-7}]{keV}\right) \left[\frac{s_N(s_N+1)}{3/4}\right] \left(\frac{v}{10^{-3}}\right)^2\left(\frac{100}{A}\right)^3\,,
\end{eqnarray}
i.e. in a range which is irrelevant for a Xe target (as well as any target presently considered) and current detector technologies. Hence the third term is the leading one, and 
when taking the ratio between the rate in \eq{dmm_dipole} and that in \eq{dmm_CR} one finds
\be
\label{dmm_ratio_recoil}
\left(\frac{dR}{dE_R}\right)_{dip, \gamma_D} \Biggm/ \left(\frac{dR}{dE_R}\right)_{CR, \gamma_D} \simeq 
\left(g_D^2\frac{d_{M,\gamma_D}^{(q)}}{\Lambda_{D,\gamma_D}^{(q)}} \Biggm/  g_D^2 \frac{c_{CR,\gamma_D}^{(q)}}{[\Lambda_{CR,\gamma_D}^{(q)}]^2}\right)^2\frac{f_T^2}{m_N^2\,f_1^2	}~\left(\frac{1}{2}+\frac{f_{T1}}{f_T}\right)^2\,.
\ee
It follows that:
\be
\left(\frac{dR}{dE_R}\right)_{dip, \gamma_D} \Biggm/ \left(\frac{dR}{dE_R}\right)_{CR, \gamma_D} \gtrsim 1
\quad \Longrightarrow \quad
g_D^2\frac{d_{M,\gamma_D}^{(q)}}{\Lambda_{D,\gamma_D}^{(q)}} \Biggm/  g_D^2 \frac{c_{CR,\gamma_D}^{(q)}}{[\Lambda_{CR,\gamma_D}^{(q)}]^2} \gtrsim 2.8\,{\rm GeV}\,,
\ee
which is the delimiter shown as a solid diagonal line in the left panel of Fig.~\ref{fig:dmm_dmcr}.

Turning now to constraints competing with direct detection results, contrary to the $\gamma m$ case, there is a strong model-independent bound impacting directly on the first operator in \eq{quarkdmeff}. In fact, the $\gamma_D m$ dipole for quarks can be responsible for enhancing the cooling rate in supernovae, allowing for nucleon-nucleon Bremsstrahlung emission of dark photons; as discussed in Section~\ref{model-constraints}, there is a tight constraint one can extrapolate from the observed neutrino flux from SN1987A. The detailed derivation of the limit is rather involved and beyond the scope of this paper; we consider instead an extrapolation from analogous scenarios. In particular, Raffelt \cite{Raffelt:1990yz} computed the energy loss rate due to nucleon-nucleon Bremsstrahlung with axion emission, with the axion entering through a derivative coupling with the nucleon axial current. More recently an improved calculation has been implemented in \cite{Carenza_2019}. For the case of $\gamma_D$ emission, Dobrescu \cite{Dobrescu:2004wz} assumed that the rate of energy loss is two times larger than in the case of axion emission, given that the dark photon has two propagating degrees of freedom. 
If one writes the effective nucleon-$\gamma_D$ interaction as
\begin{eqnarray}
\mathcal{L}_{N\gamma_D} = \frac{g_{N\gamma_D}}{4m_N}\bar{N}\sigma^{\mu\nu}N~X_{\mu\nu},
\end{eqnarray}
following Raffelt and imposing that the extra energy loss rate per unit mass induced by the novel Bremsstrahlung process cannot exceed $\unit[10^{19}]{erg~g^{-1}~s^{-1}}$, we find:  
\begin{eqnarray}
\label{dmm_limitdmgn}
g_{N\gamma_D} \lesssim 1.414 \times 10^{-9}~f^{1/2}.
\end{eqnarray}
Here, $f$ is a fudge factor accounting for the deviation from the Dobrescu assumption on the cooling rate when actually using (\ref{l5dip}) (in the following we will just set it to 1). Mapping the quantity $g_{N\gamma_D}$ to the quark dipole moment in (\ref{l5dip}), and then mapping to the $\gamma_D m$ dipole coefficient constrained by direct detection, we have
\begin{eqnarray}
\label{dmm_limitdm}
g_D^2 \frac{d_{M,\gamma_D}^{(q)}}{\Lambda_{D,\gamma_D}^{(q)}} \lesssim \left(\unit[4.18 \times 10^{-10}]{GeV^{-1}}\right)f^{1/2}\left(\frac{\alpha_D}{10^{-2}}\right)^{1/2}\left(\frac{g_{N\gamma_D}^{(lim)}}{1.414 \times 10^{-9}}\right)\left(\frac{\unit[1]{GeV}}{m_N}\right)\left(\frac{0.6}{f_T}\right).
\end{eqnarray}
This limit is shown with horizontal solid lines in the left and right panels of Fig. \ref{fig:dmm_dmcr}, for $\alpha_D = 10^{-2}$ (black) and $\alpha_D = 5 \times 10^{-2}$ (green). As it can be seen, at face value, the supernova limit is constraining $\gamma_D m$ dipole of quarks at a comparable level with respect to current direct detection data, while, regarding future sensitivities, direct detection experiments are going to be more  competitive. On the other hand, the validity of the supernova limit has been recently questioned~\cite{Bar:2019ifz} since it relies on a mainstream picture for the explosion mechanism of core-collapse supernovae which is still, to a large extent, not well-established; in alternative scenarios the limit in \eq{dmm_limitdmgn} simply does not apply. In this respect, information on the $\gamma_D m$ dipole of quarks derived from direct detection searches seem more reliable.   

While the $\gamma_D m$  CR operator does not contribute the dark photon emission via nucleon-nucleon Bremsstrahlung, a constraint can be indirectly derived within our framework implementing   
\begin{eqnarray}
g_D^2\frac{c_{CR,\gamma_D}^{(q)}}{[\Lambda_{CR,\gamma_D}^{(q)}]^2} = \frac{1}{m_Q}\frac{F_{CR,\gamma_D}^{(q)}\left(m_Q, m_{S_-}, m_{S_+}\right)}{F_{D,\gamma_D}^{(q)}\left(m_Q, m_{S_-}, m_{S_+}\right)}~g_D^2\frac{d_{M,\gamma_D}^{(q)}}{\Lambda_{D,\gamma_D}^{(q)}}.
\end{eqnarray}
In the left panel of Fig. \ref{fig:dmm_dmcr}, limits on the $\gamma_D m$ CR coefficient, as derived from the supernova limit on the $\gamma_D m$ dipole, are shown with dashed vertical lines; these projections are obtained assuming $m_Q = \unit[10]{GeV}$, $m_{S_-} = \unit[10]{TeV}$, and $m_{S_+} = \unit[10^3]{TeV}$. Note that vertical and horizontal lines cross in the dipole-dominated regime, hence the relevant comparison with direct detection rates is still in the limit of vanishing CR coefficient. Dipole dominance is typical for the parameter space in our scheme. In the left panel of Fig. \ref{fig:dmm_dmcr} we show the regions in the dipole-CR plane corresponding  to a scan with $\alpha_L = 10^{-1}$, $m_Q \in \left[\unit[10]{GeV}, \unit[50]{GeV}\right]$, $m_{S_-} \in \left[\unit[10]{TeV}, \unit[10^3]{TeV}\right]$, $m_{S_+} = \unit[1.001 \times 10^3]{TeV}$, and either $\alpha_D = 10^{-2}$  (grey region) or $\alpha_D = 5 \times 10^{-2}$ (green region); in scanning the model space, we ensured that $m_Q \leq m_{S_-} \leq m_{S_+}$. Most models are in the dipole-dominated area, with only tails extending into the CR-dominated regime in case the $\gamma_D m$ dipole gets severely suppressed when ${S_-}$ and ${S_+}$ are very close in mass and hence the mixing $\eta_s$ is very small. 

\begin{figure}[t]
\centering
\includegraphics[scale=0.4]{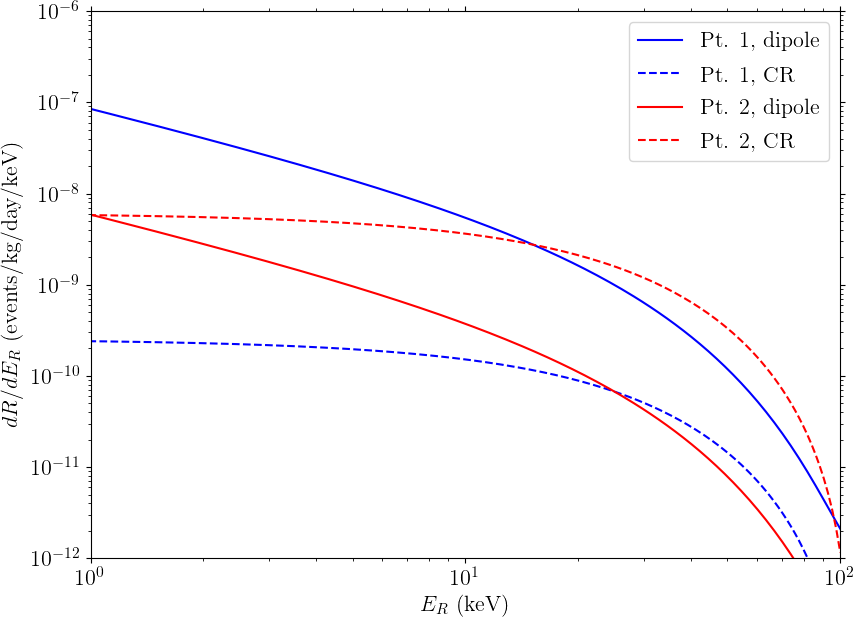}\quad\includegraphics[scale=0.4]{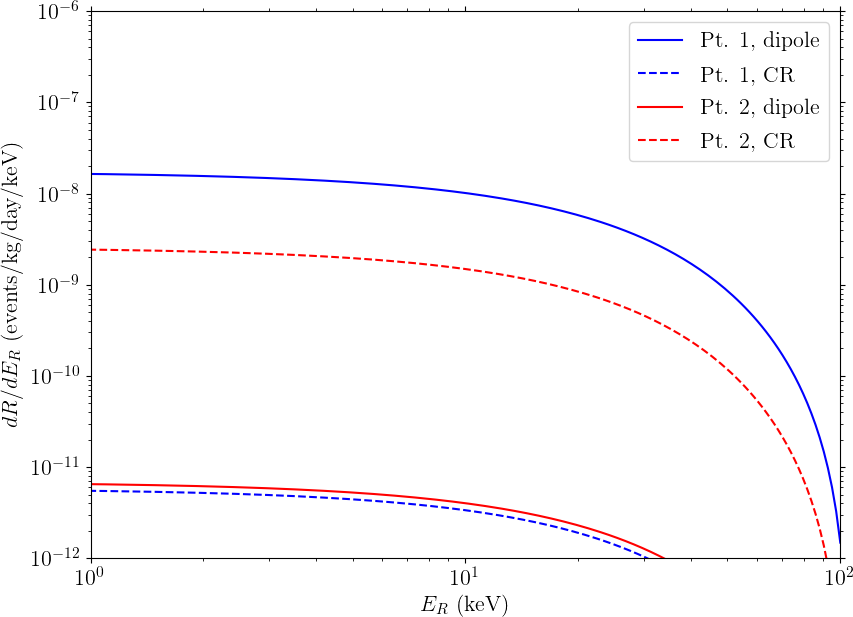}
\caption{\label{fig:recoilspex} \small Recoil spectra due to $\gamma m$ interactions (left panel) and $\gamma_D m$ interactions (right panel) for sample models in our dark sector framework. For each of the two cases, representative points in the parameter space have been chosen to have either a dipole-dominated spectrum (Pt.~1), or a CR-dominated spectrum (Pt.~2); model parameters are specified in Table \protect{\ref{table:recoilvals}}. Contributions to the rate due to the dipole operator and the CR operators are shown separately, respectively with solid and dashed lines. Note the long-range $1/E_R$ enhancement appears only in case of $\gamma m$ dipole interactions.} 
\end{figure}

In Fig. \ref{fig:recoilspex}, we plot recoil spectra in case of $\gamma m$ interactions (left panel) and $\gamma_D m$ interactions (right panel) for sample models in our dark sector framework. For each mediator, we have chosen two representative points such that ``Pt.~1" lies in the corresponding dipole-dominated region, while ``Pt.~2" in the CR-dominated regime: the corresponding model parameters are specified in Table \ref{table:recoilvals}. Contributions to the differential rate of the dipole and CR operators are shown separately. Notice the qualitatively different shapes of the dipole contribution in the two cases: the $1/E_R$ scaling due to long-range interactions can be seen in the $\gamma m$ case, while contact interactions dominate in the $\gamma_D m$ case. Notice also that Pt.~2 in the $\gamma m$ case is rather peculiar, since to find a model within the CR-dominated regime we were forced to consider a relatively small $m_\chi$, below the range considered for the scan displayed in Fig.~\ref{fig:dmm_dmcr} and what we expect typically in our framework.  

\subsection{Comparison with relic density limits}

\begin{table}[t]
\centering
\begin{tabular}{|c|c|c|c|}
\hline
Mediator & Model parameters & Dipole $\left(\unit[]{GeV^{-1}}\right)$ & CR $\left(\unit[]{GeV^{-2}}\right)$ \\ \hline
\multirow{2}{*}{$\gamma$ (Pt. 1)} & $\alpha_L = 0.1, m_\chi = \unit[1]{TeV}$ & \multirow{2}{*}{$1.45\times 10^{-7}$} & \multirow{2}{*}{$5.21 \times 10^{-10}$} \\
 & $m_{\phi_-} = \unit[10]{TeV}, m_{\phi_+} = \unit[11]{TeV}$ &  &  \\ \hline
\multirow{2}{*}{$\gamma$ (Pt. 2)} & $\alpha_L = 0.1, m_\chi = \unit[50]{GeV}$ & \multirow{2}{*}{$2.70\times 10^{-8}$} & \multirow{2}{*}{$1.81 \times 10^{-9}$} \\
 & $m_{\phi_-} = \unit[5]{TeV}, m_{\phi_+} = \unit[6]{TeV}$ &  &  \\ \hline
\multirow{2}{*}{$\gamma_D$ (Pt. 1)} & $\alpha_L = 0.1, \alpha_D = 10^{-2}, m_Q = \unit[10]{GeV}$ & \multirow{2}{*}{$1.28\times 10^{-10}$} & \multirow{2}{*}{$9.95 \times 10^{-13}$} \\
 & $m_{S_-} = \unit[50]{TeV}, m_{S_+} = \unit[10^3]{TeV}$ &  &  \\ \hline
\multirow{2}{*}{$\gamma_D$ (Pt. 2)} & $\alpha_L = 0.1, \alpha_D = 10^{-2}, m_Q = \unit[10]{GeV}$ & \multirow{2}{*}{$2.55\times 10^{-12}$} & \multirow{2}{*}{$2.09 \times 10^{-11}$} \\
 & $m_{S_-} = \unit[14]{TeV}, m_{S_+} = \unit[14.014]{TeV}$ &  &  \\ \hline
\end{tabular}
\caption{\label{table:recoilvals} \small List of representative models chosen for generating the recoil spectra in Fig. \ref{fig:recoilspex}.}
\end{table}

\begin{figure}[t]
\centering
\includegraphics[scale=0.4]{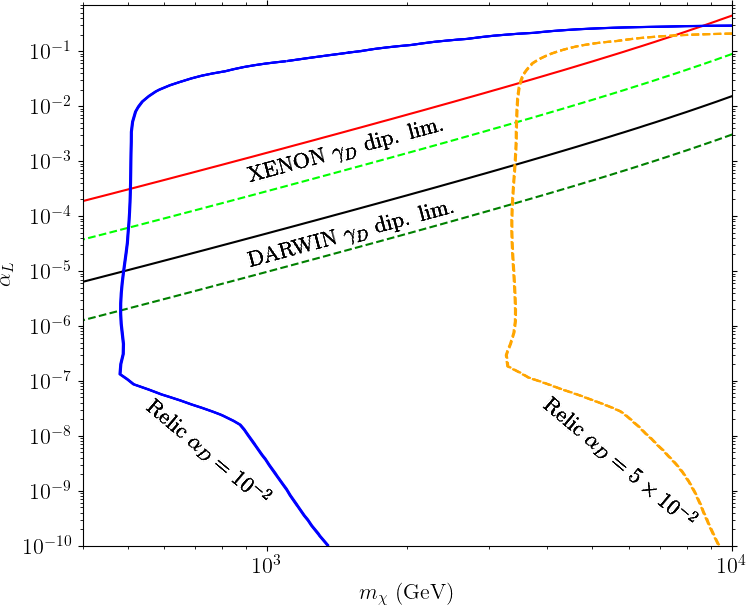}\quad\includegraphics[scale=0.4]{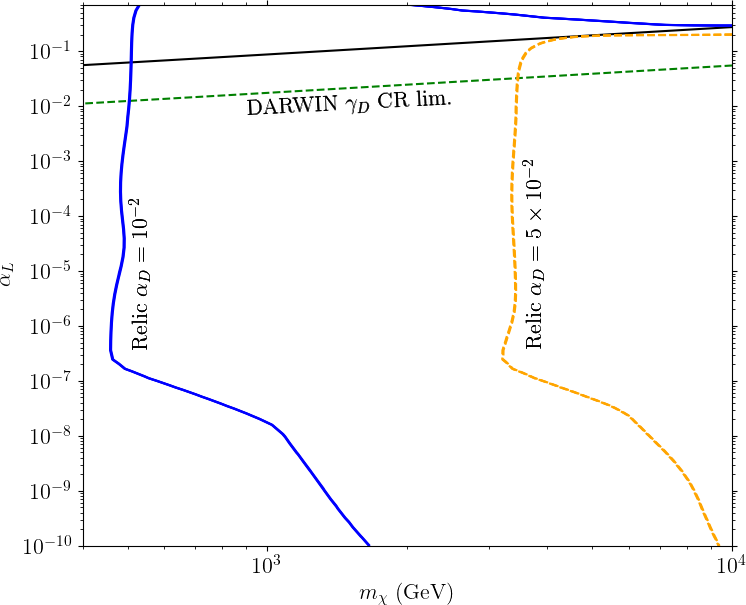}\\
~\\
\includegraphics[scale=0.42]{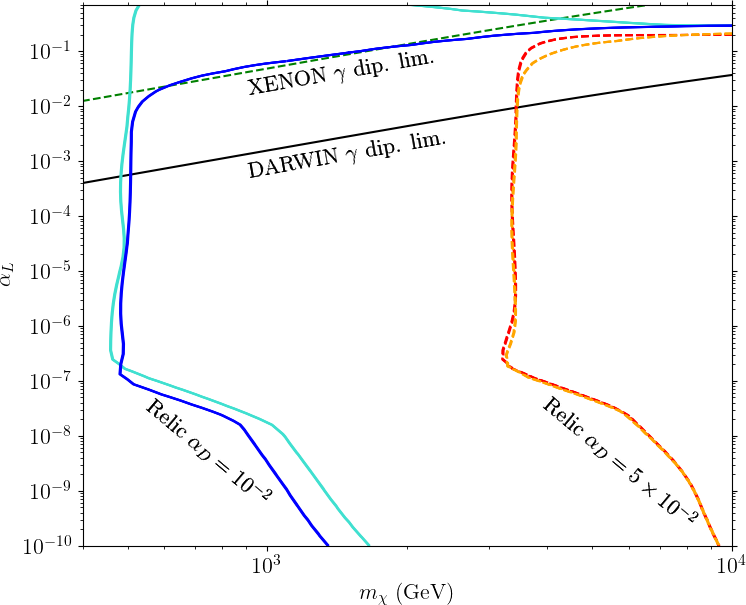}
\caption{\label{fig:relicvsdd} 
The lines labelled ``relic" correspond to model parameters $\alpha_L$ and $m_\chi$ for which the relic density of $\chi$ matches the abundance of dark matter in the Universe. In the top-left panel the case of maximal mixing for scalar messengers is considered, while in the top-right a case with $\eta_s=0$ is displayed; in these two panels solid lines refer to the choice of $\alpha_D = 10^{-2}$ and dashed lines to $\alpha_D = 5 \times 10^{-2}$, while the other parameters the model are fixed to sample values, see the text for details. Also displayed in two top panels are XENON1T limits and DARWIN sensitivity curves due to $\gamma_D m$ interactions, mainly due to the dipole operator in case of large mixing and the CR operator in case of zero mixing; solid and dashed lines refer again to the two sample values of  $\alpha_D$. In the bottom panel the four relic density isolevel curves from the top panels are reproduced to be shown against XENON1T and DARWIN results in case $\gamma m$ interactions are included while $\gamma_D m$ interactions are switched off (by e.g. raising the mass scale for scalar messengers in the quark sector); the dipole term is dominant and results do not depend on $\eta_s$ or $\alpha_D$.}
\end{figure}

We are now ready to combine direct detection results with the constraints on our framework obtained by imposing that the relic density of $\chi$ matches the observed abundance of DM in the Universe, $\Omega_{DM} h^2 = 0.1200 \pm 0.0012$ \cite{Aghanim:2018eyx}. We refer to our minimal 6-parameter setup, slicing the parameter space along the $m_\chi-\alpha_L$ plane for reference values of the dark photon coupling $\alpha_D$, of the common scalar messenger mass parameter $m_\phi = m_S$ and mixing $\eta_s$, and of the mass $m_Q$ for light quark-like dark fermions.  In Fig. \ref{fig:relicvsdd}, along the curves labelled ``relic" the dark matter relic density matches the observed dark matter density. In the ``south-east" direction, \textit{i.e.} towards larger $m_\chi$ and smaller $\alpha_L$, the $\chi$ relic density exceeds the observed dark matter density, assuming that $\alpha_D$ is fixed. In the opposite direction, the $\chi$ relic density is a fraction of the observed dark matter density. These portions of the parameter space could be, in principle, recovered referring to, \textit{e.g.}, non-thermal production of dark matter or non standard cosmological frameworks, see, e.g., \cite{Profumo:2003hq,Arcadi:2011ev}).

In the top-left panel a maximal scalar mixing $\eta_s = 1 - (m_\chi/m_\phi)^2$ has been considered, while in the top-left panel it is tuned to zero; results for two representative values of $\alpha_D$ are displayed, namely $10^{-2}$ (solid lines) and $5 \times 10^{-2}$ (dashed lines), while the other parameters are fixed to $m_Q = \unit[10]{GeV}$ and $m_{S} = \unit[10]{TeV}$. Each isolevel curve for  $\Omega_\chi h^2$ exhibits the features described by the decoupling regimes for $\chi$ discussed in Sec. \ref{sec:relic_numeric}. The upper branch corresponds to region IV, where the $\chi\bar{\chi}$ annihilation to SM leptons controls the final relic density of $\chi$. The vertical branch corresponds to region III where the annihilation to $\gamma_D$ determines the final relic density of $\chi$: note that, in order to have the same relic density, increasing $\alpha_D$ requires increasing $m_\chi$ as well, which is consistent with the expectation from Eq. (\ref{eq:cs}). The remaining branch corresponds to region II, where the final relic density is still determined by the $\gamma_D$ channel, but in the relic density regime given by Eq. (\ref{darkrelic}), where a larger $\alpha_L$ leads to larger $\xi_{\rm f.o.} $ and thus $m_\chi$ must decrease accordingly (since $\langle\sigma v\rangle_{\gamma_D}$ goes as $m_\chi^{-2}$). The branch for region I is not shown in these plots, but would simply correspond to a vertical line at lower values of $\alpha_L$. 

Concerning direct detection limits and projected sensitivities, in the top-left panel of Fig. \ref{fig:relicvsdd}, the XENON1T and DARWIN curves (solid lines corresponding again to $\alpha_D = 10^{-2}$, while dashed lines to $5 \times 10^{-2}$) are driven by the $\gamma_D m$ dipole operator, given that in the large mixing scenario this gives a larger event rate than the $\gamma_D m$ CR operator: we find that a large part of the upper branches with correct value of the relic density is already excluded by current direct detection limits, while a larger portion of region~III will be tested with DARWIN. On the other hand, in the top-right panel $\eta_s = 0$ suppresses the role of the $\gamma_D m$ dipole operator and the $\gamma_D m$ CR operator provides instead the bulk of the direct detection events: while current esperiments do not test this regime, DARWIN will be able to probe the branch with correct relic density in the region~IV snd a portion of the one in region~III. Note however that these results depend to some extent on the assumption of universality in the scalar messenger sector: the displayed direct detection curves would shift to larger values of $\alpha_L$ in case some hierarchy between $\phi$ and $S$ is assumed, with a larger $m_S$ relaxing the direct detection limits, without any significant impact on the result for the relic density of $\chi$, given that the $S$ scalars only interact with SM quarks. 

A tighter connection appears with limits and projected sensitivities when including $\gamma m$ interactions; having artificially switching off $\gamma_D m$ interactions, in the bottom panel of Fig. \ref{fig:relicvsdd} we show the curves stemming from the $\gamma m$ dipole coupling, together again with the relic density isolevel curves in case of both $\eta_s = 1 - (m_\chi/m_\phi)^2$ and $\eta_s=0$ (in this plot we are always the \textit{dipole-dominated} region in Fig. \ref{fig:smm_dmcr}, hence the $\gamma m$ CR coupling plays a minor role). Such results look less constraining than for $\gamma_D m$ interactions, however they do not depend on $m_S$ or $\eta_s$, and hence can be more solidly compared against the relic density lines: we find that XENON1T data are in fact excluding part of the upper branch at small $\eta_s$ which was not tested via the $\gamma_D m$ operators, as well as that DARWIN will have a sizable impact in probing our scenario. 

%%%%%%%%%%%%%%%%%%%%%%%%%%%%%%%%%%%%%%
\section{Summary and conclusions}
%%%%%%%%%%%%%%%%%%%%%%%%%%%%%%%%%%%%%%

It is plausible that the solution to the dark matter problem may be in a context in which, on top of one or more particles accounting for dark matter, there are several extra states and/or extra forces. In this work, we have considered a toy model realization of a multicomponent dark sector, with an additional unbroken U(1) gauge interaction, mediated by a massless dark photon, and with portal interactions between dark fermions and SM fermions through scalar messengers. The model is characterized by: (i) the dark U(1) coupling $\alpha_D$, (ii) the Yukawa-like portal couplings $\alpha_{L,R}$, and (iii) the masses of the scalar messengers and the dark fermions. Despite its simplicity, this model has a rich dynamics and several phenomenological consequences. Its stable relics can provide a significant additional radiation component, as well as match the measured dark matter density in the Universe, with dark components having sizable interactions with ordinary matter as well as non-negligible self-interactions. 

To characterize these features and have reliable estimates of final particle densities and temperatures, we have introduced a properly extended system of coupled Boltzmann equations, which track simultaneously the number density of several particle species, as well as entropy and energy exchanges between the dark and visible sectors. We have solved it numerically, implementing a few procedures allowing for fast - but very accurate - solutions. The target is to have a lepton-like dark fermion $\chi$ as the dark matter candidate, a requirement which selects viable regions in the model parameter space, without however singling out a definite scheme for the dark matter generation in the early Universe. In fact, depending on the strength of the Yukawa portal between visible and dark sector, we have identified four different regimes for the $\chi$ production, ranging from the limit of a WIMP-like scenario in a totally decoupled dark sector, to a FIMP-like generation in case of intermediate coupling, and up to a standard WIMP framework when the two sectors come and stay in kinetic equilibrium all the way through the chemical decoupling of all dark sector species. As a consequence, our framework is not very predictive regarding the mass scale of the dark matter candidate, which we can only point to be rather heavy, in the range, say, $\unit[500]{GeV}$-$\unit[10]{TeV}$, with portal couplings all the way from about $\alpha_L \simeq 10^{-2}$-$1$, down to around $10^{-9}$-$10^{-7}$.

The result on the relic density for lepton-like dark fermion $\chi$ is weakly dependent on the choice for the masses of the lighter quark-like dark fermions $Q$. The latter can have a negligible contribution to the Universe matter density, say below 1\% with respect to the heavy dark lepton contribution, if there is a sizable mass splitting between $\chi$ and $Q$, say $m_Q \lesssim \unit[100]{GeV}$ for $m_\chi \gtrsim \unit[1]{TeV}$. On the other hand, the presence of light dark fermions enters critically in setting the temperature ratio $\xi$ between dark photons and SM photons at the kinetic decoupling between the two sectors; this is one of the most critical observables in our model, since the CMB constraint on the amount of extra radiation in the Universe (usually given in terms of the effective number of neutrino-like species $N_{eff}$) limits $\xi$ to be at most 0.6 (at the 3-$\sigma$ level). For a given portal coupling, constraints on the number and masses of light quark-like dark fermions follow: E.g., in a scenario with  two light dark quarks ($N_Q = 2$) and the early-time temperature ratio initialized to $\xi_0 = 0.1$, the limit on $\xi_{\rm CMB}$ is satisfied at 1-$\sigma$ level for any dark quark mass $m_Q$ if $\alpha_L \lesssim 10^{-7}$; for $\alpha_L  \simeq 10^{-3}$, $m_Q$ lighter than $\unit[10]{GeV}$ ($\unit[2]{GeV}$) are excluded at 1-$\sigma$ (at 2-$\sigma$); if $\alpha_L  \simeq 10^{-1}$, $Q$ lighter than about  $\unit[0.5]{GeV}$ are excluded at more than 3-$\sigma$.

Regarding other constraints on our scenario, we checked its testability with direct detection searches. The elastic scattering of the dark matter candidate $\chi$ on a nucleus can be mainly driven by dipole (dimension 5) or charge radius (dimension 6) interactions mediated by either the SM photon or the dark photon. We have analyzed on general grounds the interplay among the different operators, discussing features in the recoil spectrum and enlightening that long-range effects are not always predominant (as usually assumed in this context). After deriving current limits and projected sensitivities for next-generation detectors in terms of generic dipole and charge radius couplings, we have applied the results to our specific toy model, showing, e.g., that the DARWIN experiment will cover a significant portion of the parameter space in which $\chi$ is a viable dark matter candidate, as well as it will be competitive against the tightest (but model-dependent) constraints at present, including extra contributions to the magnetic dipole moments of leptons and extra cooling of stellar systems.

This exploratory work on a particular realization of a multicomponent dark sector model, can be extended further by investigating more general early-time initial conditions as well as a further extension of the particle content or more general particle interactions. Furthermore aspects are also not discussed here, such as its mapping on precision and accelerator physics, or further cosmological and astrophysical implications, including, e.g., the level of dark matter self-interactions. Some of this directions will be investigated in future work.

\section*{Acknowledgements}
\noindent
The work of J.T.A. and P.U. was partially supported by the research grant ``The Dark Universe: A Synergic Multimessenger Approach" number 2017X7X85K under the program PRIN 2017 funded by the Ministero dell'Istruzione, Università e della Ricerca (MIUR), and by the ``Elusives" European ITN project (H2020-MSCA-ITN-2015//674896-ELUSIVES). M.F. is affiliated with the Physics Department of the University of Trieste, the \textit{Scuola Internazionale Superiore di Studi Avanzati} (SISSA), Trieste, Italy and  the Institute for Fundamental Physics of the Universe (IFPU), Trieste, Italy. The support of all these institutions is gratefully acknowledged.

\appendix

%%%%%%%%%%%%%%%%%%%%%%%%%%%%%%%%%%%%%%
\section{Matrix elements and Sommerfeld enhancement}
%%%%%%%%%%%%%%%%%%%%%%%%%%%%%%%%%%%%%%

In writing  the collision term in the right-hand side of the Boltzmann equations in section \ref{sec:relic}, we need  the amplitude squared of the relevant annihilation and elastic scattering amplitudes. Regarding the annihilation processes, the dark fermions can annihilate to dark photons or SM fermions (see Fig.~\ref{fig:dfdiagrams}). The squared amplitudes in case of $\chi$ (the expressions for $Q$ are specular) are given by
\be
\begin{aligned}
&|\mathcal{M}|^2_{\chi\bar{\chi} \rightarrow 2\gamma_D} =  32\pi^2 \alpha_D^2 Q_\chi^4 \Bigg[\frac{tu-m_\chi^2(3t+u)-m_\chi^4}{(t-m_\chi^2)^2}-\frac{2m_\chi^2(s-4m_\chi^2)}{(t-m_\chi^2)(u-m_\chi^2)} +\frac{tu-m_\chi^2(3u+t)-m_\chi^4}{(u-m_\chi^2)^2}\Bigg] \,, \\
&|\mathcal{M}|^2_{\chi\bar{\chi} \rightarrow l\bar{l}}  = 4\pi^2 \Big[\left(\alpha_L^2+\alpha_R^2\right)(m_\chi^2+ m_l^2-t)^2+8 \alpha_L \alpha_R m_\chi^2 m_l^2\Big]
\left(\frac{1}{t-m_{\phi_+}^2}+\frac{1}{t-m_{\phi_-}^2}\right)^2 \, ,\\
&|\mathcal{M}|^2_{\chi_R\bar{\chi}_R \rightarrow l_L\bar{l}_L} = 16\pi^2 \alpha_L^2 \frac{(m_\chi^2+ m_l^2-t)^2}{(t-m_{\phi}^2)^2} \, ,\\
\end{aligned}
\ee
where $s$, $t$ and $u$ are the standard Mandelstam variables. When computing the pair annihilation cross section of dark fermions we need to include the Sommerfeld enhancement induced by the long-range attractive force mediated by dark photons \cite{sommerfeld} (the importance of this non-perturbative effect in the context of dark matter annihilations was first pointed out by \cite{HisanoSommEnhancement}); for such Coulomb term, the enhancement can be computed analytically and added as a multiplicative factor to the cross section $\sigma_0$ accounting for contact interactions 
\be
\sigma = \sigma_0 S(v) \quad \quad {\rm with}  \quad \quad S(v) =  \frac{\pi\,\alpha_D}{v}  \frac{1}{1-e^{-\pi \alpha_D/v}} \,,
\ee
where $v$ is the velocity of each annihilating species in the center-of-mass frame. As for the elastic scattering processes, the dark fermions can either undergo Compton-like processes with dark photons, or scatter on SM fermions (see Fig.~\ref{fig:elastic}). The squared amplitudes in case of $\chi$ are
\be
\begin{aligned}
&|\mathcal{M}|^2_{\chi\gamma_D \rightarrow \chi\gamma_D} = -32\pi^2 \alpha_D^2 Q_\chi^4 \Bigg[\frac{su-m_\chi^2(3s+u)-m_\chi^4}{(s-m_\chi^2)^2}-\frac{2m_\chi^2(t-4m_\chi^2)}{(s-m_\chi^2)(u-m_\chi^2)} + \frac{su-m_\chi^2(3u+s)-m_\chi^4}{(u-m_\chi^2)^2} \Bigg] \, ,\\
&|\mathcal{M}|^2_{\chi l \rightarrow \chi l} = 4\pi^2 \Big[ \left(\alpha_L^2+\alpha_R^2\right)(m_\chi^2+ m_l^2-u)^2+8\alpha_L \alpha_R m_\chi^2 m_l^2\Big]
\left(\frac{1}{u-m_{\phi_+}^2}+\frac{1}{u-m_{\phi_-}^2}\right)^2 \, ,\\
&|\mathcal{M}|^2_{\chi_R \bar{l}_L \rightarrow \chi_R \bar{l}_L} = 16\pi^2 \alpha_L^2 \frac{(m_\chi^2+ m_l^2-s)^2}{(s-m_{\phi}^2)^2} \, .
\end{aligned}
\ee

%%%%%%%%%%%%%%%%%%%%%%%%%%%%%%%%%%%%%%
\section{Computation of thermal averages}
%%%%%%%%%%%%%%%%%%%%%%%%%%%%%%%%%%%%%%
\label{app:elastic}
In the Boltzmann code developed in Section~\ref{sec:relic} there are several quantities involving thermal averages. Starting with pair annihilation cross sections, a method to efficiently compute $\langle\sigma v\rangle(T)$, as defined in \eq{sigmavdef}, was detailed in \cite{Gondolo:1990dk}: Assuming equilibrium distribution functions with occupation numbers approximated by the exponential in (\ref{mb_occupation}), you can manipulate the numerator by performing a change of integration variables from the two momenta $\vec{p}_1$ and $\vec{p}_2$ to  $E_+ \equiv E_1 + E_2$,  $E_- \equiv E_1 - E_2$ and $s$, with the integral in the first two that can be performed analytically, giving 
\be
\langle\sigma v\rangle(T) \simeq \frac{1}{8m^4 T [K_2(m/T)]^2}\int_{4m^2}^\infty~ds~\sigma(s)\sqrt{s}\left(s-4m^2\right)K_1(\sqrt{s}/T) \, ,
\ee
where $K_1(z)$ and $K_2(z)$ are the modified Bessel functions of order 1 and 2, respectively. The same method can be applied to $\langle\sigma v E\rangle (T)$, see the definition in \eq{sigmavEdef}, obtaining
\be
\langle\sigma vE \rangle(T) \simeq \frac{1}{8m^4 T [K_2(m/T)]^2}\int_{4m^2}^\infty ds~\sigma(s)~s\left(s-4m^2\right)K_2(\sqrt{s}/T) \,.
\ee
Both these expressions are very convenient when coming to their numerical implementation: for any given particle physics model, you can first tabulate the cross sections $\sigma$ as a function of $s$, and then link to such tabulations for a fast computation of thermal averages at any $T$ in the temperature evolution equations.

\begin{figure}[t]
\centering
\includegraphics[scale=0.4]{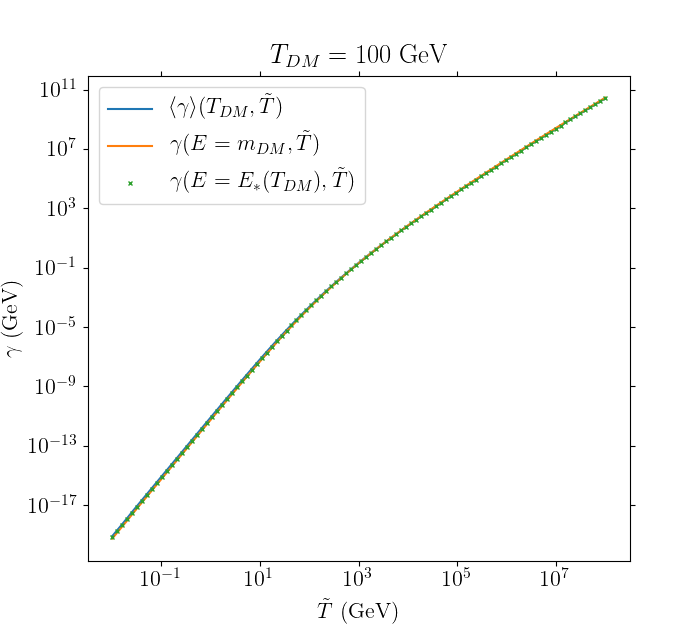}\includegraphics[scale=0.4]{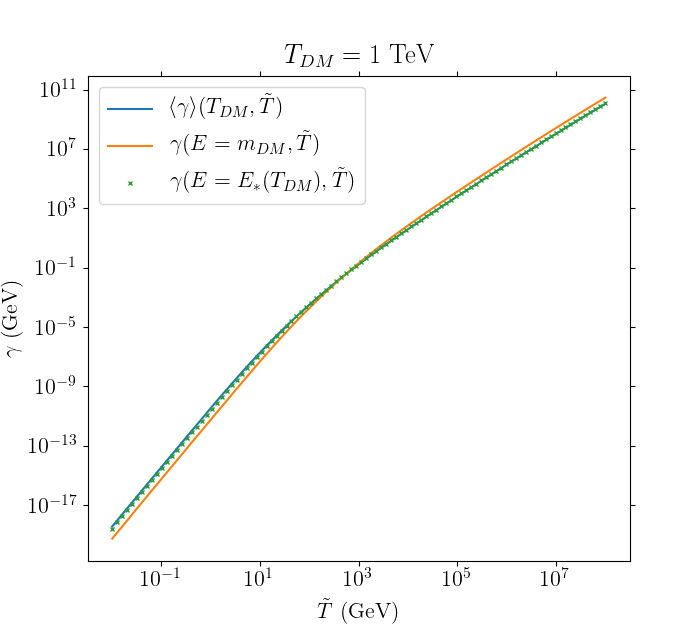}
\includegraphics[scale=0.4]{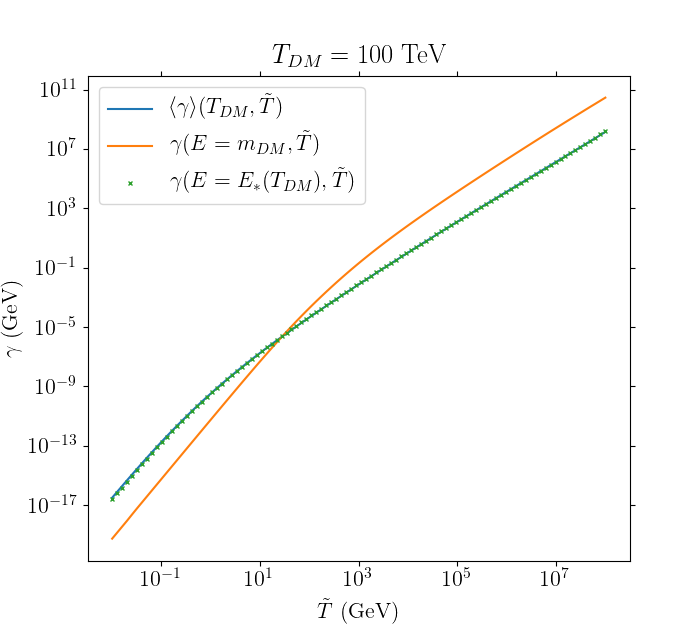}\includegraphics[scale=0.4]{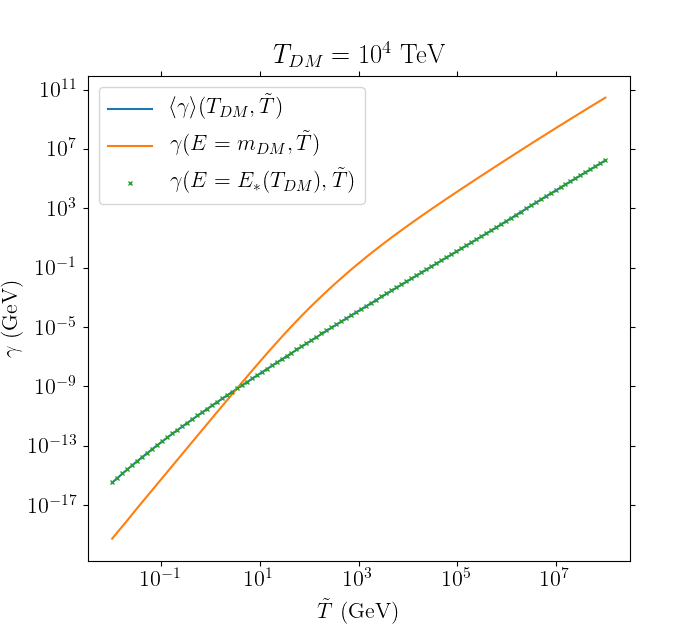}
\caption{Plots of the thermal average of the momentum transfer rate, $\langle\gamma\rangle$ (solid blue curve), the momentum transfer rate evaluated at zero momentum (solid orange curve), and the momentum transfer rate evaluated at the stationary point $E = E_*(T_{DM})$ (green crosses), as functions of the bath temperature $\tilde{T}$. The process being considered here is Compton scattering between dark fermions, with mass $m_{DM} = \unit[1]{TeV}$ and temperature $T_{DM}$, and dark photons which serve as the heat bath. The dark coupling is chosen to be $\alpha_D = 10^{-2}$. Notice the large deviation between the blue and orange curves when $T_{DM} \gg m_{DM}$. On the other hand, there is a good agreement between $\gamma(E = E_*(T_{DM}), \tilde{T})$ and $\langle\gamma\rangle(T_{DM},\tilde{T})$.}
\label{fig:gammaave}
\end{figure}

On the other hand, an analogous shortcut cannot be implemented in thermal averages for momentum transfer rates. Referring generically to the scattering process $i + B \rightarrow i + B$, the thermally averaged momentum transfer rate $\langle\gamma_{iB}\rangle (T_i,T_B)$ is a function of the temperature of both the species $i$ and bath particles $B$, see Eqs.~(\ref{gammarateave}) and (\ref{gammaeitb}), and such dependences cannot be simply factorised, making the implementation in the numerical Boltzmann code CPU-demanding. When investigating the kinetic decoupling of massive dark matter particles $i$ from the heat bath, since this typically occurs in the regime at which the temperature $T_i$ is small compared to the particle mass $m_i$, \cite{Bringmann:2006mu} and \cite{Binder:2017rgn} noticed that the dependence on the particle momentum $p_i$ in $\gamma_{iB}$ can be approximately dropped, thereby allowing to replace $\langle\gamma_{iB}\rangle(T_i,T_B)$ with $\gamma_{iB}(E_i = m_i, T_B)$. When the scattering species is relativistic,
this is not a fair estimate; on the other hand, we can still use it as a guideline for a better approximation: At $T_i \ll m_i$ the occupation number in the integrand at the numerator of the l.h.s. of \eq{gammarateave} is sharply peaked at the stationary point $E_i = m_i$ and $\gamma_{iB}(E_i, T_B)$ simply picks up the contribution coming from the stationary point. On the other hand, when $T_i \gtrsim m_i$, the occupation number has a relatively longer tail at higher energies. The pre-factor $(E_i^2 - m_i^2)^{3/2}$ in the integral cannot be neglected, and, to extract the peak contribution, one has to search for the stationary point of the function
\be
F(E_i) = \frac{E_i}{T_i} - \frac{3}{2}\ln\left(E_i^2 - m_i^2\right),
\ee
which is now at
\be
\label{esad}E_{*}(T_i) = \frac{3T_i}{2} + \sqrt{m_i^2 + \frac{9T_i^2}{4}}.
\ee
Note that going back to the limit $T_i \ll m_i$, you correctly retrieve $E_* = m_i + O(T_i)$. The agreement between $\langle\gamma\rangle(T_i,T_B)$ and $\gamma(E_i=E_*(T_i),T_B)$ is very good, as shown in a sample case in Fig. (\ref{fig:gammaave}). 

%%%%%%%%%%%%%%%%%%%%%%%%%%%%%%%%%%%%%%
%%%%%%%%%%%%%%%%%%%%%%%%%%%%%%%%%%%%%%

\section{Loop calculations}
\label{appendix:loop}
We report here a few details regarding the computation of the $\gamma$ and $\gamma_D$ vertex functions represented by the loop diagrams in Fig. \ref{fig:loopvertex}.  For the $\gamma_D$ vertex function, involving a SM quark $q$ on the external legs and a quark-like dark fermion $Q$ (with $U(1)_D$ charge $Q_Q$) and scalar messengers $S_\pm$ in the loop, we have
\be
\label{dmm_vertex}
\bar{q}(k')\,i\Gamma^\mu_{\gamma_D}q(k) = g_D Q_Q~\bar{q}(k') \sum_{\lambda = \pm}\left[\left(\frac{g_L^2+g_R^2}{2}\right)I_{a,\gamma_D}^\mu\left(m_Q, m_{S_\lambda}, k,q\right)+(\lambda)\, g_L g_R m_Q I_{b,\gamma_D}^\mu\left(m_Q, m_{S_\lambda},k,q\right)\right]q(k)\,,
\ee
where $q^\mu$ is the momentum transfer. For the $\gamma$ vertex function, with a lepton-like dark fermion $\chi$ on the external legs and the corresponding lepton (with $U(1)_{em}$ charge $Q_l$) and messengers scalars $\phi_\pm$ in the loop, we have 
\be
\label{smm_vertex}
\bar{\chi}(k')\,i\Gamma^\mu_{\gamma}\chi(k) = e Q_l~\bar{\chi}(k') \sum_{\lambda = \pm}\left[\left(\frac{g_L^2+g_R^2}{2}\right)I_{a,\gamma}^\mu\left(m_l, m_{\phi_\lambda},k,q\right)+(\lambda)\, g_L g_R m_l I_{b,\gamma}^\mu\left(m_l, m_{\phi_\lambda},k,q\right)\right]\chi(k)\,.
\ee
The functions $I_a$ and $I_b$ are loop integrals, defined as
\be
\begin{aligned}
& I_{a,V}^\mu\left(m_f, m_s,k,q\right) \equiv \int\frac{d^4l}{(2\pi)^4}\left[\frac{(\slashed{l}+\slashed{q})\gamma^\mu \slashed{l}+m_f^2 \gamma^\mu}
{D(l,m_f) \,D(l+q,m_f) \,D(k-l,m_s)}+s_V\frac{(2l+q)^\mu (\slashed{k}-\slashed{l})}{D(l,m_s) \,D(l+q,m_s)\, D(k-l,m_f)}\right]\,, \\
& I_{b,V}^\mu\left(m_f, m_s,k,q\right) \equiv \int\frac{d^4l}{(2\pi)^4} \left[\frac{\gamma^\mu \slashed{l}+(\slashed{l}+\slashed{q})\gamma^\mu}
{D(l,m_f) \,D(l+q,m_f) \,D(k-l,m_s)}+s_V\frac{(2l+q)^\mu}{D(l,m_s) \,D(l+q,m_s)\, D(k-l,m_f)}\right]\,,
\end{aligned}
\ee
where we introduced the function $D(p,m)\equiv p^2-m^2$,  while $s_\gamma = 1$ and $s_{\gamma_D} = -1$. The $s_V$ sign structure is motivated by the form of the interaction Lagrangian in Eqs. (\ref{LLRR}) and (\ref{mix}): a dark fermion and its corresponding messenger scalar must have opposite $U(1)_D$ charges, while a SM fermion and its corresponding messenger scalar must have the same $U(1)_{em}$ charge.

The additional contribution to the magnetic dipole moment of SM leptons predicted in our dark sector framework is computed from the $\gamma$ vertex function having SM leptons as external legs and a loop with $\chi$ and $\phi_\pm$. We find
\be
\bar{l}(k')\,i\Gamma^\mu_{\gamma}l(k) = e Q_l~\bar{l}(k') \sum_{\lambda = \pm}
\left[\left(\frac{g_L^2+g_R^2}{2}\right)J_a^\mu\left(m_\chi, m_{\phi_\lambda}, k,q\right)+(\lambda)\, g_L g_R m_\chi J_b^\mu\left(m_\chi, m_{\phi_\lambda}, k,q\right)\right] l(k),
\ee
where
\be
\label{Ja&b}
\begin{aligned}
& J_a^\mu\left(m_f, m_s,k,q\right) \equiv \int\frac{d^4l}{(2\pi)^4}~\frac{(2l+q)^\mu (\slashed{k}-\slashed{l})}{D(l,m_s)\, D(l+q,m_s)\, D(k-l,m_f)} \\
& J_b^\mu\left(m_f, m_s,k,q\right) \equiv \int\frac{d^4l}{(2\pi)^4}~\frac{(2l+q)^\mu}{D(l,m_s)\, D(l+q,m_s)\, D(k-l,m_f)}.
\end{aligned}
\ee
Notice that only one term appears in these loop factors; this follows from the fact that the messenger scalars have SM quantum numbers, while the dark leptons do not. 

Loop factors are computed using the standard Feynman trick to rewrite denominators. A UV cut-off needs to be introduced since $I_a$, $I_b$ and $J_a$ are logarithmically divergent;
as a renormalization condition, the vertex function at zero momentum transfer $q^\mu$ is subtracted to each vertex function. Finally, dipole and charge-radius terms are extracted at leading order in a momentum expansion of the vertex functions. The general structure is
\be
\bar{f}(k')~\Gamma_{V}^\mu f(k) = \frac{d_{M,V}^{(f)}}{\Lambda_{D,V}^{(f)}}\left[\bar{f}(k')~i\sigma^{\mu\nu}q_\nu~f(k)\right] + q^2\frac{c_{CR,V}^{(f)}}{[\Lambda_{CR,V}^{(f)}]^2}\left[\bar{f}(k')\gamma^\mu f(k)\right]\,.
\ee
Following the notation introduced in Eqs.~(\ref{eq:M&CR}) and (\ref{eq:M&CR2}), we find
\be
\begin{aligned}
& F_{D,\gamma}^{(\chi)}\left(m_l, m_{\phi_-}, m_{\phi_+}\right) = -2\left[I_1\left(\frac{m_{\phi_+}}{m_{\phi_-}}, \frac{m_l}{m_{\phi_-}}\right) + I_1\left(1, \frac{m_l}{m_{\phi_-}}\right)\right]
\\
& F_{CR,\gamma}^{(\chi)}\left(m_l, m_{\phi_-}, m_{\phi_+}\right) = -\frac{1}{6}\left[I_2\left(\frac{m_{\phi_+}}{m_{\phi_-}}, \frac{m_l}{m_{\phi_-}}\right) + I_2\left(1, \frac{m_l}{m_{\phi_-}}\right)\right] \\
& F_{D,\gamma_D}^{(q)}\left(m_Q, m_{S_-}, m_{S_+}\right) = -2\left[I_3\left(\frac{m_{S_+}}{m_{S_-}}, \frac{m_Q}{m_{S_-}}\right) - I_3\left(1, \frac{m_Q}{m_{S_-}}\right)\right]
\\
& F_{CR,\gamma_D}^{(q)}\left(m_Q, m_{S_-}, m_{S_+}\right) = \frac{1}{6}\left[I_4\left(\frac{m_{S_+}}{m_{S_-}}, \frac{m_Q}{m_{S_-}}\right) + I_4\left(1, \frac{m_Q}{m_{S_-}}\right)
\right]\\
& F_{D,\gamma}^{(l)}\left(m_\chi, m_{\phi_-}, m_{\phi_+}\right) = 2\left[I_1\left(\frac{m_{\phi_+}}{m_{\phi_-}}, \frac{m_\chi}{m_{\phi_-}}\right) - I_1\left(1, \frac{m_\chi}{m_{\phi_-}}\right)\right]\,, 
\end{aligned}
\ee
where we introduced the following functions
\begin{eqnarray}
I_1(a,b) &\equiv& \frac{1}{2\left(b^2-a^2\right)^3}\left[-2a^2b^2\ln\left(\frac{b^2}{a^2}\right) + b^4 - a^4\right] \nonumber\\
I_2(a,b) &\equiv& \frac{1}{2\left(b^2-a^2\right)^4}\left[2\left(b^6-a^6\right)\ln\left(\frac{b^2}{a^2}\right) - 3\left(b^2-a^2\right)^2\left(b^2+a^2\right)\right] \nonumber\\
I_3(a,b) &\equiv& \frac{1}{\left(b^2-a^2\right)^3}\left[\left(b^2-a^2\right)^2 - a^2\left(b^2-a^2\right)\ln\left(\frac{b^2}{a^2}\right)\right]\\
I_4(a,b) &\equiv& \frac{1}{6\left(b^2-a^2\right)^4}\left[6\left(b^6+a^6\right)\ln\left(\frac{b^2}{a^2}\right)-4\left(b^6-a^6\right) - 9\left(b^2-a^2\right)^3\right] \,. \nonumber
\end{eqnarray}

%%%%%%%%%%%%%%%%%%%%%%%%%%%%%%%%%%%%%%

\bibliographystyle{kp.bst}
\bibliography{final}

\end{document}